\documentclass{article}
\usepackage{latexsym}
\usepackage[dvipdfmx]{graphicx}
\usepackage{caption}
\usepackage{mathrsfs}
\usepackage{amsmath,amssymb}
\usepackage{amsthm}
\usepackage{mathtools}
\usepackage{breqn}
\usepackage{bm}
\usepackage{ascmac}
\usepackage{braket}
\usepackage{geometry}
\theoremstyle{plain}
\newtheorem{thm}{Theorem}[subsection]
\newtheorem{lem}[thm]{Lemma}
\newtheorem{prop}[thm]{Proposition}

\newtheorem{dfn}[thm]{Definition}
\newtheorem{asm}[thm]{Assumption}
\newtheorem{rem}[thm]{Remark}

\mathtoolsset{showonlyrefs}
\newcommand{\esssup}{\mathop{\textup{ess~sup}}\limits}

\numberwithin{equation}{subsection}
\providecommand{\keywords}[1]
{
  \small	
  \textbf{{Keywords~}} mean field game, asset pricing, optimal consumption-investment problem, exponential utility, market clearing
}
\title{\textbf{\Large{Mean field equilibrium asset pricing model with habit formation}}\thanks{
Forthcoming in {\it Asia-Pacific Financial Markets}. All content presented in this research is solely the opinion of the authors and does not represent the views of any institution. The authors disclaim any responsibility or liability for any losses or damages resulting from the use of the research.}}
\author{Masaaki Fujii\thanks{mfujii@e.u-tokyo.ac.jp, Graduate School of Economics, The University of Tokyo, 7-3-1, Hongo, Bunkyo-ku, Tokyo, Japan.},~~~~~~Masashi Sekine\thanks{sekinemasashi@g.ecc.u-tokyo.ac.jp, Graduate School of Economics, The University of Tokyo, 7-3-1, Hongo, Bunkyo-ku, Tokyo, Japan.}}
\date{November 12, 2024}

\begin{document}

\maketitle

\begin{abstract}
  This paper presents an asset pricing model in an incomplete market involving a large number of heterogeneous agents, based on the mean field game theory. The primary objective of this study is to derive the equilibrium risk premium process endogenously by considering the optimal consumption-investment problem and the market clearing condition.
  In the model, we incorporate habit formation in consumption preferences, which has been widely used to explain various phenomena in financial economics. In order to characterize the market-clearing equilibrium, we derive a quadratic-growth mean field backward stochastic differential equation (BSDE) and study its well-posedness and asymptotic behavior in the large population limit. 
  Additionally, we introduce an exponential quadratic Gaussian reformulation of the asset pricing model, in which the solution is obtained in a semi-analytic form. 
\end{abstract}

\keywords
\tableofcontents
\section{Introduction} \label{Section 1}
\subsection{Preliminary}
Asset pricing theory plays a crucial role in financial economics as it investigates how asset prices are determined through market interactions. 
The fundamental objective of the theory is to establish the equilibrium price at which the supply of assets matches its demand. 
See, for example, Back \cite{backAssetPricingPortfolio2017} and Munk \cite{Munk} for details. Karatzas \& Shereve \cite{karatzas_methods_1998} also offers comprehensive descriptions for the equilibrium asset pricing in complete markets. 
The continuous-time stochastic equilibrium pricing problems in incomplete markets are being actively researched as there are still many open issues. In recent years, numerous research efforts are devoted to show the existence of equilibrium solutions in incomplete markets under various conditions. See, for example, Christensen \& Larsen \cite{ChristensenLarsen}, Cuoco \& He \cite{CuocoHe} and \v{Z}itkovi\'{c} \cite{Zitkovic} and references therein. Let us further refer to Jarrow \cite{jarrow2018continuous} [Part III] for a well-integrated review on this subject.

    The mean field game theory, first introduced by Lasry \& Lions \cite{lasryMeanFieldGames2007} and Huang, Malhame \& Caines \cite{huangLargePopulationStochastic2006}, has emerged as a powerful framework for studying multi-agent games. Traditional approaches to such games usually result in intractable problems due to complex interactions among agents. 
The mean field game theory overcomes this challenge by replacing such problems with a stochastic control problem of a single representative agent and a fixed point problem. Lasry \& Lions \cite{lasryMeanFieldGames2007} and Huang, Malhame \& Caines \cite{huangLargePopulationStochastic2006} presented an analytic approach, in which they show that the problem can be framed as two highly coupled nonlinear partial differential equations. Meanwhile, Carmona \& Delarue \cite{carmonaProbabilisticAnalysisMeanField2013, carmonaForwardBackwardStochastic2015} introduced the probabilistic approach to the mean field problem employing forward-backward stochastic differential equations (FBSDEs) of McKean-Vlasov type in lieu of a system of partial differential equations.
The solution of these mean field equations is known to provide an $\varepsilon$-Nash equilibrium of the original game with a large but a finite number of agents. The probabilistic approach is extensively covered in two volumes of monographs Carmona \& Delarue \cite{carmonaProbabilisticTheoryMean2018,carmonaProbabilisticTheoryMean2018a}, offering thorough details and applications.
Furthermore, the mean field game theory has been applied to various studies in the field of financial economics. For instance, Fu, Su \& Zhou \cite{Fu2020MeanFE}, Fu \& Zhou \cite{fu_Mean_field_portfolio_games} and Fu \cite{fu_Mean_field_portfolio_games_consumption} propose stochastic games among multiple agents with exponential or power utility competing in a relative performance criterion. 
These examples illustrate the relevance and usefulness of mean field game theory in tackling complex interactive problems, providing valuable insights in financial economics and related fields.

    In recent years, there have been an increasing number of studies on asset pricing theory adopting the mean field game approach. They aim to determine the equilibrium price process based on the optimal behavior of the market participants under the market clearing condition. One notable area of interest has been the investigation of price formation in electricity markets.
Shrivats, Firoozi \& Jaimungal \cite{SREC} employs FBSDEs of McKean-Vlasov type to study pricing model in Solar Renewable Energy Certificate (SREC) markets and Firoozi, Shrivats \& Jaimungal \cite{Firoozi_et_al} studies principal agent mean field games in REC markets. Gomes \& S\'{a}ude \cite{Gomes_MFG} develops a deterministic price formation model in which agents can both store and trade electricity. Gomes, Gutierrez \& Ribeiro \cite{Gomes_rand_MFG} extends this model by considering the randomness on the supply side and \cite{Gomes_MFG_noise} deals with a price formation of commodities with stochastic production.
In the realm of financial economics, Evangelista, Saporito \& Thamsten \cite{evangelista2022price} develops a mean field game theoretic model of asset pricing with consideration of liquidity issues. Fujii \& Takahashi \cite{fujiiMeanFieldGame2022, Fujii-Takahashi_strong} present a mean field pricing model for securities under stochastic order flows and \cite{fujii2022equilibrium} provide its extension with a major player.  Fujii \cite{Fujii-equilibrium-pricing}  develops a price formation model in which the market participants consist of two groups: cooperative and non-cooperative ones.
Moreover, Fujii \& Sekine \cite{fujiiMeanFieldEquilibriumPrice2023a} studies an mean field equilibrium pricing model in an incomplete market participated by heterogeneous agents with exponential utility, but without considering agents' consumption. 

    The main contribution of this paper is an extension of the aforementioned work \cite{fujiiMeanFieldEquilibriumPrice2023a}. This paper aims to further explore the equilibrium pricing model in an incomplete market with heterogeneous agents, taking the agents' consumption behavior and habit formation into account. The research of consumption habit formation has been a fundamental and classical subject in financial economics. The existence of the habit formation relaxes the assumption of time-separable utility functions by making the utility dependent not only on the current level of consumption but also on the agent's accumulated stock of past consumption. 
Early studies include, for instance,  \cite{constantinides_habit_1990,detemple_asset_1991,Pollak1970HabitFA,RyderHeal}. Our model specifically incorporates heterogeneity among agents in various aspects, including their initial wealths, initial consumption habits, liabilities and coefficients of risk aversion.
In this paper, we start from the utility maximization problem of a single agent, which draws inspiration from the work Hu, Imkeller \& M\"{u}ller \cite{huUtilityMaximizationIncomplete2005a}, and derive the relevant BSDE. After proving its well-posedness, we construct the market risk premium process endogenously under the market clearing condition by introducing the mean field BSDE. As we have done in \cite{fujiiMeanFieldEquilibriumPrice2023a}, we prove its well-posedness using the method proposed by Tevzadze \cite{tevzadzeSolvabilityBackwardStochastic2008} with additional assumptions on the size of the parameters. We then verify that the risk premium process, expressed by its solution, indeed clears the market in the large population limit. Another contribution of this paper is to offer an exponential quadratic Gaussian (EQG) formulation of the model, in which a solution to the mean field BSDE can be characterized by a system of ordinary differential equations.
Since the EQG model provides a semi-analytic solution, it would allow detailed numerical studies in the future works.

    This paper consists of five sections and an appendix. In Section \ref{Section 1}, after providing the introduction, we give the notations for frequently used sets and spaces. 
In Section \ref{Section 2}, we offer a mathematical formulation of the financial market and solve the optimal consumption-investment problem for a single agent. 
In Section \ref{Section 3}, we derive a mean field BSDE whose driver has a quadratic growth in both stochastic integrand and its conditional expectation and prove that it has a bounded solution under additional assumptions. We also verify that its solution does characterize the financial market in equilibrium in the large population limit. 
Furthermore, in Section \ref{Section 4}, we introduce the EQG framework and prove each result corresponding to Section 3.
We conclude the paper with a brief summary and a suggestion for possible extensions in Section \ref{Section 5}. 

\subsection{Notations}
In this paper, we shall work on a finite time interval $[0,T]$ for some $T>0$. For a given filtered probability space with usual conditions $(\Omega,\mathcal{F},\mathbb{P},\mathbb{F}~(:=(\mathcal{F}_t)_{t\in[0,T]}))$ and a vector space $E$ over $\mathbb{R}$, we use the following notations to describe frequently used sets and function spaces.\\

\noindent
(1) $\mathcal{T}(\mathbb{F})$ is a set of all $\mathbb{F}$-stopping times with values in $[0,T]$.\\

\noindent
(2) $\mathbb{L}^0(\mathcal{F},E)$ is a set of $E$-valued $\mathcal{F}$-measurable random variables. \\

\noindent
(3) $\mathbb{L}^2(\mathbb{P},\mathcal{F},E)$ is a set of $E$-valued $\mathcal{F}$-measurable random variables $\xi$ satisfying $\|\xi\|_2:=\mathbb{E}^{\mathbb{P}}[|\xi|^2]^{\frac{1}{2}}<\infty$.\\

\noindent
(4) $\mathbb{L}^\infty(\mathbb{P},\mathcal{F},E)$ is a set of $E$-valued $\mathcal{F}$-measurable random variables $\xi$ satisfying $\|\xi\|_\infty:=\esssup_{\omega\in\Omega}|\xi(\omega)|<\infty$.\\

\noindent
(5) $\mathbb{L}^0(\mathbb{F},E)$ is a set of $E$-valued $\mathbb{F}$-progressively measurable stochastic processes. \\

\noindent
(6) $\mathbb{H}^2(\mathbb{P},\mathbb{F},E)$ is a set of $E$-valued $\mathbb{F}$-progressively measurable stochastic processes $X$ satisfying
\[
\|X\|_{\mathbb{H}^2}:=\mathbb{E}^{\mathbb{P}}\Bigl[\int_0^T |X_t|^2 dt\Bigr]^{\frac{1}{2}}<\infty.
\]

\noindent
(7) $\mathbb{L}^\infty(\mathbb{P},\mathbb{F},E)$ is a set of $E$-valued $\mathbb{F}$-progressively measurable stochastic processes $X$ satisfying
\[
\|X\|_{\mathbb{L}^\infty}:=\esssup_{(t,\omega)\in[0,T]\times\Omega}|X_t(\omega)| ~<\infty.
\]

\noindent
(8) $\mathbb{H}^2_{\mathrm{BMO}}(\mathbb{P},\mathbb{F},E)$ is a set of $E$-valued $\mathbb{F}$-progressively measurable stochastic processes $X$ satisfying
\[
\|X\|_{\mathbb{H}^2_{\mathrm{BMO}}}:=\sup_{\tau\in\mathcal{T}(\mathbb{F})} \Bigl\| \mathbb{E}^{\mathbb{P}}\Bigl[\int_\tau^T |X_t|^2dt | \mathcal{F}_\tau\Bigr]^{\frac{1}{2}}\Bigr\|_\infty <\infty,
\]
where $\|\cdot\|_{\infty}$ denotes the $\mathbb{P}$-essential supremum as in (4).\\

\noindent
(9) $\mathbb{S}^2(\mathbb{P},\mathbb{F},E)$ is a set of $E$-valued $\mathbb{F}$-progressively measurable continuous stochastic processes $X$ satisfying
\[
\|X\|_{\mathbb{S}^2}:=\mathbb{E}^{\mathbb{P}}\Bigl[\sup_{t\in[0,T]}|X_t|^2\Bigr]^{\frac{1}{2}}<\infty.
\]

\noindent
(10) $\mathbb{S}^\infty(\mathbb{P},\mathbb{F},E)$ is a set of $E$-valued $\mathbb{F}$-progressively measurable continuous stochastic processes $X$ satisfying
\[
\|X\|_{\mathbb{S}^\infty}:=\esssup_{(t,\omega)\in[0,T]\times\Omega}|X_t(\omega)| ~< \infty.
\]

\noindent
(11) $\mathcal{C}([0,T],E)$ is a set of continuous functions $f:[0,T]\to E$. \\

\noindent
(12) $\mathcal{C}^1([0,T],E)$ is a set of once continuously differentiable functions $f:[0,T]\to E$. \\

\noindent
(13) We set $\mathbb{R}^n_+:=\{x\in\mathbb{R}^n; x\geq 0\}$ and $\mathbb{R}^n_{++}:=\{x\in\mathbb{R}^n; x> 0\}$ for $n\in\mathbb{N}$. Also, $\mathbb{M}_n$ is a set of real symmetric matrices of size $n\times n$.\par
For (1) to (12), we may omit the arguments such as $(\mathbb{P},\mathcal{F},\mathbb{F},E)$ if obvious. Throughout the paper, the symbol $C$ represents a general nonnegative constant which may change line by line. Also, the argument $\omega\in\Omega$ is usually omitted when there is no risk of misinterpretation.

\section{Optimal consumption-investment problem for a single agent} \label{Section 2}
In this section, we investigate the optimal consumption-investment problem for a single agent (whom we shall call ``agent-1'' hereafter). 
We basically follow the same line of arguments as in Fujii \& Sekine \cite{fujiiMeanFieldEquilibriumPrice2023a} and adopt the technique developed by Hu, Imkeller \& M\"{u}ller \cite{huUtilityMaximizationIncomplete2005a}. In this work, however, we take an agent's consumption and habit formation into consideration. As we shall see, this extension requires a clever choice of supermartingale processes that are needed to verify the optimality.

\subsection{The market and the utility function}

To formulate the optimization problem for agent-1, let us first introduce the relevant probability spaces.\\

\noindent
(1) We denote by $(\Omega^0,\mathcal{F}^0,\mathbb{P}^0)$ a complete probability space with complete and right-continuous filtration $\mathbb{F}^0:=(\mathcal{F}^0_t)_{t\in[0,T]}$ generated by a $d_0$-dimensional standard Brownian motion $W^0:=(W^0_t)_{t\in[0,T]}$ with $\mathcal{F}^0 := \mathcal{F}^0_T$. $(\Omega^0,\mathcal{F}^0,\mathbb{P}^0)$ is used to describe the randomness of the financial market. 
Moreover, we denote by $(\Omega^1,\mathcal{F}^1,\mathbb{P}^1)$ a complete probability space with complete and right-continuous filtration $\mathbb{F}^1:=(\mathcal{F}^1_t)_{t\in[0,T]}$ generated by a $d$-dimensional standard Brownian motion $W^1:=(W^1_t)_{t\in[0,T]}$ and a $\sigma$-algebra $\sigma(\xi^1,\gamma^1,\beta^1,X^1_0,F^1_0)$, where the completion of the latter gives $\mathcal{F}^1_0$. We set $\mathcal{F}^1 := \mathcal{F}^1_T$.
Here, $\xi^1$, $X^1_0$ and $F^1_0$ are $\mathbb{R}$-valued, bounded random variables and $\gamma^1$ and $\beta^1$ are $\mathbb{R}_{++}$-valued bounded random variables. $(\Omega^1,\mathcal{F}^1,\mathbb{P}^1)$ is used to describe the idiosyncratic environment for agent-1.\\
 
\noindent
(2) We denote by $(\Omega^{0,1},\mathcal{F}^{0,1},\mathbb{P}^{0,1})$ a complete probability space over $\Omega^{0,1} := \Omega^0 \times \Omega^1$. Here, $(\mathcal{F}^{0,1},\mathbb{P}^{0,1})$ is the completion of $(\mathcal{F}^0 \otimes \mathcal{F}^1,\mathbb{P}^{0}\otimes \mathbb{P}^{1})$ and $\mathbb{F}^{0,1}:=(\mathcal{F}^{0,1}_t)_{t\in[0,T]}$ denotes the complete and right continuous augmentation of $(\mathcal{F}_t^0 \otimes \mathcal{F}_t^1)_{t\in[0,T]}$.\\ 

We set $\mathcal{T}^{0,1}:=\mathcal{T}(\mathbb{F}^{0,1})$ and $\mathcal{T}^{0}:=\mathcal{T}(\mathbb{F}^{0})$ for notational simplicity. The market dynamics and the idiosyncratic environment of agent-1 are modelled on the filtered probability space $(\Omega^{0,1},\mathcal{F}^{0,1},\mathbb{P}^{0,1},\mathbb{F}^{0,1})$. 
Whenever we introduce random variables on a marginal probability space, we identify them with their natural extension to the product space. For example, we use the same symbol $X$ for a random variable $X(\omega^0)$ defined on $(\Omega^0,\mathcal{F}^0,\mathbb{P}^0)$ and its natural extension $X(\omega^0,\omega^1):=X(\omega^0)$ defined on $(\Omega^{0,1},\mathcal{F}^{0,1},\mathbb{P}^{0,1})$.
In this section, we write $\mathbb{E}[\cdot]$ instead of $\mathbb{E}^{\mathbb{P}^{0,1}}[\cdot]$ unless otherwise stated. \par
We now introduce the market dynamics and its properties in the following assumption.
\begin{asm}~\\
    \label{asm1}
    \textup{(i)} The risk-free interest rate is zero.\\
    \textup{(ii)} There are $n\in\mathbb{N}$ non-dividend paying risky stocks whose price dynamics, represented by an $n$-dimensional vector, is given by
    \begin{equation}
        \begin{split}
            \label{stock price}
            S_t&= S_0 + \int_0^t \mathrm{diag}(S_r)(\mu_rdr + \sigma_r dW^0_r),~~~t\in[0,T],
        \end{split}
    \end{equation}
    where $S_0\in\mathbb{R}^n_{++}$, $\mu := (\mu_t)_{t\in[0,T]}\in\mathbb{H}^2_{\mathrm{BMO}}(\mathbb{P}^{0},\mathbb{F}^0,\mathbb{R}^n)$ and $\sigma :=(\sigma_t)_{t\in[0,T]}\in\mathbb{L}^\infty(\mathbb{P}^0,\mathbb{F}^0,\mathbb{R}^{n\times d_0})$. $S_0$ is an $n$-dimensional vector representing the initial stock prices. Moreover, we assume that the process $\sigma$ is of full rank and satisfies
    \[
        \underline{\lambda}I_n\leq (\sigma_t\sigma_t^\top)\leq\overline{\lambda}I_n,~~~~dt\otimes \mathbb{P}^0\text{-}\mathrm{a.e.}
    \]
    for some positive constants $0<\underline{\lambda}<\overline{\lambda}$ and an identity matrix of size $n$, denoted by $I_n$. We set $n\leq d_0$ so that the financial market is incomplete in general.
\end{asm}

Under this assumption, the process $(\sigma_t\sigma_t^\top)_{t\in[0,T]}$ is regular and the risk premium process $\theta:=(\theta_t)_{t\in[0,T]}$ is defined by $\theta_t = \sigma_t^\top(\sigma_t\sigma_t^\top)^{-1}\mu_t\in\mathbb{H}^2_{\mathrm{BMO}}(\mathbb{P}^{0},\mathbb{F}^0,\mathbb{R}^{d_0})$. Note that $\theta_t\in\mathrm{Range}(\sigma_t^\top)=\mathrm{Ker}(\sigma_t)^\perp$. It is worth mentioning that by having $\theta\in\mathbb{H}^2_{\mathrm{BMO}}$, we can change the probability measure $\mathbb{P}^0$ to the risk-neutral measure $\mathbb{Q}$, which is defined by 
\begin{equation}
    \label{risk-neutral}
    \Bigl.\frac{d\mathbb{Q}}{d\mathbb{P}^{0,1}}\Bigr|_{\mathcal{F}_t} = \mathcal{E}\Bigl(-\int_0^\cdot \theta_s^\top dW^0_s \Bigr)_t,~~t\in[0,T].
\end{equation}
This ensures the well-posedness of the stock price process \eqref{stock price}, even though $\mu$ is unbounded (See Kazamaki \cite{kazamaki_sufficient_1979}).

\begin{dfn}
    \label{subspace-L}
    For each $s\in[0,T]$, let us denote by $L_s := \{u^\top\sigma_s ; u\in\mathbb{R}^n\}$ the linear subspace of $\mathbb{R}^{1\times d_0}$ spanned by the $n$ row vectors of $\sigma_s$. Furthermore, we define a map $\Pi_s:\mathbb{R}^{1\times d_0}\to L_s$ as an orthogonal projection onto $L_s$.
\end{dfn}

By its construction, we have $\theta_s^\top\in L_s$ for every $s\in[0,T]$.
\begin{rem}
    For notational convenience, we shall write
    \[
        Z^\|_s := \Pi_s(Z_s) ,~~~Z^\perp_s := Z_s-\Pi_s(Z_s),~~~s\in[0,T]
    \]
    for an $\mathbb{R}^{1\times d_0}$-valued progressively measurable process $Z$. Note that the process $(Z_s^\|)_{s\in[0,T]}$ is also progressively measurable by Karatzas \& Shreve \cite{karatzas_methods_1998} [Lemma 4.4].
\end{rem}

Now, we shall model the idiosyncratic environment of agent-1 through a 5-tuple $(\xi^1,\gamma^1,\beta^1,X^1_0,F^1)$.
\begin{asm}~\\
    \label{asm2}
    \textup{(i)} $\xi^1$ is an $\mathbb{R}$-valued, bounded, and $\mathcal{F}^1_0$-measurable random variable representing the initial wealth of agent-1.\\
    \textup{(ii)} $\gamma^1$ is an $\mathbb{R}$-valued, bounded, and $\mathcal{F}^1_0$-measurable random variable satisfying $\underline{\gamma}\leq\gamma^1\leq\overline{\gamma}$ with some positive constants $0<\underline{\gamma}\leq\overline{\gamma}$. $\gamma^1$ is the coefficient of absolute risk aversion of agent-1 with respect to his/her net wealth. \\
    \textup{(iii)} $\beta^1$ is an $\mathbb{R}$-valued, bounded, and $\mathcal{F}^1_0$-measurable random variable satisfying $\underline{\beta}\leq\beta^1\leq\overline{\beta}$ with some positive constants $0<\underline{\beta}\leq\overline{\beta}$. $\beta^1$ is the coefficient of absolute risk aversion of agent-1 with respect to his/her consumption level. \\
    \textup{(iv)} $X^1_0$ is an $\mathbb{R}$-valued, bounded, and $\mathcal{F}^1_0$-measurable random variable representing agent-1's initial stock of habits. \\
    \textup{(v)} $F^1:=(F^1_t)_{t\in[0,T]}$ is an $\mathbb{R}$-valued, bounded, and $\mathbb{F}^{0,1}$-progressively measurable process. For each $t\in[0,T]$, $F^1_t$ represents the amount of liability at time $t$ of agent-1.\\
    \textup{(vi)} $\rho:=(\rho_t)_{t\in[0,T]}$ is an $\mathbb{R}$-valued, bounded, and $\mathbb{F}^{0}$-progressively measurable process. The process $\rho$ represents the habit trend influenced by the market shocks. \\
    \textup{(vii)} Agent-1 is a price taker; agent-1 must accept the prevailing prices as he/she lacks the market share to impact the market price.
\end{asm}

The trading and consumption strategies of agent-1 are denoted by $(\pi,c)$, where $\pi:=(\pi_t)_{t\in[0,T]}$ is an $\mathbb{R}^n$-valued, $\mathbb{F}^{0,1}$-progressively measurable process representing the amount of money invested in $n$ stocks and $c:=(c_t)_{t\in[0,T]}$ is an $\mathbb{R}$-valued, $\mathbb{F}^{0,1}$-progressively measurable process representing agent-1's consumption\footnote{We do not forbid the process $c$ having negative values in order to make the analysis simple. The negative $c$ may be interpreted as, for example, ``net consumption'', i.e. consumption minus labour income.} process. 
The wealth process of agent-1 with strategy $(\pi,c)$ is then given by
\[
    \mathcal{W}^{1,(\pi,c)}_t = \xi^1 + \sum_{j=1}^n \int_0^t \frac{\pi_{j,s}}{S^j_s}dS^j_s - \int_0^t c_sds =\xi^1 + \int_0^t (\pi_s^\top\sigma_s\theta_s - c_s)ds + \int_0^t \pi_s^\top\sigma_s dW_s^0.
\]

We now formulate the utility maximization problem of agent-1 as follows: agent-1 solves
\[
    \sup_{(\pi,c)\in\mathbb{A}^1} U^1(\pi,c)
\]
subject to
\[
    \mathcal{W}^{1,(\pi,c)}_t =\xi^1 + \int_0^t (\pi_s^\top\sigma_s\theta_s - c_s)ds + \int_0^t \pi_s^\top\sigma_s dW_s^0,~~~t\in[0,T],
\]
where $\mathbb{A}^1$ is a set of admissible strategies for agent-1, whose definition is to be given and $U^1:\mathbb{A}^1\to\mathbb{R}$ is the utility function defined by
\begin{equation}
    \begin{split}
        \label{utility}
        U^1(\pi,c) := \mathbb{E}\Bigl[-\exp\Bigl(-\delta T-\gamma^1(\mathcal{W}^{1,(\pi,c)}_T-F^1_T)\Bigr) -a \int_0^T \exp\Bigl(-\delta t-\gamma^1(\mathcal{W}^{1,(\pi,c)}_t-F^1_t)-\beta^1(c_t-X_t^{1,c})\Bigr)dt\Bigr]
    \end{split}
\end{equation}
for some constants $a,\delta>0$ representing the weight of the running utility with respect to the terminal utility and the discount rate, respectively. Here, $X^{1,c}$ represents the agent-1's consumption habits defined by a mean-reverting process
\begin{equation}
    \begin{split}
        \label{habit}   
        X_t^{1,c} = X^1_0 + \int_0^t \{-\kappa(X^{1,c}_s-\rho_s) + b(c_s-\rho_s)\}ds,~~~t\in[0,T]
    \end{split}
\end{equation}
for some constants $b, \kappa>0$. By a simple calculation, we can write it in an explicit form as
\[
    X_t^{1,c} = e^{-\kappa t}X^1_0 + \int_0^t e^{-\kappa(t-s)} \{b c_s + (\kappa-b)\rho_s\}ds,~~~t\in[0,T].
\]

\begin{rem}
    The economic interpretation of the habit process $X^{1,c}$ and the utility function $U^1$ is as follows.\\
    \textup{(i)} The consumption habit $X^{1,c}_t$ is determined by the accumulation of past private consumption $(c_s)_{s\in[0,t]}$ and the given consumption trend $(\rho_s)_{s\in[0,t]}$ in the market. In particular, a higher level of past consumption increases the agent's current habit by having $b>0$. The size of the parameter $\kappa>0$ determines the rate at which the consumption habit decays. \\
    \textup{(ii)} The term $c-X^{1,c}$ in the running utility means that the agent evaluates the current consumption level relative to his/her habit. Importantly, the agent's preference is no longer time-separable as the past consumption level has an effect on the current consumption choice. \\
    \textup{(iii)}  The amount of net asset $\mathcal{W}^{1,(\pi,c)}-F^1$ enters both in the agent-1's running and terminal preferences. Note that the low portfolio performance $\mathcal{W}^{1,(\pi,c)}_t-F^1_t<0$ is heavily punished whereas the high performance $\mathcal{W}^{1,(\pi,c)}_t-F^1_t>0$ is only weakly valued in this type of utility function.
\end{rem}

The admissible strategy for agent-1 is defined as follows.
\begin{dfn} (Admissible space for agent-1)\\
    \label{Admissible space for agent-1}
        The admissible space $\mathbb{A}^1$ is the set of trading and consumption strategies $(\pi,c)\in \mathbb{H}^2(\mathbb{P}^{0,1},\mathbb{F}^{0,1},\mathbb{R}^{n})\times \mathbb{H}^2(\mathbb{P}^{0,1},\mathbb{F}^{0,1},\mathbb{R})$ such that a family
        \[
            \Bigl\{\exp\Bigl(-\gamma^1\mathcal{W}^{1,(\pi,c)}_\tau + \beta^1 |c_\tau| + K^1 |X_\tau^{1,c}|  \Bigr) ; \tau\in\mathcal{T}^{0,1}\Bigr\}
        \]
        is uniformly integrable for some $K^1>\gamma^1(A_1+\sqrt{A_1^2+B_1})^{-1} \lor \beta^1 $ where
        \[
            A_1:=\frac{1}{2}\Bigl(\kappa-b+\frac{\gamma^1}{\beta^1}\Bigr),~~~B_1:=\frac{\gamma^1 b}{\beta^1}.
        \]
        Moreover, we define $\mathcal{A}^1:=\{(p,c)=(\pi^\top\sigma,c) ; (\pi,c)\in\mathbb{A}^1\}$.
\end{dfn}
By writing $p_s:=\pi_s^\top\sigma_s$ for each $s\in[0,T]$, the utility maximization problem can be equivalently written as
\begin{equation}
    \begin{split}
        \label{p-utility}
        \sup_{(p,c)\in\mathcal{A}^1} \widetilde{U}^1(p,c)
    \end{split}
\end{equation}
subject to
\[
    \mathcal{W}^{1,(p,c)}_t =\xi^1 + \int_0^t (p_s\theta_s - c_s)ds + \int_0^t p_s dW_s^0,
\]
where $\widetilde{U}^1:\mathcal{A}^1\to\mathbb{R}$ is defined by
\[
    \widetilde{U}^1(p,c):=\mathbb{E}\Bigl[-\exp\Bigl(-\delta T-\gamma^1(\mathcal{W}^{1,(p,c)}_T-F^1_T)\Bigr) -a \int_0^T \exp\Bigl(-\delta t-\gamma^1(\mathcal{W}^{1,(p,c)}_t-F^1_t)-\beta^1(c_t-X_t^{1,c})\Bigr)dt\Bigr].
\]
Note further that for each $s\in[0,T]$ and $(p,c)\in\mathcal{A}^1$, we have $p_s\in L_s$. 

\begin{rem}
    If $(p,c)\in\mathcal{A}^1$, we have $\widetilde{U}^1(p,c)>-\infty$. Indeed, for any $(p,c)\in\mathcal{A}^1$, the uniform integrability implies
    \[
        \sup_{t\in[0,T]}\mathbb{E}\Bigl[\exp\Bigl(-\gamma^1\mathcal{W}^{1,(p,c)}_t - \beta^1 (c_t -  X_t^{1,c}) \Bigr)\Bigr] < \infty.
    \]
    We can also see that the family
    \[
        \Bigl\{\int_0^\tau \exp \Bigl(-\gamma^1\mathcal{W}^{1,(p,c)}_s - \beta^1 (c_s -  X_s^{1,c})\Bigr)ds;\tau\in\mathcal{T}^{0,1} \Bigr\}
    \]
    is uniformly integrable.
\end{rem}

\subsection{Optimization}
Based on Hu, Imkeller \& M\"{u}ller \cite{huUtilityMaximizationIncomplete2005a}, we derive a BSDE which characterizes the optimality. To begin with, we consider a family of stochastic processes satisfying the following conditions.  
\begin{dfn} (Condition-R)\\
    \label{condition-R}
    A family of stochastic processes $\Bigl\{R^{1,(p,c)}:=(R^{1,(p,c)}_t)_{t\in[0,T]}; (p,c)\in\mathcal{A}^1\Bigr\}\subset\mathbb{L}^0(\mathbb{F}^{0,1},\mathbb{R})$ is said to satisfy the condition-R if the following properties are met.\\
    \textup{(i)} For all $(p,c)\in\mathcal{A}^1$, $R^{1,(p,c)}$ satisfies
    \[
        R^{1,(p,c)}_T = -\exp\Bigl(-\delta T-\gamma^1(\mathcal{W}^{1,(p,c)}_T-F^1_T)\Bigr) -a \int_0^T \exp\Bigl(-\delta t-\gamma^1(\mathcal{W}^{1,(p,c)}_t-F^1_t)-\beta^1(c_t-X_t^{1,c})\Bigr)dt,~~\mathbb{P}^{0,1}\text{-}\mathrm{a.s.}
    \] 
    \textup{(ii)} There exists some $\mathcal{F}^{0,1}_0$-measurable random variable $R^1_0$ such that the equality $R^{1,(p,c)}_0 = R^1_0$ holds $\mathbb{P}^{0,1}$-almost surely for all $(p,c)\in\mathcal{A}^1$.\\
    \textup{(iii)} $R^{1,(p,c)}$ is an $(\mathbb{F}^{0,1},\mathbb{P}^{0,1})$-supermartingale for all $(p,c)\in\mathcal{A}^1$ and there exists some $(p^*,c^*)\in\mathcal{A}^1$ such that $R^{1,(p^*,c^*)}$ is an $(\mathbb{F}^{0,1},\mathbb{P}^{0,1})$-martingale.
\end{dfn}
Once such a family is identified, we have, for all $(p,c)\in\mathcal{A}^1$,
\[
    \widetilde{U}^1(p,c) = \mathbb{E}[R^{1,(p,c)}_T]\leq\mathbb{E}[R^1_0] = \mathbb{E}[R^{1,(p^*,c^*)}_T] =\widetilde{U}^1(p^*,c^*),
\]
which indicates that $(p^*,c^*)$ is an optimal strategy for agent-1. To find an appropriate family of processes $\{R^{1,(p,c)}\}$, we suppose that each $R^{1,(p,c)}$ has the following form: for $t\in[0,T]$,
\begin{equation}
    \begin{split}
        \label{ansatz}
        R^{1,(p,c)}_t = -\exp\Bigl(-\delta t-\gamma^1(\mathcal{W}^{1,(p,c)}_t-Y^1_t-\zeta^1_tX_t^{1,c})\Bigr) -a \int_0^t \exp\Bigl(-\delta s-\gamma^1(\mathcal{W}^{1,(p,c)}_s-F^1_s)-\beta^1(c_s-X_s^{1,c})\Bigr)ds.~~~
    \end{split}
\end{equation}
Here, $\zeta^1$ is an $\mathcal{F}^{1}_0$-measurable and continuously differentiable process with $\zeta^1_T=0$ satisfying an ordinary differential equation (ODE) specified later. $Y^1$ is a solution to the following BSDE whose driver $f^1$ is to be determined:
\begin{equation}
    \begin{split}
       Y^1_t = F^1_T + \int_t^T f^1(s,Y^1_s,Z^{1,0}_s,Z^1_s) ds -\int_t^T Z^{1,0}_s dW^0_s  -\int_t^T Z^1_s dW^1_s ,~~~t\in[0,T].
    \end{split}
\end{equation}

For notational simplicity, we may suppress the superscript ``1" when obvious. By Ito formula,
\begin{equation*}
    \begin{split}
        dR^{(p,c)}_t 
        &=
        -\exp\Bigl(-\delta t-\gamma(\mathcal{W}^{(p,c)}_t-Y_t-\zeta_tX^c_t)\Bigr)\Bigl\{-\delta dt -\gamma d(\mathcal{W}^{(p,c)}_t-Y_t) + \frac{\gamma^2}{2}d\langle\mathcal{W}^{(p,c)}-Y\rangle_t +\gamma \dot\zeta_t X^c_tdt \Bigr.\\
        &~~~~~~\Bigl. +\gamma\zeta_tdX^c_t + a \exp\Bigl(-\gamma(Y_t-F_t+\zeta_tX^c_t)-\beta(c_t-X^c_t)\Bigr)dt \Bigr\}\\
        &=
        -\exp\Bigl(-\delta t-\gamma(\mathcal{W}^{(p,c)}_t-Y_t-\zeta_tX^c_t)\Bigr)\Bigl\{-\delta  -\gamma (p_t\theta_t-c_t+f(t,Y_t,Z^0_t,Z^1_t)) + \frac{\gamma^2}{2}(|p_t-Z^0_t|^2+|Z^1_t|^2)\\
        &~~~+\gamma (\dot\zeta_t-\kappa\zeta_t) X^c_t +\gamma\zeta_tb c_t +\gamma\zeta_t\rho_t(\kappa-b) + a \exp\Bigl(-\gamma(Y_t-F_t+\zeta_tX^c_t)-\beta(c_t-X^c_t)\Bigr)\Bigr\}dt\\
        &~~~+\gamma \exp\Bigl(-\delta t-\gamma(\mathcal{W}^{(p,c)}_t-Y_t-\zeta_tX^c_t)\Bigr)\Bigl((p_t-Z^0_t)dW^0_t-Z^1_t dW^1_t\Bigr),
    \end{split}
\end{equation*}
where $\dot\zeta_t:=\dfrac{d}{dt}\zeta_t$. In order to make $R^{(p,c)}$ a supermartingale for all $(p,c)\in\mathcal{A}^1$, we need
\begin{equation*}
    \begin{split}
    &f(t,Y_t,Z^0_t,Z^1_t)\\
    &\leq 
    -\frac{\delta}{\gamma}  -(p_t\theta_t-c_t) + \frac{\gamma}{2}(|p_t-Z^0_t|^2+|Z^1_t|^2) + (\dot\zeta_t-\kappa\zeta_t) X^c_t +\zeta_tb c_t +\zeta_t\rho_t(\kappa-b) + \frac{a}{\gamma} \exp\Bigl(-\gamma(Y_t-F_t+\zeta_tX^c_t)-\beta(c_t-X^c_t)\Bigr).
    \end{split}
\end{equation*}
Moreover, $R^{(p,c)}$ is a true martingale for some $(p^*,c^*)$ only if
\begin{equation*}
    \begin{split}
    &f(t,Y_t,Z^0_t,Z^1_t)\\
    &=
    -\frac{\delta}{\gamma}  -(p^*_t\theta_t-c^*_t) + \frac{\gamma}{2}(|p^*_t-Z^0_t|^2+|Z^1_t|^2) + (\dot\zeta_t-\kappa\zeta_t) X^{c^*}_t +\zeta_tb c^*_t +\zeta_t\rho_t(\kappa-b) + \frac{a}{\gamma} \exp\Bigl(-\gamma(Y_t-F_t+\zeta_tX^{c^*}_t)-\beta(c^*_t-X^{c^*}_t)\Bigr).
    \end{split}
\end{equation*}
Combining these observations, we deduce that
\begin{equation}
    \begin{split}
        \label{driver-f}
    &f(t,Y_t,Z^0_t,Z^1_t)\\
    &=
    -\frac{\delta}{\gamma} + (\dot\zeta_t-\kappa\zeta_t) X^c_t  +\zeta_t\rho_t(\kappa-b) \\
    &~~~+ \inf_{p\in L_t}\Bigl\{ -p\theta_t + \frac{\gamma}{2}(|p-Z^0_t|^2+|Z^1_t|^2)\Bigr\}+ \inf_{c\in\mathbb{R}} \Bigl\{ (1+\zeta_tb) c  + \frac{a}{\gamma} \exp\Bigl(-\gamma(Y_t-F_t+\zeta_tX^c_t)-\beta(c-X^c_t)\Bigr)\Bigr\}.
    \end{split}
\end{equation}
Assuming $1+b\zeta_t>0$ for all $t\in[0,T]$ temporarily, the candidate for the optimal strategy reads: for $t\in[0,T]$,
\begin{equation}
    \begin{split}
        \label{optimal}
        p^*_t &= Z^{0\|}_t + \frac{\theta^\top_t}{\gamma},\\
        c^*_t &= X^{c^*}_t + \frac{1}{\beta}\Bigl\{\log\Bigl(\frac{a\beta}{\gamma(1+b\zeta_t)}\Bigr)-\gamma(Y_t-F_t+\zeta_tX^{c^*}_t)\Bigr\},
    \end{split}
  \end{equation}
whose admissibility, namely $(p^*,c^*)\in\mathcal{A}^1$, needs to be verified later. Now we obtain
\begin{equation*}
    \begin{split}
    f(t,Y_t,Z^0_t,Z^1_t)
    &=
    -Z^{0\|}_t\theta_t - \frac{|\theta_t|^2}{2\gamma} + \frac{\gamma}{2}(|Z^{0\perp}_t|^2 + |Z^1_t|^2) - \frac{\delta}{\gamma} + (\kappa-b)\zeta_t\rho_t + \frac{1 + b\zeta_t}{\beta}\Bigl\{1 + \log\Bigl(\frac{a\beta}{\gamma(1 + b\zeta_t)}\Bigr)+\gamma(F_t-Y_t)\Bigr\}\\
    &~~~~~~~~~+X^c_t\Bigl\{\frac{1}{\beta}(1+b\zeta_t)(\beta-\gamma\zeta_t)+(\dot\zeta_t-\kappa\zeta_t)\Bigr\}.
    \end{split}
\end{equation*}

In order to make $R^{(p,c)}$ satisfy (ii) of the condition-R, we need
\[
    \frac{1}{\beta}(1+b\zeta_t)(\beta-\gamma\zeta_t)+(\dot\zeta_t-\kappa\zeta_t)=0
\]
for every $t\in[0,T]$ so that the process $Y$ is independent of $c$. To be specific, it is necessary to solve the following ordinary differential equation of Riccati type:
\begin{equation}
    \begin{split}
        \label{zeta-ODE}
        &\dot\zeta_t = \Bigl(\kappa-b+\frac{\gamma}{\beta}\Bigr)\zeta_t + \frac{\gamma b}{\beta}\zeta_t^2-1, ~~t\in[0,T],\\
        &\zeta_T = 0.
    \end{split}
\end{equation}
This is actually explicitly solvable (See, for example, Carmona \& Delarue \cite{carmonaProbabilisticTheoryMean2018} [Equation (2.50)]) as
\begin{equation}
    \label{zeta}
        \zeta_t = \frac{e^{(\delta^+-\delta^-)(T-t)}-1}{\delta^+-\delta^-e^{(\delta^+-\delta^-)(T-t)}},~~~t\in[0,T],
\end{equation}
where
\[
    \delta^{\pm}:=-A\pm\sqrt{A^2+B},~~~A:=\frac{1}{2}\Bigl(\kappa-b+\frac{\gamma}{\beta}\Bigr),~~~B:=\frac{\gamma b}{\beta}.
\]
Note that $\zeta$ satisfies 
\[
    0\leq \zeta_t\leq \frac{1}{\delta^+} e^{(\delta^+-\delta^-)T}\land \frac{1}{|\delta^-|},
\]
and in particular, $1+b\zeta_t>0$ for all $t\in[0,T]$. 

Consequently, we have derived a BSDE for the optimality:
\begin{equation}
    \begin{split}
        \label{qgBSDE}
        Y^1_t&= F^1_T + \int_t^T f^1(s,Y^1_s,Z^{1,0}_s,Z^1_s)ds - \int_t^T Z^{1,0}_s dW^0_s - \int_t^T Z^1_s dW^1_s ,~~~t\in[0,T]
    \end{split}
  \end{equation}
with
\begin{equation*}
    \begin{split}
        f^1(s,Y^1_s,Z^{1,0}_s,Z^1_s) = -Z^{1,0\|}_s\theta_s - \frac{|\theta_s|^2}{2\gamma^1} + \frac{\gamma^1}{2}(|Z^{1,0\perp}_s|^2 + |Z^1_s|^2) -\frac{\gamma^1(1+b\zeta^1_s)}{\beta^1}Y^1_s + g^1_s,
    \end{split}
  \end{equation*}
  where
  \[
      g^1_s := - \frac{\delta}{\gamma^1} + (\kappa-b)\zeta^1_s\rho_s + \frac{1 + b\zeta^1_s}{\beta^1}\Bigl\{1 + \log\Bigl(\frac{a\beta^1}{\gamma^1(1 + b\zeta^1_s)}\Bigr)+\gamma^1 F^1_s\Bigr\}.
  \]

\subsection{Well-posedness and verification}
We now study the well-posedness of \eqref{qgBSDE}. Let us begin with the \textit{a priori} estimation.
\begin{lem}
    \label{a-priori1}
    Let Assumptions \ref{asm1} and \ref{asm2} be in force. If the BSDE \eqref{qgBSDE} has a bounded solution $(Y,Z^{1,0},Z^1)\in\mathbb{S}^\infty(\mathbb{P}^{0,1},\mathbb{F}^{0,1},\mathbb{R})\times\mathbb{H}^2(\mathbb{P}^{0,1},\mathbb{F}^{0,1},\mathbb{R}^{1\times d_0})\times\mathbb{H}^2(\mathbb{P}^{0,1},\mathbb{F}^{0,1},\mathbb{R}^{1\times d})$, then $(Z^{1,0},Z^1)\in\mathbb{H}^2_{\mathrm{BMO}}(\mathbb{P}^{0,1},\mathbb{F}^{0,1},\mathbb{R}^{1\times d_0})\times\mathbb{H}^2_{\mathrm{BMO}}(\mathbb{P}^{0,1},\mathbb{F}^{0,1},\mathbb{R}^{1\times d})$ and such a solution is unique.
\end{lem}
\noindent
\textbf{\textit{proof}}\\
In the proof, we may omit the superscript ``1'' when obvious for notational simplicity. First of all, we have
\begin{equation*}
    \begin{split}
     f(s,Y_s,Z^0_s,Z^1_s) 
     = 
     -Z^{0\|}_s\theta_s - \frac{|\theta_s|^2}{2\gamma} + \frac{\gamma}{2}(|Z^{0\perp}_s|^2 + |Z^1_s|^2) -\frac{\gamma(1+b\zeta_s)}{\beta}Y_s + g_s
     \leq
     \frac{\gamma}{2}(|Z^{0}_s|^2 + |Z^1_s|^2) + C(\|Y\|_{\mathbb{S}^\infty} + \|g\|_{\mathbb{L}^\infty}).
    \end{split}
\end{equation*}
Then, by Ito formula,
\begin{equation*}
    \begin{split}
     de^{2\gamma Y_t}
     =
     2\gamma e^{2\gamma Y_t}( dY_t +\gamma d\langle Y \rangle_t)
     =
     2\gamma e^{2\gamma Y_t}\Bigl\{(- f(t,Y_t,Z^0_t,Z^1_t) + \gamma|Z^{0}_t|^2 + \gamma|Z^1_t|^2)dt + Z^0_t dW^0_t + Z^1_t dW^1_t\Bigr\}.
    \end{split}
\end{equation*}
Hence,
\begin{equation*}
    \begin{split}
     e^{2\gamma Y_T} - e^{2\gamma Y_t} 
     &=
     2\gamma \int_t^T e^{2\gamma Y_s}\Bigl\{(- f(s,Y_s,Z^0_s,Z^1_s) + \gamma|Z^{0}_s|^2 + \gamma|Z^1_s|^2)ds + Z^0_s dW^0_s + Z^1_s dW^1_s\Bigr\}\\
     &\geq
     2\gamma \int_t^T e^{2\gamma Y_s}\Bigl\{- \frac{\gamma}{2}(|Z^{0}_s|^2 + |Z^1_s|^2) - C(\|Y\|_{\mathbb{S}^\infty} + \|g\|_{\mathbb{L}^\infty}) + \gamma|Z^{0}_s|^2 + \gamma|Z^1_s|^2\Bigr\}ds +  2\gamma \int_t^T e^{2\gamma Y_s}\Bigl\{Z^0_s dW^0_s + Z^1_s dW^1_s\Bigr\}\\
     &=
     2\gamma \int_t^T e^{2\gamma Y_s}\Bigl\{\frac{\gamma}{2}(|Z^{0}_s|^2 + |Z^1_s|^2) - C(\|Y\|_{\mathbb{S}^\infty} + \|g\|_{\mathbb{L}^\infty}) \Bigr\} ds  + 2\gamma \int_t^T e^{2\gamma Y_s}\Bigl\{Z^0_s dW^0_s + Z^1_s dW^1_s\Bigr\}.
    \end{split}
\end{equation*}
Thus, for any $t\in[0,T]$
\begin{equation*}
    \begin{split}
     \mathbb{E}\Bigl[ \int_t^T (|Z^{0}_s|^2 + |Z^1_s|^2)ds |\mathcal{F}^{0,1}_t\Bigr]
     \leq
     Ce^{4\overline{\gamma} \|Y\|_{\mathbb{S}^\infty}}(1 + \|Y\|_{\mathbb{S}^\infty} + \|g\|_{\mathbb{L}^\infty})
     <\infty.
    \end{split}
\end{equation*}
Clearly, $(Z^0,Z^1)\in\mathbb{H}^2_{\mathrm{BMO}}\times\mathbb{H}^2_{\mathrm{BMO}}$. \par
Next, suppose that there exists two solutions $(Y,Z^0,Z^1)$ and $(\acute Y, \acute Z^0, \acute Z^1)$ both of which are in $\mathbb{S}^\infty\times\mathbb{H}^2_{\mathrm{BMO}}\times\mathbb{H}^2_{\mathrm{BMO}}$. Let us write $\Delta Y = Y-\acute Y,~~~\Delta Z^i = Z^i -\acute Z^i~~~(i=0,1)$. Then we have
\begin{equation*}
    \begin{split}
    f(s,Y_s,Z^0_s,Z^1_s) - f(s,\acute Y_s,\acute Z^0_s,\acute Z^1_s)
    &=
    -\Delta Z^{0\|}_s\theta_s + \frac{\gamma}{2}\Delta Z^{0\perp}_s (Z^{0\perp}_s + \acute Z^{0\perp}_s)^{\top} + \frac{\gamma}{2}\Delta Z^{1}_s (Z^{1}_s + \acute Z^{1}_s)^{\top} - \frac{\gamma(1 + b\zeta_s)}{\beta}\Delta Y_s\\
    &=
    -\Delta Z^{0}_s\theta_s + \frac{\gamma}{2}\Delta Z^{0}_s (Z^{0\perp}_s + \acute Z^{0\perp}_s)^{\top} + \frac{\gamma}{2}\Delta Z^{1}_s (Z^{1}_s + \acute Z^{1}_s)^{\top} - \frac{\gamma(1 + b\zeta_s)}{\beta}\Delta Y_s.
    \end{split}
\end{equation*}
Now, we define a new probability measure $\widetilde{\mathbb{P}}$($\sim \mathbb{P}^{0,1}$) by
\[
    \Bigl.\frac{d\widetilde{\mathbb{P}}}{d\mathbb{P}^{0,1}}\Bigr|_{\mathcal{F}^{0,1}_t} = \mathcal{E}\Bigl(\int_0^\cdot \Bigl\{-\theta_s^\top + \frac{\gamma}{2}(Z^{0\perp}_s + \acute Z^{0\perp}_s) \Bigr\}dW^0_s + \int_0^\cdot \frac{\gamma}{2}(Z^{1}_s + \acute Z^{1}_s)dW^1_s \Bigr)_t,~~t\in[0,T].
\]
By Kazamaki \cite{kazamaki_sufficient_1979} and Kazamaki \cite{kazamakiContinuousExponentialMartingales1994} [Remark 3.1], the right hand side is a martingale of class $\mathcal{D}$ and hence the new probability measure $\widetilde{\mathbb{P}}$ is well-defined. Then, the Girsanov's theorem implies that the processes
\begin{equation*}
    \begin{split}
        \widetilde W^0_t := W^0_t + \int_0^t \{\theta_s - \frac{\gamma}{2}(Z^{0\perp}_s + \acute Z^{0\perp}_s)^\top\} ds,~~~\widetilde W^1_t := W^1_t  - \int_0^t \frac{\gamma}{2}(Z^{1}_s + \acute Z^{1}_s)^\top ds,~~~t\in[0,T]
    \end{split}
\end{equation*}
are the standard $(\mathbb{F}^{0,1},\widetilde{\mathbb{P}})$-Brownian motions. Now we have:
\begin{equation}
    \begin{split}
        \label{tilde_W_BSDE}
     \Delta Y_t 
     &=
     \int_t^T \Bigl\{-\Delta Z^{0}_s\theta_s + \frac{\gamma}{2}\Delta Z^{0}_s (Z^{0\perp}_s + \acute Z^{0\perp}_s)^{\top} + \frac{\gamma}{2}\Delta Z^{1}_s (Z^{1}_s + \acute Z^{1}_s)^{\top} - \frac{\gamma(1 + b\zeta_s)}{\beta}\Delta Y_s\Bigr\}ds - \int_t^T \Delta Z^0_s dW^0_s - \int_t^T \Delta Z^1_s dW^1_s\\
     &=
     -\int_t^T  \frac{\gamma(1 + b\zeta_s)}{\beta}\Delta Y_s ds - \int_t^T \Delta Z^0_s d\widetilde W^0_s - \int_t^T \Delta Z^1_s d\widetilde W^1_s,~~t\in[0,T].
    \end{split}
\end{equation}
Then, it follows that $\Delta Y=0,\Delta Z^0=0$ and $\Delta Z^1 = 0$ for $\widetilde{\mathbb{P}}$ (and thus $\mathbb{P}^{0,1}$)-almost surely since they obviously satisfy \eqref{tilde_W_BSDE} and the solution of \eqref{tilde_W_BSDE} is unique due to the standard result for Lipschitz BSDEs (See, e.g. Zhang \cite{ZhangBSDE} [Chapter 4]). $\square$\par

For the risk neutral measure $\mathbb{Q}(\sim\mathbb{P}^{0,1})$ defined by \eqref{risk-neutral}, the Girsanov's theorem implies that the processes
\begin{equation}
    \begin{split}
    W^{0,\mathbb{Q}}_t := W^{0,\mathbb{P}}_t + \int_0^t \theta_s ds,~~~W^{1,\mathbb{Q}}_t := W^{1,\mathbb{P}}_t,~~t\in[0,T]
\end{split}
\end{equation}
form the standard $(\mathbb{F}^0,\mathbb{Q})$ and $(\mathbb{F}^1,\mathbb{Q})$-Brownian motions, respectively. Under this measure, the BSDE \eqref{qgBSDE} becomes
\begin{equation}
    \begin{split}
        \label{qgBSDE-Q}
        Y^1_t&= F^1_T + \int_t^T \Bigl\{- \frac{|\theta_s|^2}{2\gamma^1} + \frac{\gamma^1}{2}(|Z^{1,0\perp}_s|^2 + |Z^1_s|^2) -\frac{\gamma^1(1+b\zeta^1_s)}{\beta^1}Y^1_s + g^{1}_s\Bigr\}ds - \int_t^T Z^{1,0}_s dW^{0,\mathbb{Q}}_s - \int_t^T Z^1_s dW^{1,\mathbb{Q}}_s
    \end{split}
\end{equation}
for $t\in[0,T]$. Moreover, by Kazamaki \cite{kazamakiContinuousExponentialMartingales1994} [Theorem 3.3], we have $\theta\in \mathbb{H}^2_{\mathrm{BMO}}(\mathbb{Q},\mathbb{F}^{0})$. Since $\theta$ is unbounded in general, the standard technique cannot be applied directly to prove the well-posedness of the equation \eqref{qgBSDE-Q}. We adopt the same regularization used in Fujii \& Sekine \cite{fujiiMeanFieldEquilibriumPrice2023a}.

\begin{thm}
    \label{sec2-well-posed}
    Let Assumptions \ref{asm1} and \ref{asm2} be in force. Then, the BSDE \eqref{qgBSDE} has a unique solution $(Y,Z^{1,0},Z^1)\in\mathbb{S}^\infty(\mathbb{P}^{0,1},\mathbb{F}^{0,1},\mathbb{R})\times\mathbb{H}^2_{\mathrm{BMO}}(\mathbb{P}^{0,1},\mathbb{F}^{0,1},\mathbb{R}^{1\times d_0})\times\mathbb{H}^2_{\mathrm{BMO}}(\mathbb{P}^{0,1},\mathbb{F}^{0,1},\mathbb{R}^{1\times d})$.
\end{thm}
\noindent
\textbf{\textit{proof}}\\
Obviously, $W^{0,\mathbb{Q}}, W^{1,\mathbb{Q}}$ are adapted to $\mathbb{F}^{0,1}$, but they do not necessarily generate $\mathbb{F}^{0,1}$. However, due to the equivalence of $\mathbb{Q}$ and $\mathbb{P}^{0,1}$, Jeanblanc, Yor \& Chesney \cite{jeanblanc_mathematical_2009} [Proposition 1.7.7.1] shows that every $(\mathbb{F}^{0,1},\mathbb{Q})$-local martingale has a representation through a stochastic integral with respect to $(W^{0,\mathbb{Q}},W^{1,\mathbb{Q}})$. 
This fact allows us to use the standard approach for BSDEs to deal with the equation \eqref{qgBSDE-Q}. 
In addition, if there exists a bounded solution $(Y,Z^0,Z^1)\in\mathbb{S}^\infty(\mathbb{Q},\mathbb{F}^{0,1})\times\mathbb{H}^2_{\mathrm{BMO}}(\mathbb{Q},\mathbb{F}^{0,1})\times\mathbb{H}^2_{\mathrm{BMO}}(\mathbb{Q},\mathbb{F}^{0,1})$ to the equation \eqref{qgBSDE-Q}, it obviously solves the BSDE \eqref{qgBSDE} under the original measure $\mathbb{P}^{0,1}$. The uniqueness follows from Lemma \ref{a-priori1}.
Thus, it suffices to find a bounded solution of the BSDE \eqref{qgBSDE-Q}. \par
For the remainder of the proof, we may omit the superscript ``1'' if obvious. We consider the next truncated BSDE:
\begin{equation}
    \begin{split}
        \label{qgBSDE-Q-trunc}
        Y^n_t&= F_T + \int_t^T \Bigl\{- \frac{|\theta_s|^2 \land n}{2\gamma} + \frac{\gamma}{2}(|Z^{n,0\perp}_s|^2 + |Z^{n,1}_s|^2) -\frac{\gamma(1+b\zeta_s)}{\beta}Y^n_s + g_s\Bigr\}ds - \int_t^T Z^{n,0}_s dW^{0,\mathbb{Q}}_s - \int_t^T Z^{n,1}_s dW^{1,\mathbb{Q}}_s~~~~~
    \end{split}
\end{equation}
for $t\in[0,T]$. By the standard result of Kobylanski \cite{Kobylanski2000BackwardSD}, we deduce that the truncated BSDE \eqref{qgBSDE-Q-trunc} has a unique solution $(Y^n,Z^{n,0},Z^{n,1})\in\mathbb{S}^\infty\times\mathbb{H}^2_{\mathrm{BMO}}\times\mathbb{H}^2_{\mathrm{BMO}}$ for all $n\in\mathbb{N}$. In addition, the comparison principle presented in the same work shows that $Y^{n+1}\leq Y^n$ holds for all $n\in\mathbb{N}$.
In particular, this principle gives an estimate $\sup_{n\in\mathbb{N}}\|Y^n\|_{\mathbb{S}^\infty} <\infty$ by considering the following two BSDEs. For $t\in[0,T]$,
\begin{equation}
    \begin{split}
    \overline{Y}_t &= \|F\|_{\mathbb{L}^\infty} +\int_t^T \Bigl\{\frac{\overline{\gamma}}{2}(|\overline{Z}^{0\perp}_s|^2 + |\overline{Z}^1_s|^2)+ \frac{\overline{\gamma}(1+b\|\zeta\|_{\mathbb{L}^\infty})}{\underline{\beta}}|\overline{Y}_s| + \|g\|_{\mathbb{L}^\infty}\Bigr\} ds - \int_t^T \overline{Z}^0_s dW^{0,\mathbb{Q}}_s - \int_t^T \overline{Z}^1_s dW^{1,\mathbb{Q}}_s, \\
    \underline{Y}_t &= -\|F\|_{\mathbb{L}^\infty} -\int_t^T \Bigl\{\frac{|\theta_s|^2}{2\underline{\gamma}}+ \frac{\overline{\gamma}(1+b\|\zeta\|_{\mathbb{L}^\infty})}{\underline{\beta}}|\underline{Y}_s| + \|g\|_{\mathbb{L}^\infty}\Bigr\} ds - \int_t^T \underline{Z}^0_s dW^{0,\mathbb{Q}}_s - \int_t^T \underline{Z}^1_s dW^{1,\mathbb{Q}}_s.
\end{split}
\end{equation}
Then, $\underline{Y}_t\leq Y^n_t\leq \overline{Y}_t, ~~\mathbb{Q}$-a.s. for all $t\in[0,T]$ by the comparison principle, and it is also easy to see $\overline{Z}^0=0$ and $\overline{Z}^1=0$. The backward Gronwall's inequality (See, for example, Pardoux \& R{\u {a}}{\c s}canu \cite{pardouxStochasticDifferentialEquations2014} [Corollary 6.61]) yields $\overline{Y}_t \leq C(\|F\|_{\mathbb{L}^\infty}+\|g\|_{\mathbb{L}^\infty})$ for all $t\in[0,T]$. For $\underline{Y}$, it is obvious that $\underline{Y}\leq 0~~\mathbb{Q}$-a.s. and thus $|\underline{Y}_t|=-\underline{Y}_t$. Then, it is straightforward to see
\begin{equation*}
    \begin{split}
    \underline{Y}_t 
    &= 
    -\exp\Bigl(\frac{\overline{\gamma}(1+b\|\zeta\|_{\mathbb{L}^\infty})}{\underline{\beta}}(T-t)\Bigr)\|F\|_{\mathbb{L}^\infty} - \mathbb{E}\Bigl[\int_t^T \exp\Bigl(\frac{\overline{\gamma}(1+b\|\zeta\|_{\mathbb{L}^\infty})}{\underline{\beta}}(s-t)\Bigr)\Bigl(\frac{|\theta_s|^2}{2\underline{\gamma}}+\|g\|_{\mathbb{L}^\infty}\Bigr)ds|\mathcal{F}^{0,1}_t\Bigr]\\
    &\geq
    -C(\|F\|_{\mathbb{L}^\infty}+\|g\|_{\mathbb{L}^\infty}+\|\theta\|^2_{\mathbb{H}^2_{\mathrm{BMO}}}).
    \end{split}
\end{equation*}
Therefore, $(Y^n)_{n\in\mathbb{N}}\subset\mathbb{S}^\infty$ is a bounded and monotonically decreasing sequence. 

We then define a bounded process $Y$ by $Y_t(\omega) := \lim_{n\to\infty} Y^n_t(\omega)$ for almost all $(t,\omega)\in[0,T] \times\Omega$. (for $(t,\omega)$ in $dt\otimes \mathbb{Q}$-negligible sets, we may put $Y_t(\omega) = 0$.) In addition, by following the same argument as in Lemma \ref{a-priori1}, we deduce that
\[
    \sup_{n\in\mathbb{N}}\|(Z^{n,0},Z^{n,1})\|^2_{\mathbb{H}^2_{\mathrm{BMO}}}\leq C(1 + \|F\|_{\mathbb{L}^\infty}+\|\theta\|^2_{\mathbb{H}^2_{\mathrm{BMO}}}+\|g\|_{\mathbb{L}^\infty})\exp(C(\|F\|_{\mathbb{L}^\infty}+\|\theta\|^2_{\mathbb{H}^2_{\mathrm{BMO}}}+\|g\|_{\mathbb{L}^\infty}))<\infty,
\]
which means $(Z^{n,0},Z^{n,1})_{n\in\mathbb{N}}$ is weakly relatively compact in $\mathbb{H}^2$. Choosing a subsequence if necessary, there exists $(Z^{0},Z^{1})\in\mathbb{H}^2\times\mathbb{H}^2$ such that
\[
    Z^{n,0}\rightharpoonup Z^0,~~~~~Z^{n,1}\rightharpoonup Z^1~~~(n\to\infty)
\]
in the sense of weak convergence in $\mathbb{H}^2$. Finally, we shall prove that $(Y,Z^0,Z^1)\in\mathbb{S}^\infty\times\mathbb{H}^2_{\mathrm{BMO}}\times\mathbb{H}^2_{\mathrm{BMO}}$ and that it actually solves the BSDE \eqref{qgBSDE}. Since the remaining arguments are basically the same as Fujii \& Sekine \cite{fujiiMeanFieldEquilibriumPrice2023a}, they are given in Appendix A. $\square$\par

We now verify the admissibility of \eqref{optimal} and the condition-R.
\begin{thm} (Verification)\\
    \label{verification}
    Let Assumptions \ref{asm1} and \ref{asm2} be in force. Moreover, let $(Y,Z^{1,0},Z^1)\in\mathbb{S}^\infty(\mathbb{P}^{0,1},\mathbb{F}^{0,1},\mathbb{R})\times\mathbb{H}^2_{\mathrm{BMO}}(\mathbb{P}^{0,1},\mathbb{F}^{0,1},\mathbb{R}^{1\times d_0})\times\mathbb{H}^2_{\mathrm{BMO}}(\mathbb{P}^{0,1},\mathbb{F}^{0,1},\mathbb{R}^{1\times d})$ be the solution to the BSDE \eqref{qgBSDE}. Then, the process $(p^{1,*},c^{1,*})$ defined by \eqref{optimal}, that is,
    \begin{equation*}
        \begin{split}
        p^{1,*}_t &:= Z^{1,0\|}_t + \frac{\theta^\top_t}{\gamma^1},~~~t\in[0,T],\\
        c^{1,*}_t &:= X^{1,c^{1,*}}_t + \frac{1}{\beta^1}\Bigl\{\log\Bigl(\frac{a\beta^1}{\gamma^1(1+b\zeta^1_t)}\Bigr)-\gamma^1(Y^1_t-F^1_t+\zeta^1_tX^{1,c^{1,*}}_t)\Bigr\},~~~t\in[0,T]
        \end{split}
    \end{equation*}
    is a unique optimal strategy for agent-1.
\end{thm}
\noindent
\textbf{\textit{proof}}\\
As usual, we omit the superscript ``1'' if there is no risk of confusion. We first show $(p^*,c^*)$ is admissible. It is straightforward to see that $c^*$ is bounded by using the Gronwall's inequality:
\begin{equation*}
    \begin{split}
    |c^*_t|
    \leq
    C(1+|X^{c^*}_t| )
    \leq
    C + C \int_0^t |c^*_s|ds
    \end{split}
\end{equation*}
and thus $\sup_{t\in[0,T]}|c^*_t| < \infty$. This also implies $X^{c^*}\in\mathbb{S}^\infty$. Thus, it suffices to show the uniform integrability of the family
\[
    \Bigl\{\exp\Bigl(-\gamma\mathcal{W}^{(p^*,c^*)}_\tau\Bigr); \tau\in\mathcal{T}^{0,1}\Bigr\}.
\]
Let us introduce a process $\psi$ by
\[
    \psi_t:=\exp\Bigl(-\delta t -\gamma(\mathcal{W}^{(p^*,c^*)}_t - Y_t - \zeta_t X^{c^*}_t)\Bigr),~~ t\in[0,T].
\]
By the definition of the process $R^{(p,c)}$, we have
\begin{equation*}
    \begin{split}
        R^{(p^*,c^*)}_t = -\psi_t -a\int_0^t \exp\Bigl(-\gamma(Y_s-F_s)-\gamma\zeta_s X^{c^*}_s -\beta(c^*_s-X^{c^*}_s)\Bigr)\psi_sds,
    \end{split}
\end{equation*}
then it holds that 
\begin{equation*}
    \begin{split}
        d\psi_t = -dR^{(p^*,c^*)}_t -a \exp\Bigl(-\gamma(Y_t-F_t)-\gamma\zeta_t X^{c^*}_t -\beta(c^*_t-X^{c^*}_t)\Bigr)\psi_t dt.
    \end{split}
\end{equation*}
Recalling how we have chosen $(p^*,c^*)$, we have
\begin{equation*}
    \begin{split}
    dR^{(p^*,c^*)}_t 
    &= 
    \gamma \exp\Bigl(-\delta t-\gamma(\mathcal{W}^{(p^*,c^*)}_t-Y_t-\zeta_tX^{c^*}_t)\Bigr)\Bigl((p^*_t-Z^0_t)dW^0_t-Z^1_t dW^1_t\Bigr) \\
    &= 
    \psi_t\Bigl\{\Bigl(\theta_t^\top-\gamma Z^{0,\perp}_t\Bigr)dW^0_t-\gamma Z^1_t dW^1_t\Bigr\}.
\end{split}
\end{equation*}
From these observations, we obtain
\[
    d\psi_t = -a \exp\Bigl(-\gamma(Y_t-F_t)-\gamma\zeta_t X^{c^*}_t -\beta(c^*_t-X^{c^*}_t)\Bigr)\psi_t dt - \psi_t\Bigl\{\Bigl(\theta_t^\top-\gamma Z^{0,\perp}_t\Bigr)dW^0_t-\gamma Z^1_t dW^1_t\Bigr\},
\]
and thus
\begin{equation*}
    \begin{split}
    &\psi_t = \exp\Bigl(-\gamma(\xi-Y_0-\zeta_0 X_0)-a\int_0^t\exp(-\gamma(Y_s-F_s)-\gamma\zeta_s X^{c^*}_s -\beta(c^*_s-X_s))ds \Bigr)\\
    &~~~~~~~~~~~~~~~~~~~~~~~\times\mathcal{E}\Bigl(-\int_0^\cdot \Bigl(\theta_s-\gamma Z^{0,\perp}_s\Bigr)dW^0_s + \int_0^\cdot \gamma Z^1_s dW^1_s\Bigr)_t.
    \end{split}
\end{equation*}
Since $\theta, Z^0, Z^1\in\mathbb{H}^2_{\mathrm{BMO}}$ and $\xi, Y, \zeta, F, X^{c^*}$ and $c^*$ are all bounded, we deduce that $\{\psi_\tau;\tau\in\mathcal{T}^{0,1}\}$ is uniformly integrable. Therefore, given the boundedness of $Y$ and $X^{c^*}$, so is the family $\Bigl\{\exp\Bigl(-\gamma\mathcal{W}^{(p^*,c^*)}_\tau\Bigr); \tau\in\mathcal{T}^{0,1}\Bigr\}$. Hence $(p^*,c^*)\in\mathcal{A}^1$.\par
Now we check that the family $\{R^{(p,c)} ; (p,c)\in\mathcal{A}^1\}$ defined by \eqref{ansatz} satisfies the condition-R. The first condition is obviously satisfied. Also, for all $(p,c)\in\mathcal{A}^1$, we have $R^{(p,c)}_0 =  -\exp(-\gamma(\xi-Y_0-\zeta_0 X_0))$, which is $\mathcal{F}^{0,1}_0$-measurable and clearly independent of $(p,c)$. Thus, condition (ii) is fulfilled. Now we move on to (iii). For any $(p,c)\in\mathcal{A}^1$, the family $\{R^{(p,c)}_\tau ; \tau\in\mathcal{T}^{0,1}\}$ is uniformly integrable due to the definition of the set $\mathcal{A}^1$, the boundedness of $Y$ and $|\gamma\zeta_t|\leq K$. Recalling how we have chosen the driver $f$, the process $R^{(p,c)}$ has a nonpositive drift for all $(p,c)\in\mathcal{A}^1$.
Indeed, from \eqref{driver-f} the drift term of $R^{(p,c)}$ reads, for all $(p,c)\in\mathcal{A}^1$,
\begin{equation*}
    \begin{split}
        &-\exp\Bigl(-\delta t-\gamma(\mathcal{W}^{(p,c)}_t-Y_t-\zeta_tX^c_t)\Bigr)\Bigl\{-\delta  -\gamma (p_t\theta_t-c_t+f(t,Y_t,Z^0_t,Z^1_t)) + \frac{\gamma^2}{2}(|p_t-Z^0_t|^2+|Z^1_t|^2)\Bigr.\\
        &~~~~~~\Bigl. +\gamma (\dot\zeta_t-\kappa\zeta_t) X^c_t +\gamma\zeta_tb c_t +\gamma\zeta_t\rho_t(\kappa-b) + a \exp\Bigl(-\gamma(Y_t-F_t+\zeta_tX^c_t)-\beta(c_t-X^c_t)\Bigr)\Bigr\}\\
        &=
        -\gamma \exp\Bigl(-\delta t-\gamma(\mathcal{W}^{(p,c)}_t-Y_t-\zeta_tX^c_t)\Bigr)\Bigl[\Bigl\{- p_t\theta_t+ \frac{\gamma}{2}(|p_t-Z^0_t|^2+|Z^1_t|^2)\Bigr\}-\Bigl\{- p^*_t\theta_t+ \frac{\gamma}{2}(|p^*_t-Z^0_t|^2+|Z^1_t|^2)\Bigr\}\Bigr.\\
        &~~~~~~+ \Bigl\{(1+\zeta_tb) c_t  + \frac{a}{\gamma} \exp\Bigl(-\gamma(Y_t-F_t+\zeta_tX^c_t)-\beta(c_t-X^c_t)\Bigr)\Bigr\}\\
        &~~~~~~- \Bigl\{(1+\zeta_tb) c^*_t  + \frac{a}{\gamma} \exp\Bigl(-\gamma(Y_t-F_t+\zeta_tX^{c^*}_t)-\beta(c^*_t-X^{c^*}_t)\Bigr)\Bigr\} + (\dot\zeta_t-\kappa\zeta_t) (X^c_t - X^{c^*}_t)\Bigr]\\
        &\leq
        -\gamma \exp\Bigl(-\delta t-\gamma(\mathcal{W}^{(p,c)}_t-Y_t-\zeta_tX^c_t)\Bigr)\Bigl[\inf_{\varrho\in\mathbb{R}}\Bigl\{(1+\zeta_tb) \varrho  + \frac{a}{\gamma}\exp\Bigl(-\gamma(Y_t-F_t+\zeta_tX^c_t)-\beta(\varrho-X^c_t)\Bigr)\Bigr\} \\
        &~~~~~~- \Bigl\{(1+\zeta_tb) c^*_t  + \frac{a}{\gamma} \exp\Bigl(-\gamma(Y_t-F_t+\zeta_tX^{c^*}_t)-\beta(c^*_t-X^{c^*}_t)\Bigr)\Bigr\}+ (\dot\zeta_t-\kappa\zeta_t) (X^c_t - X^{c^*}_t)\Bigr]\\
        &=
        -\gamma \exp\Bigl(-\delta t-\gamma(\mathcal{W}^{(p,c)}_t-Y_t-\zeta_tX^c_t)\Bigr)\Bigl\{\frac{1}{\beta}(1+\zeta_tb)(\beta-\gamma\zeta_t)+ (\dot\zeta_t-\kappa\zeta_t)\Bigr\} (X^c_t - X^{c^*}_t)\\
        &=
        0.
    \end{split}
\end{equation*}
Here, we have used the equalities
\begin{equation*}
    \begin{split}
        &\inf_{\varrho\in\mathbb{R}}\Bigl\{(1+\zeta_tb) \varrho  + \frac{a}{\gamma}\exp\Bigl(-\gamma(Y_t-F_t+\zeta_tX^c_t)-\beta(\varrho-X^c_t)\Bigr)\Bigr\}= \frac{1+\zeta_tb}{\beta}\Bigl\{(\beta-\gamma\zeta_t)X_t^c+1+\log\Bigl(\frac{a\beta}{\gamma(1+\zeta_tb)}\Bigr)+\gamma(F_t-Y_t)\Bigr\},\\
        &(1+\zeta_tb) c^*_t  + \frac{a}{\gamma} \exp\Bigl(-\gamma(Y_t-F_t+\zeta_tX^{c^*}_t)-\beta(c^*_t-X^{c^*}_t)\Bigr) = \frac{1+\zeta_tb}{\beta}\Bigl\{(\beta-\gamma\zeta_t)X_t^{c^*}+1+\log\Bigl(\frac{a\beta}{\gamma(1+\zeta_tb)}\Bigr)+\gamma(F_t-Y_t)\Bigr\},
    \end{split}
\end{equation*}
and the ODE \eqref{zeta-ODE} in the last equality. Together with the uniform integrability, the supermartingale property is now clear. 
In addition, since the process $R^{(p^*,c^*)}$ is uniformly integrable and follows
\[
    dR^{(p^*,c^*)}_t = \gamma \exp\Bigl(-\delta t-\gamma(\mathcal{W}^{(p^*,c^*)}_t-Y_t-\zeta_tX^{c^*}_t)\Bigr)\Bigl((p^*_t-Z^0_t)dW^0_t-Z^1_t dW^1_t\Bigr),
\]
it is a true martingale. Finally, the strict convexity of $p\mapsto  -p\theta_t + \frac{\gamma}{2}(|p-Z^0_t|^2+|Z^1_t|^2)$ and $c\mapsto (1+\zeta_tb) c  + \frac{a}{\gamma} e^{-\gamma(Y_t-F_t+\zeta_tX_t)-\beta(c-X_t)}$ shows that such $(p^*,c^*)$ is unique. Consequently, $R^{(p,c)}$ is a true martingale if and only if $(p,c)=(p^*,c^*)$ thereby satisfying (iii). $\square$

\section{Mean field equilibrium model under the market clearing condition} \label{Section 3}
Based on the results of the previous section, we construct a financial market with multiple agents. 
With the help of the mean field game theory, we are going to determine the risk premium process $\theta$ endogenously so that the resultant stock prices satisfy the market-clearing condition, i.e. buy and sell orders among the agents are always balanced for the period $[0,T]$.\par
This section first provides a heuristic derivation of a mean field BSDE which characterizes the financial market in a state of equilibrium and then proves its well-posedness under certain conditions. Finally, we verify that the solution of the mean field BSDE does indeed provide the risk premium process in the large population limit.

\subsection{Multi-agent problem and the relevant BSDE}
Suppose there are $N\in\mathbb{N}$ agents in the common financial market. In order to study the equilibrium state, let us first introduce the relevant probability spaces as in Section 2.\\

\noindent
(1) We denote by $(\Omega^0,\mathcal{F}^0,\mathbb{P}^0)$ a complete probability space with complete and right-continuous filtration $\mathbb{F}^0:=(\mathcal{F}^0_t)_{t\in[0,T]}$ generated by a $d_0$-dimensional standard Brownian motion $W^0:=(W^0_t)_{t\in[0,T]}$ with $\mathcal{F}^0:=\mathcal{F}^0_T$. The space $(\Omega^0,\mathcal{F}^0,\mathbb{P}^0)$ is used to describe the randomness of the financial market and the market-wide information common to all agents. 
Moreover, we denote by $(\Omega^i,\mathcal{F}^i,\mathbb{P}^i)$ ($i=1,\ldots,N$) a complete probability space with complete and right-continuous filtration $\mathbb{F}^i:=(\mathcal{F}^i_t)_{t\in[0,T]}$, generated by a $d$-dimensional standard Brownian motion $W^i:=(W^i_t)_{t\in[0,T]}$ and a $\sigma$-algebra $\sigma(\xi^i,\gamma^i,\beta^i,X^i_0,F^i_0)$, where the completion of the latter gives $\mathcal{F}^i_0$. We set $\mathcal{F}^i:=\mathcal{F}^i_T$. 
Here, $(\xi^i,X^i_0,F^i_0)$ are $\mathbb{R}$-valued bounded random variables and $(\gamma^i,\beta^i)$ are $\mathbb{R}_{++}$-valued bounded random variables. Each space $(\Omega^i,\mathcal{F}^i,\mathbb{P}^i)$ is used to describe the idiosyncratic environment of agent-$i$.\\

\noindent
(2) We denote by $(\Omega^{0,i},\mathcal{F}^{0,i},\mathbb{P}^{0,i})$ ($i=1,\ldots,N$) a complete probability space over $\Omega^{0,i} := \Omega^0 \times \Omega^i$. Here, $(\mathcal{F}^{0,i},\mathbb{P}^{0,i})$ is the completion of $(\mathcal{F}^0 \otimes \mathcal{F}^i,\mathbb{P}^{0}\otimes \mathbb{P}^{i})$ and $\mathbb{F}^{0,i}:=(\mathcal{F}^{0,i}_t)_{t\in[0,T]}$ denotes the complete and right-continuous augmentation of $(\mathcal{F}_t^0 \otimes \mathcal{F}_t^i)_{t\in[0,T]}$.
 Moreover, we set $\mathcal{T}^{0,i}:=\mathcal{T}(\mathbb{F}^{0,i})$ for notational simplicity.\\ 

\noindent
(3) Let $(\Omega,\mathcal{F},\mathbb{P})$ be an enlarged complete probability space defined on $\Omega:=\prod_{i=0}^N\Omega^i$. $(\mathcal{F},\mathbb{P})$ is the completion of $\Bigl(\bigotimes_{i=0}^N\mathcal{F}^i,\bigotimes_{i=0}^N\mathbb{P}^i\Bigr)$ and the filtration $\mathbb{F}=(\mathcal{F}_t)_{t\in[0,T]}$ is the complete and right-continuous augmentation of $(\bigotimes_{i=0}^N\mathcal{F}^{i}_t)_{t\in[0,T]}$. \\

In this section, we make the following assumptions on heterogeneity of agents.
\begin{asm}~\\
    \label{asm3}
    \textup{(i)} For each $i\in\{1,\ldots,N\}$, all statements of Assumption \ref{asm2} hold with ``1" replaced by $``i"$.\\
    \textup{(ii)} $(\xi^i,\gamma^i, \beta^i,X^i_0)_{i\in\{1,\ldots,N\}}$ have the same distribution, i.e. they are independently and identically distributed on $(\Omega,\mathbb{F},\mathbb{P})$.\\
    \textup{(iii)} The liability processes $(F^i_t;t\in[0,T])_{i\in\{1,\ldots,N\}}$ are $\mathcal{F}^0$-conditionally independent and identically distributed on $(\Omega,\mathbb{F},\mathbb{P})$.
\end{asm}
The multi-agent problem is formulated on the filtered probability space $(\Omega,\mathcal{F},\mathbb{P},\mathbb{F})$ in the following way. Each agent-$i$ solves an optimal consumption-investment problem:
\begin{equation*}
    \begin{split}
        \sup_{(\pi,c)\in\mathbb{A}^i} U^i(\pi,c)
    \end{split}
\end{equation*}
subject to
\[
    \mathcal{W}^{i,(\pi,c)}_t =\xi^i + \int_0^t (\pi_s^\top\sigma_s\theta_s - c_s)ds + \int_0^t \pi_s^\top\sigma_s dW_s^0,~~~t\in[0,T],
\]
where $\mathbb{A}^i$ is an admissible space for agent-$i$, whose definition is to be given later. $U^i:\mathbb{A}^i\to\mathbb{R}$ is a utility function of agent-$i$ defined similarly to Section 2 by
\[
    U^i(\pi,c):=\mathbb{E}\Bigl[-\exp\Bigl(-\delta T-\gamma^i(\mathcal{W}^{i,(\pi,c)}_T-F^i_T)\Bigr) -a \int_0^T \exp\Bigl(-\delta t-\gamma^i(\mathcal{W}^{i,(\pi,c)}_t-F^i_t)-\beta^i(c_t-X_t^{i,c})\Bigr)dt\Bigr].
\]
Here, $X^{i,c}$ is agent-$i$'s consumption habits defined by
\[
    X_t^{i,c} = X^i_0 + \int_0^t \{-\kappa X^{i,c}_s + b c_s + (\kappa-b)\rho_s\}ds,~~~t\in[0,T].
\]
In this model, the parameters $\delta,a,\kappa,b>0$ and the habit trend $\rho\in\mathbb{L}^\infty(\mathbb{P}^0,\mathbb{F}^0,\mathbb{R})$ are common to all agents.\footnote{It would be possible to make the variables $(\delta,a,\kappa,b,\rho)$ different for each agent as we have done so for $(\xi^i,\gamma^i,\beta^i,X^i_0,F^i)$. For simplicity, we assume that they are common across the agents in this work.}
In the same manner as Section 2, we define the admissible space $(\mathbb{A}^i)_{i=1,\ldots,N}$ as follows.
\begin{dfn} (Admissible space: a multi-agent version)\\
    For each $i=1,\ldots,N$, the admissible space $\mathbb{A}^i$ is the set of strategies $(\pi,c)\in\mathbb{H}^2(\mathbb{P}^{0,i},\mathbb{F}^{0,i},\mathbb{R}^{n})\times\mathbb{H}^2(\mathbb{P}^{0,i},\mathbb{F}^{0,i},\mathbb{R})$ such that a family
    \[
        \Bigl\{\exp\Bigl(-\gamma^i\mathcal{W}^{i,(\pi,c)}_\tau + \beta^i |c_\tau| + K^i |X_\tau^{i,c}|  \Bigr) ; \tau\in\mathcal{T}^{0,i}\Bigr\}
    \]
    is uniformly integrable for some $K^i>\gamma^i(A_i+\sqrt{A_i^2+B_i})^{-1} \lor \beta^i $ where
    \[
        A_i:=\frac{1}{2}\Bigl(\kappa-b+\frac{\gamma^i}{\beta^i}\Bigr),~~~B_i:=\frac{\gamma^i b}{\beta^i}.
    \]
    Moreover, we define $\mathcal{A}^i:=\{(p,c)=(\pi^\top\sigma,c) ; (\pi,c)\in\mathbb{A}^i\}$.
\end{dfn}
In the same way as in Section 2, we restate the problem by writing $p_t:=\pi_t^\top\sigma_t$ for $t\in[0,T]$.
\begin{equation}
    \begin{split}
        \sup_{(p,c)\in\mathcal{A}^i} \widetilde{U}^i(p,c)
    \end{split}
\end{equation}
subject to
\[
    \mathcal{W}^{i,(p,c)}_t =\xi^i + \int_0^t (p^i_s\theta_s - c^i_s)ds + \int_0^t p^i_s dW_s^0,~~~t\in[0,T],
\]
where the objective function $\widetilde{U}^i:\mathcal{A}^i\to\mathbb{R}$ is defined by
\[
    \widetilde{U}^i(p,c):=\mathbb{E}\Bigl[-\exp\Bigl(-\delta T-\gamma^i(\mathcal{W}^{i,(p,c)}_T-F^i_T)\Bigr) -a \int_0^T \exp\Bigl(-\delta t-\gamma^i(\mathcal{W}^{i,(p,c)}_t-F^i_t)-\beta^i(c_t-X_t^{i,c})\Bigr)dt\Bigr].
\]

We also introduce an $\mathcal{F}^{i}_0$-measurable continuously differentiable process $\zeta^i$ for $i\in\{1,\ldots,N\}$ by
\begin{equation*}
        \zeta^i_t = \frac{e^{(\delta_i^+-\delta_i^-)(T-t)}-1}{\delta_i^+-\delta_i^-e^{(\delta_i^+-\delta_i^-)(T-t)}},~~~\delta_i^{\pm}:=-A_i\pm\sqrt{A_i^2+B_i},~~~t\in[0,T].
\end{equation*}
As before, $\zeta^i$ satisfies 
\[
    0\leq \zeta^i_t\leq \frac{1}{\delta_i^+} e^{(\delta_i^+-\delta_i^-)T}\land \frac{1}{|\delta_i^-|}.
\]
By the same argument as in Section 2, the unique optimal strategy for each agent-$i$ is
\begin{equation*}
    \begin{split}
    p^{i,*}_t  &:=  (\pi_t^{i,*})^\top\sigma_t  = Z^{i,0\|}_t + \frac{\theta^\top_t}{\gamma^i},~~~t\in[0,T],\\
    c^{i,*}_t &:= X^{i,c^{i,*}}_t + \frac{1}{\beta^i}\Bigl\{\log\Bigl(\frac{a\beta^i}{\gamma^i(1+b\zeta^i_t)}\Bigr)-\gamma^i(Y^i_t-F^i_t+\zeta^i_tX^{i,c^{i.*}}_t)\Bigr\},~~~t\in[0,T],
    \end{split}
\end{equation*}
where the triple $(Y^i,Z^{i,0},Z^i)\in\mathbb{S}^\infty\times\mathbb{H}^2_{\mathrm{BMO}}\times\mathbb{H}^2_{\mathrm{BMO}}$ solves the BSDE \eqref{qgBSDE} with the superscript ``1'' replaced by ``$i$''. To derive the relevant mean field BSDE, let us define the market-clearing condition.
\begin{dfn}~(Market clearing condition)\\
    \label{market-clearing}
    The financial market satisfies the market-clearing condition if the equality
    \begin{equation}
        \label{MC-eqn}
        \frac{1}{N}\sum_{i=1}^N \pi_t^{i,*} = 0
    \end{equation}
    holds $dt\otimes \mathbb{P}$-almost everywhere. Here, $\pi_t^{i,*}$ denotes the optimal trading strategy of the agent-$i$.
\end{dfn}

As in Fujii \& Sekine \cite{fujiiMeanFieldEquilibriumPrice2023a} [Section 4], this condition motivates us to study the following mean field BSDE defined on $(\Omega^{0,i},\mathcal{F}^{0,i},\mathbb{P}^{0,i},\mathbb{F}^{0,i})$ for each $i\in\{1,\ldots,N\}$:
\begin{equation}
    \begin{split}
        \label{MF-BSDE1}
            Y^i_t&= F^i_T + \int_t^T f^i(s,Y^i_s,Z^{i,0}_s,Z^i_s)ds - \int_t^T Z^{i,0}_s dW^0_s - \int_t^T Z^i_s dW^i_s,~~~t\in[0,T]
    \end{split}
  \end{equation}
with
\begin{equation*}
    \begin{split}
        f^i(s,Y^i_s,Z^{i,0}_s,Z^i_s)
        = 
        \hat\gamma Z^{i,0\|}_s\mathbb{E}[Z^{i,0\|}_s|\mathcal{F}^0]^{\top} - \frac{\hat\gamma^2}{2\gamma^i}|\mathbb{E}[Z^{i,0\|}_s|\mathcal{F}^0]|^2 + \frac{\gamma^i}{2}(|Z^{i,0\perp}_s|^2 + |Z^i_s|^2) -\frac{\gamma^i(1+b\zeta^i_s)}{\beta^i}Y^i_s + g^i_s,
    \end{split}
  \end{equation*}
where
\[
    g^i_s := - \frac{\delta}{\gamma^i} + (\kappa-b)\zeta^i_s\rho_s + \frac{1 + b\zeta^i_s}{\beta^i}\Bigl\{1 + \log\Bigl(\frac{a\beta^i}{\gamma^i(1 + b\zeta^i_s)}\Bigr)+\gamma^i F^i_s\Bigr\}.
\]
We shall see later that this BSDE has a bounded solution $(Y^i,Z^{i,0},Z^i)\in\mathbb{S}^\infty \times\mathbb{H}^2_{\mathrm{BMO}}\times\mathbb{H}^2_{\mathrm{BMO}}$ under some conditions and the process $\theta^{\mathrm{mfg}} \in \mathbb{H}^2_{\mathrm{BMO}}(\mathbb{F}^0,\mathbb{R}^{d_0})$ defined by
\footnote{The market clearing condition requires the risk premium process $\theta$ to satisfy
\[
    \frac{1}{N}\sum_{i=1}^N Z_t^{i,0\|} + \Bigl(\frac{1}{N}\sum_{i=1}^N \frac{1}{\gamma^i}\Bigr)\theta_t^\top = 0.
\]
By the mutual independence of $(\mathcal{F}_t^i)_{i\geq 1}$ and symmetry among agents, it is anticipated that $\theta^{\mathrm{mfg}}$ is the market-clearing risk premium process in the large population limit.  See Fujii \& Sekine \cite{fujiiMeanFieldEquilibriumPrice2023a} [Section 4] for details.}
\begin{equation}
    \label{mf-premium}
    \theta^{\mathrm{mfg}}_t = -\hat\gamma\mathbb{E}[Z^{1,0\|}_t|\mathcal{F}^0]^{\top},~~~t\in[0,T]
\end{equation}
with $\hat\gamma:=\mathbb{E}\Bigl[\frac{1}{\gamma^1}\Bigr]^{-1}$ in fact clears the financial market in the large population limit.
\subsection{Well-posedness of the mean field BSDE}
We are now going to investigate the well-posedness of the equation \eqref{MF-BSDE1}.
\begin{lem}
    Let Assumptions \ref{asm1} and \ref{asm3} be in force. If the BSDE \eqref{MF-BSDE1} has a bounded solution $(Y^i,Z^{i,0},Z^i)\in\mathbb{S}^\infty(\mathbb{P}^{0,i},\mathbb{F}^{0,i},\mathbb{R})\times\mathbb{H}^2(\mathbb{P}^{0,i},\mathbb{F}^{0,i},\mathbb{R}^{1\times d_0})\times\mathbb{H}^2(\mathbb{P}^{0,i},\mathbb{F}^{0,i},\mathbb{R}^{1\times d})$, then $(Z^{i,0},Z^i)\in\mathbb{H}^2_{\mathrm{BMO}}(\mathbb{P}^{0,i},\mathbb{F}^{0,i},\mathbb{R}^{1\times d_0})\times\mathbb{H}^2_{\mathrm{BMO}}(\mathbb{P}^{0,i},\mathbb{F}^{0,i},\mathbb{R}^{1\times d})$.
\end{lem}
\noindent
\textbf{\textit{proof}}\\
Without loss of generality, we choose the agent-1 as a representative agent and suppress the superscript ``1'' when obvious. We have, by Young's inequality,
\begin{equation*}
    \begin{split}
     f(s,Y_s,Z^0_s,Z^1_s)
     &= 
     \hat\gamma Z^{0\|}_s\mathbb{E}[Z^{0\|}_s|\mathcal{F}^0]^{\top} - \frac{\hat\gamma^2}{2\gamma}|\mathbb{E}[Z^{0\|}_s|\mathcal{F}^0]|^2 + \frac{\gamma}{2}(|Z^{0\perp}_s|^2 + |Z^1_s|^2) -\frac{\gamma(1+b\zeta_s)}{\beta}Y_s + g_s\\
     &\leq
     \frac{\gamma}{2}(|Z^{0}_s|^2 + |Z^1_s|^2) + C(\|Y\|_{\mathbb{S}^\infty} + \|g\|_{\mathbb{L}^\infty}).
    \end{split}
\end{equation*}
Then, by Ito formula,
\begin{equation*}
    \begin{split}
     de^{2\gamma Y_t}
     =
     2\gamma e^{2\gamma Y_t}( dY_t +\gamma d\langle Y \rangle_t)
     =
     2\gamma e^{2\gamma Y_t}\Bigl\{(- f(t,Y_t,Z^0_t,Z^1_t) + \gamma|Z^{0}_t|^2 + \gamma|Z^1_t|^2)dt + Z^0_t dW^0_t + Z^1_t dW^1_t\Bigr\}.
    \end{split}
\end{equation*}
This yields:
\begin{equation*}
    \begin{split}
     e^{2\gamma Y_T} - e^{2\gamma Y_t} 
     &=
     2\gamma \int_t^T e^{2\gamma Y_s}\Bigl\{(- f(s,Y_s,Z^0_s,Z^1_s) + \gamma|Z^{0}_s|^2 + \gamma|Z^1_s|^2)ds + Z^0_s dW^0_s + Z^1_s dW^1_s\Bigr\}\\
     &\geq
     2\gamma \int_t^T e^{2\gamma Y_s}\Bigl\{- \frac{\gamma}{2}(|Z^{0}_s|^2 + |Z^1_s|^2) - C(\|Y\|_{\mathbb{S}^\infty} + \|g\|_{\mathbb{L}^\infty}) + \gamma|Z^{0}_s|^2 + \gamma|Z^1_s|^2\Bigr\}ds +  2\gamma \int_t^T e^{2\gamma Y_s}\Bigl\{Z^0_s dW^0_s + Z^1_s dW^1_s\Bigr\}\\
     &\geq
     2\gamma \int_t^T e^{2\gamma Y_s}\Bigl\{\frac{\gamma}{2}(|Z^{0}_s|^2 + |Z^1_s|^2) - C(\|Y\|_{\mathbb{S}^\infty} + \|g\|_{\mathbb{L}^\infty}) \Bigr\}ds  + 2\gamma \int_t^T e^{2\gamma Y_s}\Bigl\{Z^0_s dW^0_s + Z^1_s dW^1_s\Bigr\}.
    \end{split}
\end{equation*}
Thus, for all $t\in[0,T]$, we get
\begin{equation*}
    \begin{split}
     \mathbb{E}\Bigl[ \int_t^T (|Z^{0}_s|^2 + |Z^1_s|^2)ds |\mathcal{F}^{0,1}_t\Bigr]
     \leq
     Ce^{4\overline{\gamma} \|Y\|_{\mathbb{S}^\infty}}(1+\|Y\|_{\mathbb{S}^\infty} + \|g\|_{\mathbb{L}^\infty})
     <\infty,
    \end{split}
\end{equation*}
and clearly, $(Z^0,Z^1)\in\mathbb{H}^2_{\mathrm{BMO}}\times\mathbb{H}^2_{\mathrm{BMO}}$. $\square$\par
To show the well-posedness, we need to make an additional assumption on the size of the terminal liability $F^i_T$ and the process $g^i$.
\begin{asm}
    \label{asm4}
    Assume that, for each $i\in\{1,\ldots,N\}$, the random variable $F^i_T$ and the process $(g^i_t)_{t\in[0,T]}$ satisfy
    \begin{equation}
        \label{small}
        \sqrt{\|F^i_T\|^2_{\infty} + 4\Bigl\|\int_0^T |g^i_s| ds\Bigr\|^2_{\infty}} <   \frac{1}{16 c_\gamma}\land \frac{1}{32 C_\gamma},
    \end{equation}  
    where
    \begin{equation*}
        \begin{split}
            c_\gamma:=\frac{\overline{\gamma}}{2} \lor \frac{\hat\gamma^2}{\underline{\gamma}},~~~C_\gamma:=\hat\gamma + \Bigl(\frac{\hat\gamma^2}{2\underline{\gamma}} \lor \frac{\overline{\gamma}}{2}\Bigr).
        \end{split}
    \end{equation*}
\end{asm}
\begin{rem}~\\
    For each $s\in[0,T]$, we have $|g^i_s|\to 0$ when, for instance, $\delta\to0$, $\|\rho\|_{\mathbb{L}^\infty}\to 0$ and $\beta^i\to\infty$. This observation allows us to find appropriate parameters that fulfils the condition \eqref{small} if $F^i_T$ is sufficiently small.
\end{rem}
Here is our first main result of this section. The method is inspired by the work Tevzadze \cite{tevzadzeSolvabilityBackwardStochastic2008}.
\begin{thm}
    \label{MF-BSDE-wellposed1}
    Let Assumptions \ref{asm1}, \ref{asm3} and \ref{asm4} be in force. Then, the mean field BSDE \eqref{MF-BSDE1} has a bounded solution $(Y^i,Z^{i,0},Z^i)\in\mathbb{S}^\infty(\mathbb{P}^{0,i},\mathbb{F}^{0,i},\mathbb{R}) \times\mathbb{H}^2_{\mathrm{BMO}}(\mathbb{P}^{0,i},\mathbb{F}^{0,i},\mathbb{R}^{1\times d_0})\times\mathbb{H}^2_{\mathrm{BMO}}(\mathbb{P}^{0,i},\mathbb{F}^{0,i},\mathbb{R}^{1\times d})$. 
\end{thm}

\noindent
\textbf{\textit{proof}}\\
Again, we choose the agent-1 as a representative agent and omit the superscript ``1'' for simplicity.\\
(Step I)\\
By completing the square, we have
\begin{equation*}
    \begin{split}
        \hat\gamma Z^{0\|}_s\mathbb{E}[Z^{0\|}_s|\mathcal{F}^0]^{\top} - \frac{\hat\gamma^2}{2\gamma}|\mathbb{E}[Z^{0\|}_s|\mathcal{F}^0]|^2 + \frac{\gamma}{2}(|Z^{0\perp}_s|^2 + |Z^1_s|^2) 
        &=
        -\Bigl|\frac{\hat\gamma}{\sqrt{2\gamma}}\mathbb{E}[Z^{0\|}_s|\mathcal{F}^0] - \frac{\sqrt{\gamma}}{\sqrt{2}}Z^{0\|}_s\Bigr|^2 +\frac{\gamma}{2}|Z^{0\|}_s|^2+\frac{\gamma}{2}(|Z^{0\perp}_s|^2 + |Z^1_s|^2) \\
        &=
        -\Bigl|\frac{\hat\gamma}{\sqrt{2\gamma}}\mathbb{E}[Z^{0\|}_s|\mathcal{F}^0] - \frac{\sqrt{\gamma}}{\sqrt{2}}Z^{0\|}_s\Bigr|^2 +\frac{\gamma}{2}(|Z^{0}_s|^2 + |Z^1_s|^2).
    \end{split}
\end{equation*}
It then follows that
\begin{equation*}
    \begin{split}
        \hat\gamma Z^{0\|}_s\mathbb{E}[Z^{0\|}_s|\mathcal{F}^0]^{\top} - \frac{\hat\gamma^2}{2\gamma}|\mathbb{E}[Z^{0\|}_s|\mathcal{F}^0]|^2 + \frac{\gamma}{2}(|Z^{0\perp}_s|^2 + |Z^1_s|^2) 
        \leq
        \frac{\gamma}{2}(|Z^{0}_s|^2 + |Z^1_s|^2)
    \end{split}
\end{equation*}
and
\begin{equation*}
    \begin{split}
        \hat\gamma Z^{0\|}_s\mathbb{E}[Z^{0\|}_s|\mathcal{F}^0]^{\top} - \frac{\hat\gamma^2}{2\gamma}|\mathbb{E}[Z^{0\|}_s|\mathcal{F}^0]|^2 + \frac{\gamma}{2}(|Z^{0\perp}_s|^2 + |Z^1_s|^2) 
        &\geq
        -\Bigl|\frac{\hat\gamma}{\sqrt{2\gamma}}\mathbb{E}[Z^{0\|}_s|\mathcal{F}^0] - \frac{\sqrt{\gamma}}{\sqrt{2}}Z^{0\|}_s\Bigr|^2 +\frac{\gamma}{2}|Z^{0}_s|^2 \\
        &\geq
        -\frac{\hat\gamma^2}{\gamma}|\mathbb{E}[Z^{0\|}_s|\mathcal{F}^0]|^2 - \gamma |Z^{0\|}_s|^2 +\frac{\gamma}{2}|Z^{0}_s|^2 \\
        &\geq
        -\frac{\hat\gamma^2}{\gamma}|\mathbb{E}[Z^{0\|}_s|\mathcal{F}^0]|^2 - \frac{\gamma}{2}|Z^{0}_s|^2.
    \end{split}
\end{equation*}
Putting these together, we obtain
\begin{equation*}
    \begin{split}
        \Bigl|\hat\gamma Z^{0\|}_s\mathbb{E}[Z^{0\|}_s|\mathcal{F}^0]^{\top} - \frac{\hat\gamma^2}{2\gamma}|\mathbb{E}[Z^{0\|}_s|\mathcal{F}^0]|^2 + \frac{\gamma}{2}(|Z^{0\perp}_s|^2 + |Z^1_s|^2) \Bigr|
        &\leq
        \frac{\hat\gamma^2}{\gamma}|\mathbb{E}[Z^{0\|}_s|\mathcal{F}^0]|^2 + \frac{\gamma}{2}(|Z^{0}_s|^2 + |Z^1_s|^2)\\
        &\leq
        c_\gamma (|\mathbb{E}[Z^{0\|}_s|\mathcal{F}^0]|^2 + |Z^{0}_s|^2 + |Z^1_s|^2)
    \end{split}
\end{equation*}
for
\[
    c_\gamma=\frac{\overline{\gamma}}{2} \lor \frac{\hat\gamma^2}{\underline{\gamma}}.
\]
This yields
\begin{equation*}
    \begin{split}
        Y_s f(s,Y_s,Z^0_s,Z^1_s)
        &\leq
        |Y_s|\Bigl|\hat\gamma Z^{0\|}_s\mathbb{E}[Z^{0\|}_s|\mathcal{F}^0]^{\top} - \frac{\hat\gamma^2}{2\gamma}|\mathbb{E}[Z^{0\|}_s|\mathcal{F}^0]|^2 + \frac{\gamma}{2}(|Z^{0\perp}_s|^2 + |Z^1_s|^2)\Bigr|- \frac{\gamma(1+b\zeta_s)}{\beta}|Y_s|^2 + |Y_s||g_s|\\
        &\leq
        c_\gamma |Y_s|(|\mathbb{E}[Z^{0\|}_s|\mathcal{F}^0]|^2 + |Z^{0}_s|^2 + |Z^1_s|^2) + |Y_s||g_s|.
    \end{split}
\end{equation*}

Let us now consider a BSDE
\begin{equation*}
    \begin{split}
        Y_t&= F_T + \int_t^T f(s,Y_s,z^0_s,z^1_s)ds - \int_t^T Z^0_s dW^0_s - \int_t^T Z^1_s dW^1_s,~~~t\in[0,T]
    \end{split}
\end{equation*}
for an arbitrary $(z^0,z^1)\in\mathbb{H}^2_{\mathrm{BMO}}\times\mathbb{H}^2_{\mathrm{BMO}}$ as an input. From the standard result for Lipschitz BSDEs, there exists a unique solution $(Y,Z^0,Z^1)\in\mathbb{S}^\infty\times\mathbb{H}^2_{\mathrm{BMO}}\times\mathbb{H}^2_{\mathrm{BMO}}$ for every given $(z^0,z^1)\in\mathbb{H}^2_{\mathrm{BMO}}\times\mathbb{H}^2_{\mathrm{BMO}}$. 
In this manner, we define a map $\Gamma:\mathbb{H}^2_{\mathrm{BMO}}(\mathbb{P}^{0,1},\mathbb{F}^{0,1},\mathbb{R}^{1\times d_0}\times \mathbb{R}^{1\times d})\to \mathbb{H}^2_{\mathrm{BMO}}(\mathbb{P}^{0,1},\mathbb{F}^{0,1},\mathbb{R}^{1\times d_0}\times \mathbb{R}^{1\times d})$ by $\Gamma(z^0,z^1)=(Z^0,Z^1)$.
By Ito formula,
\begin{equation*}
    \begin{split}
        &|Y_t|^2 + \mathbb{E}\Bigl[\int_t^T (|Z_s^0|^2 + |Z_s^1|^2 )ds | \mathcal{F}^{0,1}_t\Bigr]\\
        &=
        \mathbb{E}\Bigl[|F_T|^2 + 2\int_t^T Y_sf(s,Y_s,z^0_s,z^1_s)ds | \mathcal{F}^{0,1}_t\Bigr]\\
        &\leq
        \|F_T\|^2_{\infty} + 2c_\gamma\mathbb{E}\Bigl[ \int_t^T |Y_s|(|\mathbb{E}[z^{0\|}_s|\mathcal{F}^0]|^2 + |z^{0}_s|^2 + |z^1_s|^2 )ds | \mathcal{F}^{0,1}_t\Bigr] + 2\mathbb{E}\Bigl[ \int_t^T |Y_s||g_s|ds | \mathcal{F}^{0,1}_t\Bigr]\\
        &\leq
        \|F_T\|^2_{\infty} + 2c_\gamma\|Y\|_{\mathbb{S}^\infty}\mathbb{E}\Bigl[ \int_t^T (|\mathbb{E}[z^{0\|}_s|\mathcal{F}^0]|^2 + |z^{0}_s|^2 + |z^1_s|^2)ds | \mathcal{F}^{0,1}_t\Bigr]+ 2\|Y\|_{\mathbb{S}^\infty}\mathbb{E}\Bigl[ \int_t^T|g_s|ds | \mathcal{F}^{0,1}_t\Bigr]\\
        &\leq
        \|F_T\|^2_{\infty} + 4c_\gamma\|Y\|_{\mathbb{S}^\infty}\|(z^0,z^1)\|^2_{\mathbb{H}^2_{\mathrm{BMO}}} + 2\|Y\|_{\mathbb{S}^\infty}\mathbb{E}\Bigl[ \int_t^T|g_s|ds | \mathcal{F}^{0,1}_t\Bigr] \\
        &\leq
        \|F_T\|^2_{\infty} + \frac{1}{2}\|Y\|^2_{\mathbb{S}^\infty} + 16c_\gamma^2\|(z^0,z^1)\|^4_{\mathbb{H}^2_{\mathrm{BMO}}} + 4 \Bigl\|\int_0^T |g_s| ds\Bigr\|^2_{\infty}.
    \end{split}
\end{equation*}
Here, we have used the fact that
\begin{equation*}
    \begin{split}
        \sup_{\tau\in\mathcal{T}^{0,1}}\Bigl\|\mathbb{E}\Bigl[ \int_\tau^T |\mathbb{E}[z^{0\|}_s|\mathcal{F}^0]|^2 ds | \mathcal{F}^{0,1}_\tau\Bigr]\Bigr\|_\infty
        &\leq
        \|z^0\|^2_{\mathbb{H}^2_{\mathrm{BMO}}},
    \end{split}
\end{equation*}
which is shown in Fujii \& Sekine \cite{fujiiMeanFieldEquilibriumPrice2023a} [Lemma 4.2]. Taking the essential supremum on both sides, we get:
\begin{equation*}
    \begin{split}
        \esssup_{(t,\omega)\in[0,T]\times\Omega}\Bigl(|Y_t|^2 + \mathbb{E}\Bigl[\int_t^T (|Z_s^0|^2 + |Z_s^1|^2 )ds | \mathcal{F}^{0,1}_t\Bigr]\Bigr)
        &\leq
        \|F_T\|^2_{\infty} + \frac{1}{2}\|Y\|^2_{\mathbb{S}^\infty} + 16c_\gamma^2\|(z^0,z^1)\|^4_{\mathbb{H}^2_{\mathrm{BMO}}} + 4 \Bigl\|\int_0^T |g_s| ds\Bigr\|^2_{\infty}.
    \end{split}
\end{equation*}
Using the fact that
\begin{equation*}
    \begin{split}
        \frac{1}{2}(\|Y\|^2_{\mathbb{S}^\infty} + \|(Z^0,Z^1)\|^2_{\mathbb{H}^2_{\mathrm{BMO}}})
        \leq
        \max\{\|Y\|^2_{\mathbb{S}^\infty} , \|(Z^0,Z^1)\|^2_{\mathbb{H}^2_{\mathrm{BMO}}} \}
        \leq
        \esssup_{(t,\omega)\in[0,T]\times\Omega}\Bigl(|Y_t|^2 + \mathbb{E}\Bigl[\int_t^T (|Z_s^0|^2 + |Z_s^1|^2 )ds | \mathcal{F}^{0,1}_t\Bigr]\Bigr),
    \end{split}
\end{equation*}
we obtain
\begin{equation*}
    \begin{split}
       \|(Z^0,Z^1)\|^2_{\mathbb{H}^2_{\mathrm{BMO}}}
        &\leq
        2\|F_T\|^2_{\infty} + 8\Bigl\|\int_0^T |g_s| ds\Bigr\|^2_{\infty} + 32c_\gamma^2 \|(z^0,z^1)\|^4_{\mathbb{H}^2_{\mathrm{BMO}}}.
    \end{split}
\end{equation*}
Since we have assumed that 
\begin{equation*}
    \begin{split}
        \|F_T\|^2_{\infty} + 4\Bigl\|\int_0^T |g_s| ds\Bigr\|^2_{\infty} \leq \frac{1}{256c_\gamma^2},
    \end{split}
\end{equation*}
there exists $R>0$ such that the inequality
\begin{equation*}
    \begin{split}
        2\|F_T\|^2_{\infty} + 8\Bigl\|\int_0^T |g_s| ds\Bigr\|^2_{\infty} + 32c_\gamma^2 R^4\leq R^2
    \end{split}
\end{equation*}
holds true. We can choose, for instance,
\begin{equation}
    \label{R-choice}
    R = 2\sqrt{\|F_T\|^2_{\infty} + 4\Bigl\|\int_0^T |g_s| ds\Bigr\|^2_{\infty}}\leq \frac{1}{8c_\gamma}.
\end{equation}
(Step II)\\
From the results of (Step I), we have $\Gamma(\mathcal{B}_R)\subset \mathcal{B}_R$, where
\[
    \mathcal{B}_R:=\{(z^0,z^1)\in \mathbb{H}^2_{\mathrm{BMO}}(\mathbb{F}^{0,1},\mathbb{R}^{1\times d_0}\times \mathbb{R}^{1\times d}) ;  \|(z^0,z^1)\|_{\mathbb{H}^2_{\mathrm{BMO}}}\leq R\}.
\]
Our objective is now to prove that $\Gamma|_{\mathcal{B}_R}:\mathcal{B}_R\to \mathcal{B}_R$ is a strict contraction. For $(z^0,z^1), (\acute{z}^0,\acute{z}^1)\in \mathcal{B}_R$, we set $(Z^0,Z^1):=\Gamma(z^0,z^1)$ and $(\acute{Z}^0,\acute{Z}^1):=\Gamma(\acute{z}^0,\acute{z}^1)$. Also, let $Y$ and $\acute{Y}$ be corresponding solutions and set $\Delta Y := Y-\acute{Y}$ and $\Delta Z^i := Z^i-\acute{Z}^i$. Notice that
\begin{equation*}
    \begin{split}
        &\Delta Y_s\{f(s,Y_s,z^0_s,z^1_s)-f(s,\acute{Y}_s,\acute{z}^0_s,\acute{z}^1_s)\}\\
        &\leq
        |\Delta Y_s|\Bigl\{\hat\gamma(|\mathbb{E}[z^{0\|}_s |\mathcal{F}^0]| + |\acute{z}^{0\|}_s|)(|\Delta z^{0\|}_s| + |\mathbb{E}[\Delta z^{0\|}_s |\mathcal{F}^0]|) + \frac{\hat\gamma^2}{2\underline{\gamma}}(|\mathbb{E}[z^{0\|}_s |\mathcal{F}^0]| + |\mathbb{E}[\acute{z}^{0\|}_s |\mathcal{F}^0]|)|\mathbb{E}[\Delta z^{0\|}_s |\mathcal{F}^0]|\Bigr.\\
        &~~~~~~~~~~~~~\Bigl. +\frac{\overline{\gamma}}{2}(|z^{0\perp}_s| + |\acute{z}^{0\perp}_s| + |z^1_s| + |\acute{z}^1_s| )(|\Delta z^{0\perp}_s| + |\Delta z^1_s|)\Bigr\} -\frac{\gamma(1+b\zeta_s)}{\beta}|\Delta Y_s|^2 \\
        &\leq
        |\Delta Y_s|\Bigl\{\Bigl(\Bigl(\hat\gamma + \frac{\hat\gamma^2}{2\underline{\gamma}}\Bigr)|\mathbb{E}[z^{0\|}_s |\mathcal{F}^0]| + \frac{\hat\gamma^2}{2\underline{\gamma}}|\mathbb{E}[\acute{z}^{0\|}_s |\mathcal{F}^0]| + \hat\gamma|\acute{z}^{0\|}_s |\Bigr) |\mathbb{E}[\Delta z^{0\|}_s |\mathcal{F}^0]|  \Bigr.\\
        &~~~~~~~~~~~~~\Bigl. +\Bigl(\hat\gamma + \frac{\overline{\gamma}}{2}\Bigr)(|\mathbb{E}[z^{0\|}_s |\mathcal{F}^0]| + |\acute{z}^{0}_s| +|z^{0}_s| + |z^1_s| + |\acute{z}^1_s| )(|\Delta z^{0}_s| + |\Delta z^1_s|)\Bigr\}.
    \end{split}
\end{equation*}
Applying Ito formula to $|\Delta Y_t|^2$ , we have
\begin{equation*}
    \begin{split}
       &|\Delta Y_t|^2 + \mathbb{E}\Bigl[\int_t^T (|\Delta Z_s^0|^2 + |\Delta Z_s^1|^2) ds | \mathcal{F}^{0,1}_t\Bigr]\\
       &=
       2\mathbb{E}\Bigl[\int_t^T \Delta Y_s\{f(s,Y_s,z^0_s,z^1_s)-f(s,\acute{Y}_s,\acute{z}^0_s,\acute{z}^1_s)\} ds | \mathcal{F}^{0,1}_t\Bigr]\\
       &\leq
       2  \|\Delta Y\|_{\mathbb{S}^\infty}\mathbb{E}\Bigl[\int_t^T \Bigl(\Bigl(\hat\gamma + \frac{\hat\gamma^2}{2\underline{\gamma}}\Bigr)|\mathbb{E}[z^{0\|}_s |\mathcal{F}^0]| + \frac{\hat\gamma^2}{2\underline{\gamma}}|\mathbb{E}[\acute{z}^{0\|}_s |\mathcal{F}^0]| + \hat\gamma|\acute{z}^{0\|}_s|\Bigr) |\mathbb{E}[\Delta z^{0\|}_s |\mathcal{F}^0]| ds | \mathcal{F}^{0,1}_t\Bigr]\\
       &~~~~~~+2\Bigl(\hat\gamma + \frac{\overline{\gamma}}{2}\Bigr) \|\Delta Y\|_{\mathbb{S}^\infty}\mathbb{E}\Bigl[\int_t^T (|\mathbb{E}[z^{0\|}_s |\mathcal{F}^0]| + |z^{0}_s| + |\acute{z}^{0}_s| + |z^1_s| + |\acute{z}^1_s| )(|\Delta z^{0}_s| + |\Delta z^1_s|) ds | \mathcal{F}^{0,1}_t\Bigr]\\
       &\leq
       2\|\Delta Y\|_{\mathbb{S}^\infty}\mathbb{E}\Bigl[\int_t^T  \Bigl(\Bigl(\hat\gamma + \frac{\hat\gamma^2}{2\underline{\gamma}}\Bigr)|\mathbb{E}[z^{0\|}_s |\mathcal{F}^0]| + \frac{\hat\gamma^2}{2\underline{\gamma}}|\mathbb{E}[\acute{z}^{0\|}_s |\mathcal{F}^0]| + \hat\gamma|\acute{z}^{0\|}_s|\Bigr)^2ds | \mathcal{F}^{0,1}_t\Bigr]^{\frac{1}{2}}\times\mathbb{E}\Bigl[\int_t^T|\mathbb{E}[\Delta z^{0\|}_s |\mathcal{F}^0]|^2 ds | \mathcal{F}^{0,1}_t\Bigr]^{\frac{1}{2}}\\
       &~~~+ 2\Bigl(\hat\gamma + \frac{\overline{\gamma}}{2}\Bigr)\|\Delta Y\|_{\mathbb{S}^\infty}\mathbb{E}\Bigl[\int_t^T (|\mathbb{E}[z^{0\|}_s |\mathcal{F}^0]| + |z^{0}_s| + |\acute{z}^{0}_s| + |z^1_s| + |\acute{z}^1_s| )^2ds | \mathcal{F}^{0,1}_t\Bigr]^{\frac{1}{2}}\times\mathbb{E}\Bigl[\int_t^T(|\Delta z^{0}_s| + |\Delta z^1_s|)^2 ds | \mathcal{F}^{0,1}_t\Bigr]^{\frac{1}{2}}\\
       &\leq
       2\sqrt{3}\|\Delta Y\|_{\mathbb{S}^\infty}\mathbb{E}\Bigl[\int_t^T  \Bigl(\Bigl(\hat\gamma + \frac{\hat\gamma^2}{2\underline{\gamma}}\Bigr)^2|\mathbb{E}[z^{0\|}_s |\mathcal{F}^0]|^2 + \frac{\hat\gamma^4}{4\underline{\gamma}^2}|\mathbb{E}[\acute{z}^{0\|}_s |\mathcal{F}^0]|^2 + \hat\gamma^2|\acute{z}^{0\|}_s|^2\Bigr)ds | \mathcal{F}^{0,1}_t\Bigr]^{\frac{1}{2}}\times\mathbb{E}\Bigl[\int_t^T|\mathbb{E}[\Delta z^{0\|}_s |\mathcal{F}^0]|^2 ds | \mathcal{F}^{0,1}_t\Bigr]^{\frac{1}{2}}\\
       &~~~+ 2\sqrt{10}\Bigl(\hat\gamma + \frac{\overline{\gamma}}{2}\Bigr)\|\Delta Y\|_{\mathbb{S}^\infty}\mathbb{E}\Bigl[\int_t^T (|\mathbb{E}[z^{0\|}_s |\mathcal{F}^0]|^2  +|z^{0}_s|^2 + |\acute{z}^{0}_s|^2 + |z^1_s|^2 + |\acute{z}^1_s|^2 )ds | \mathcal{F}^{0,1}_t\Bigr]^{\frac{1}{2}}\times\mathbb{E}\Bigl[\int_t^T(|\Delta z^{0}_s|^2 + |\Delta z^1_s|^2) ds | \mathcal{F}^{0,1}_t\Bigr]^{\frac{1}{2}}\\
       &\leq
       2\sqrt{6}\Bigl(\hat\gamma + \frac{\hat\gamma^2}{2\underline{\gamma}}\Bigr)\|\Delta Y\|_{\mathbb{S}^\infty} R\|(\Delta z^0, \Delta z^1)\|_{\mathbb{H}^2_{\mathrm{BMO}}} + 2\sqrt{30}\Bigl(\hat\gamma+\frac{\overline{\gamma}}{2}\Bigr)\|\Delta Y\|_{\mathbb{S}^\infty} R \|(\Delta z^0, \Delta z^1)\|_{\mathbb{H}^2_{\mathrm{BMO}}}\\
       &\leq
       2(\sqrt{6} + \sqrt{30})C_\gamma\|\Delta Y\|_{\mathbb{S}^\infty} R\|(\Delta z^0, \Delta z^1)\|_{\mathbb{H}^2_{\mathrm{BMO}}} \\
       &\leq 
       \frac{1}{2}\|\Delta Y\|_{\mathbb{S}^\infty}^2 + 2(\sqrt{6} + \sqrt{30})^2C_\gamma^2 R^2 \|(\Delta z^0, \Delta z^1)\|^2_{\mathbb{H}^2_{\mathrm{BMO}}} \\
       &\leq
       \frac{1}{2}\|\Delta Y\|_{\mathbb{S}^\infty}^2 + 128C_\gamma^2 R^2 \|(\Delta z^0, \Delta z^1)\|^2_{\mathbb{H}^2_{\mathrm{BMO}}},
    \end{split}
\end{equation*}
where
\[
    C_\gamma = \hat\gamma + \Bigl(\frac{\hat\gamma^2}{2\underline{\gamma}} \lor \frac{\overline{\gamma}}{2}\Bigr).
\]
Taking the essential supremum, we get
\begin{equation*}
    \begin{split}
       \esssup_{(t,\omega)\in[0,T]\times\Omega}\Bigl(|\Delta Y_t|^2 + \mathbb{E}\Bigl[\int_t^T (|\Delta Z_s^0|^2 + |\Delta Z_s^1|^2) ds | \mathcal{F}^{0,1}_t\Bigr]\Bigr)
       &\leq
       \frac{1}{2}\|\Delta Y\|_{\mathbb{S}^\infty}^2 + 128C_\gamma^2 R^2 \|(\Delta z^0, \Delta z^1)\|^2_{\mathbb{H}^2_{\mathrm{BMO}}}.
    \end{split}
\end{equation*}
Using the fact that
\begin{equation*}
    \begin{split}
        \frac{1}{2}(\|\Delta Y\|^2_{\mathbb{S}^\infty} + \|(\Delta Z^0,\Delta Z^1)\|^2_{\mathbb{H}^2_{\mathrm{BMO}}})
        &\leq
        \max\{\|\Delta Y\|^2_{\mathbb{S}^\infty} , \|(\Delta Z^0,\Delta Z^1)\|^2_{\mathbb{H}^2_{\mathrm{BMO}}} \}\\
        &\leq
        \esssup_{(t,\omega)\in[0,T]\times\Omega}\Bigl(|\Delta Y_t|^2 + \mathbb{E}\Bigl[\int_t^T (|\Delta Z_s^0|^2 + |\Delta Z_s^1|^2 )ds | \mathcal{F}^{0,1}_t\Bigr]\Bigr),
    \end{split}
\end{equation*}
we obtain
\begin{equation*}
    \begin{split}
        \|(\Delta Z^0, \Delta Z^1)\|^2_{\mathbb{H}^2_{\mathrm{BMO}}}
       &\leq
       256C_\gamma^2R^2\|(\Delta z^0, \Delta z^1)\|^2_{\mathbb{H}^2_{\mathrm{BMO}}}.
    \end{split}
\end{equation*}
Under \eqref{small}, $\Gamma|_{\mathcal{B}_R}$ becomes a strict contraction. Indeed, having chosen $R$ by \eqref{R-choice}, we clearly have $256 C_\gamma^2R^2 < 1$. This yields that there exists a unique fixed point of $\Gamma|_{\mathcal{B}_R}$, which represents a bounded solution of the BSDE \eqref{MF-BSDE1}. $\square$

\begin{rem}
    Due to the uniqueness of the fixed point, the mean field BSDE \eqref{MF-BSDE1} has a unique solution if we restrict the domain to $\mathbb{S}^\infty\times\mathcal{B}_R$.
\end{rem}

\subsection{Asymptotic equilibrium in the large population limit}
We now prove that the optimal trading strategy in the market with risk premium process $\theta^{\mathrm{mfg}}$ defined by \eqref{mf-premium} satisfies the market clearing condition \eqref{MC-eqn} in the large population limit. 
To deal with the large population limit, we need to enlarge our probability space in the following way. Let $(\overline{\Omega},\overline{\mathcal{F}},\overline{\mathbb{P}})$ be an complete probability space defined on $\overline{\Omega}:=\prod_{i=0}^\infty\Omega^i$. Here, $(\overline{\mathcal{F}},\overline{\mathbb{P}})$ is the completion of $\Bigl(\bigotimes_{i=0}^\infty\mathcal{F}^i,\bigotimes_{i=0}^\infty\mathbb{P}^i\Bigr)$ and the filtration $\overline{\mathbb{F}}:=(\overline{\mathcal{F}}_t)_{t\in[0,T]}$ is the complete and right-continuous augmentation of $(\bigotimes_{i=0}^\infty\mathcal{F}^{i}_t)_{t\in[0,T]}$. In the remainder of this section, $\mathbb{E}[\cdot]$ denotes the expectation with respect to $\overline{\mathbb{P}}$.

Let us arbitrarily choose one bounded solution $(\mathcal{Y}^1,\mathcal{Z}^0,\mathcal{Z}^1)\in\mathbb{S}^\infty\times\mathbb{H}^2_{\mathrm{BMO}}\times\mathbb{H}^2_{\mathrm{BMO}}$ of the mean field BSDE
\begin{equation*}
    \begin{split}
            \mathcal{Y}^1_t
            &= 
            F^1_T + \int_t^T \Bigl\{\hat\gamma \mathcal{Z}^{0\|}_s\mathbb{E}[\mathcal{Z}^{0\|}_s|\mathcal{F}^0]^{\top} - \frac{\hat\gamma^2}{2\gamma^1}|\mathbb{E}[\mathcal{Z}^{0\|}_s|\mathcal{F}^0]|^2 + \frac{\gamma^1}{2}(|\mathcal{Z}^{0\perp}_s|^2 + |\mathcal{Z}^1_s|^2) -\frac{\gamma^1(1+b\zeta^1_s)}{\beta^1}\mathcal{Y}^1_s + g^1_s\Bigr\}ds\\
            &~~~~~~~~~~~~~~~~~ - \int_t^T \mathcal{Z}^{0}_s dW^0_s - \int_t^T \mathcal{Z}^1_s dW^1_s,~~~t\in[0,T],
    \end{split}
  \end{equation*}
where
\[
    g^1_s := - \frac{\delta}{\gamma^1} + (\kappa-b)\zeta^1_s\rho_s + \frac{1 + b\zeta^1_s}{\beta^1}\Bigl\{1 + \log\Bigl(\frac{a\beta^1}{\gamma^1(1 + b\zeta^1_s)}\Bigr)+\gamma^1 F^1_s\Bigr\},~~~s\in[0,T],
\]
and fix it. Theorem \ref{MF-BSDE-wellposed1} provides one such example under the appropriate conditions. Using this solution, we define the process $\theta^{\mathrm{mfg}}\in\mathbb{H}^2_{\mathrm{BMO}}(\mathbb{F}^0,\mathbb{R}^{d_0})$ by $\theta^{\mathrm{mfg}}_t := -\hat\gamma\mathbb{E}[\mathcal{Z}^{0\|}_t | \mathcal{F}^0]^\top$ for $t\in[0,T]$ as in \eqref{mf-premium}.\footnote{Notice that the process $\theta^{\mathrm{mfg}}$ is consistent with Assumption \ref{asm1} as a risk premium process.} \par
Recalling Section 2.2 and 3.1, if the market risk premium process is $\theta^{\mathrm{mfg}}$, the optimal trading strategy for agent-$i$ is given by
\[
    p^{i,*}_t := (\pi^{i,*}_t)^\top\sigma_t = Z^{i,0\|}_t + \frac{(\theta^{\mathrm{mfg}}_t)^\top}{\gamma^i} = Z^{i,0\|}_t - \frac{\hat\gamma}{\gamma^i}\mathbb{E}[\mathcal{Z}^{0\|}_t | \mathcal{F}^0],~~t\in[0,T].
\]
Here, $Z^{i,0}$ is a solution of the following \textit{non}-mean field BSDE:
\begin{equation}
    \begin{split}
        \label{asymp-non-MF-BSDE}
        Y^i_t
        &= 
        F^i_T + \int_t^T \Bigl\{-Z^{i,0\|}_s\theta^{\mathrm{mfg}}_s - \frac{|\theta^{\mathrm{mfg}}_s|^2}{2\gamma^i} + \frac{\gamma^i}{2}(|Z^{i,0\perp}_s|^2 + |Z^i_s|^2) -\frac{\gamma^i(1+b\zeta^i_s)}{\beta^i}Y^i_s + g^i_s\Bigr\}ds - \int_t^T Z^{i,0}_s dW^0_s - \int_t^T Z^i_s dW^i_s \\
        &=
        F^i_T + \int_t^T \Bigl\{\hat\gamma Z^{i,0\|}_s\mathbb{E}[\mathcal{Z}^{0\|}_s|\mathcal{F}^0]^{\top} - \frac{\hat\gamma^2}{2\gamma^i}|\mathbb{E}[\mathcal{Z}^{0\|}_s|\mathcal{F}^0]|^2 + \frac{\gamma^i}{2}(|Z^{i,0\perp}_s|^2 + |Z^i_s|^2) -\frac{\gamma^i(1+b\zeta^i_s)}{\beta^i}Y^i_s + g^i_s\Bigr\}ds\\
            &~~~~~~~~~~~~~~~~~ - \int_t^T Z^{i,0}_s dW^0_s - \int_t^T Z^i_s dW^i_s,~~t\in[0,T]
    \end{split}
  \end{equation}
with
\[
    g^i_s := - \frac{\delta}{\gamma^i} + (\kappa-b)\zeta^i_s\rho_s + \frac{1 + b\zeta^i_s}{\beta^i}\Bigl\{1 + \log\Bigl(\frac{a\beta^i}{\gamma^i(1 + b\zeta^i_s)}\Bigr)+\gamma^i F^i_s\Bigr\},~~~s\in[0,T].
\]
This equation has a unique bounded solution $(Y^i,Z^{i,0},Z^i)\in\mathbb{S}^\infty\times\mathbb{H}^2_{\mathrm{BMO}}\times\mathbb{H}^2_{\mathrm{BMO}}$ by Theorem \ref{sec2-well-posed}. With the above setup, we have the main result of this section. 

\begin{thm}~(Asymptotic equilibrium)\\
    \label{Asymp}
    Let Assumptions \ref{asm1} and \ref{asm3} be in force. Suppose that the mean field BSDE \eqref{MF-BSDE1} has a bounded solution, and that we arbitrarily choose and fix one such solution $(\mathcal{Y}^1,\mathcal{Z}^0,\mathcal{Z}^1)\in\mathbb{S}^\infty(\mathbb{P}^{0,1},\mathbb{F}^{0,1},\mathbb{R})\times \mathbb{H}^2_{\mathrm{BMO}}(\mathbb{P}^{0,1},\mathbb{F}^{0,1},\mathbb{R}^{1\times d_0})\times \mathbb{H}^2_{\mathrm{BMO}}(\mathbb{P}^{0,1},\mathbb{F}^{0,1},\mathbb{R}^{1\times d})$.
    Then, the process $\theta^{\mathrm{mfg}}$, defined by $\theta^{\mathrm{mfg}}_t := -\hat\gamma\mathbb{E}[\mathcal{Z}^{0\|}_t | \mathcal{F}^0]^\top$ for $t\in[0,T]$, clears the financial market in the large population limit in the sense that
    \begin{equation}
        \lim_{N\to\infty}\mathbb{E}\int_0^T \Bigl|\frac{1}{N}\sum_{i=1}^N \pi^{i,*}_t\Bigr|^2dt = 0,
    \end{equation}
    where $(\pi^{i,*}_t;t\in[0,T])_{i\in\mathbb{N}}$ are the agents' optimal trading strategies.
\end{thm}
\noindent
\textbf{\textit{proof}}\\
(Step I)\\
As mentioned above, the BSDE \eqref{asymp-non-MF-BSDE} with $i=1$ has a unique solution $(Y^1,Z^{1,0},Z^1)\in\mathbb{S}^\infty\times\mathbb{H}^2_{\mathrm{BMO}}\times\mathbb{H}^2_{\mathrm{BMO}}$ by Theorem \ref{sec2-well-posed}. Since $(\mathcal{Y}^1,\mathcal{Z}^0,\mathcal{Z}^1)$ obviously solves the same equation, the uniqueness implies $(Y^1,Z^{1,0},Z^1)=(\mathcal{Y}^1,\mathcal{Z}^0,\mathcal{Z}^1)$. In particular, we have $\theta^{\mathrm{mfg}}_t = -\hat\gamma\mathbb{E}[Z^{1,0\|}_t | \mathcal{F}^0]^\top$ for $t\in[0,T]$.

Moreover, the (strong) uniqueness implies that, for each $i\in\mathbb{N}$, there exists a measurable function $\Phi$ such that 
\[
    (Y_t^i,Z_t^{i,0},Z_t^i)_{t\in[0,T]}=\Phi (W^0,W^i,\xi^i,\gamma^i,\beta^i,X^i_0,F^i,\theta^{\mathrm{mfg}}),~~~ \mathbb{P}^{0,1}\text{-}\mathrm{a.s.},
\] 
by Yamada-Watanabe's theorem (See, for example, Carmona \& Delarue \cite{carmonaProbabilisticTheoryMean2018a} [Theorem 1.33]). It then follows that $\{(Y_t^i,Z_t^{i,0},Z_t^i);t\in[0,T]\}_{i\in\mathbb{N}}$ are $\mathcal{F}^0$-conditionally independently and identically distributed.\\

\noindent
(Step II)\\
Since $\pi^{i,*}_t=(\sigma_t\sigma_t^\top)^{-1}\sigma_t (p^{i,*}_t)^\top$ for $t\in[0,T]$ and $|(\sigma_t\sigma_t^\top)^{-1}\sigma_t|\leq C$ uniformly in $t$ by Assumption \ref{asm1}, we have
\begin{equation}
    \begin{split}
        \label{pi-and-p}
        \mathbb{E}\int_0^T \Bigl|\frac{1}{N}\sum_{i=1}^N \pi^{i,*}_t\Bigr|^2dt
        &\leq
        C\mathbb{E}\int_0^T \Bigl|\frac{1}{N}\sum_{i=1}^N p^{i,*}_t\Bigr|^2dt.
    \end{split}
\end{equation}
for all $N\in\mathbb{N}$. Moreover, it is clear that
\begin{equation*}
    \begin{split}
       \frac{1}{N}\sum_{i=1}^N p^{i,*}_t
       &=
       \frac{1}{N}\sum_{i=1}^N \Bigl(Z^{i,0\|}_t- \mathbb{E}[Z^{1,0\|}_t | \mathcal{F}^0]\Bigr) + \frac{1}{N}\sum_{i=1}^N \Bigl(1-\frac{\hat\gamma}{\gamma^i}\Bigr)\mathbb{E}[Z^{1,0\|}_t | \mathcal{F}^0].
    \end{split}
\end{equation*}
Then, we have the following estimate:
\begin{equation*}
    \begin{split}
       \mathbb{E}\int_0^T \Bigl|\frac{1}{N}\sum_{i=1}^N p^{i,*}_t\Bigr|^2dt 
       &\leq
       2\mathbb{E}\int_0^T \Bigl|\frac{1}{N}\sum_{i=1}^N \Bigl(Z^{i,0\|}_t- \mathbb{E}[Z^{1,0\|}_t | \mathcal{F}^0]\Bigr)\Bigr|^2 dt + 2\mathbb{E}\int_0^T \Bigl|\frac{1}{N}\sum_{i=1}^N \Bigl(1-\frac{\hat\gamma}{\gamma^i}\Bigr)\mathbb{E}[Z^{1,0\|}_t | \mathcal{F}^0]\Bigr|^2 dt\\
       &=
       2\mathbb{E}\int_0^T \Bigl|\frac{1}{N}\sum_{i=1}^N \Bigl(Z^{i,0\|}_t- \mathbb{E}[Z^{1,0\|}_t | \mathcal{F}^0]\Bigr)\Bigr|^2 dt + 2\mathbb{E}\Bigl[\Bigl|\frac{1}{N}\sum_{i=1}^N \Bigl(1-\frac{\hat\gamma}{\gamma^i}\Bigr)\Bigr|^2\Bigr]\mathbb{E}\int_0^T\Bigl|\mathbb{E}[Z^{1,0\|}_t | \mathcal{F}^0]\Bigr|^2 dt\\
       &\leq
       \frac{2}{N^2}\sum_{i=1}^N\mathbb{E}\int_0^T \Bigl|Z^{i,0\|}_t- \mathbb{E}[Z^{1,0\|}_t | \mathcal{F}^0]\Bigr|^2 dt + \frac{2}{N^2}\sum_{i=1}^N \mathbb{E}\Bigl[\Bigl|1-\frac{\hat\gamma}{\gamma^i}\Bigr|^2\Bigr]\mathbb{E}\int_0^T|Z^{1,0\|}_t |^2 dt\\
       &\leq
       \frac{4}{N}\Bigl(1+\frac{\hat\gamma^2}{\underline{\gamma}^2}\Bigr)\|Z^{1,0\|}\|^2_{\mathbb{H}^2}\\
       &\to
       0~~~~(N\to\infty).
    \end{split}
\end{equation*}
Here, we used the fact that $(\gamma^i)_{i\in\mathbb{N}}$ are i.i.d. random variables and that $(Z_t^{i,0})_{i\in\mathbb{N}}$ are $\mathcal{F}^0$-conditionally i.i.d. Together with \eqref{pi-and-p}, we get the desired result. $\square$

\section{Special solution for the exponential quadratic Gaussian model} \label{Section 4}
In this section, we reformulate the equilibrium model via the exponential quadratic Gaussian (EQG) framework. In the previous section, we made several strong assumptions to prove the existence of bounded solutions to the mean field BSDE \eqref{MF-BSDE1}. 
The EQG framework, on the other hand, provides a good example where unbounded solutions can be obtained under certain conditions. Since this framework allows us to have a semi-explicit representation of the solutions, it will help us to carry out detailed numerical analysis in the future works.
\subsection{Reformulation of the equilibrium model}
Suppose there are infinitely many agents in the common financial market. In this section, we assume that the coefficients of absolute risk aversion $(\gamma^i,\beta^i)_{i\in\mathbb{N}}$ are common to all agents and we hereafter denote their common values by $(\gamma,\beta)\in\mathbb{R}_{++}\times\mathbb{R}_{++}$. 
Since they are no longer random variables, we need a slight modification of the definition of the relevant probability spaces.\\

\noindent
(1) We denote by $(\Omega^0,\mathcal{F}^0,\mathbb{P}^0)$ a complete probability space with complete and right-continuous filtration $\mathbb{F}^0:=(\mathcal{F}^0_t)_{t\in[0,T]}$ generated by a $d_0$-dimensional standard Brownian motion $W^0:=(W^0_t)_{t\in[0,T]}$ with $\mathcal{F}^0:=\mathcal{F}^0_T$.
Also, we denote by $(\Omega^i,\mathcal{F}^i,\mathbb{P}^i)$ ($i\in\mathbb{N}$) a complete probability space with complete and right-continuous filtration $\mathbb{F}^i:=(\mathcal{F}^i_t)_{t\in[0,T]}$, generated by a $d$-dimensional standard Brownian motion $W^i:=(W^i_t)_{t\in[0,T]}$ and a $\sigma$-algebra $\sigma(\xi^i,X^i_0,x^i_0)$, where the completion of the latter gives $\mathcal{F}^i_0$. We set $\mathcal{F}^i:=\mathcal{F}^i_T$. 
Here, $(\xi^i,X^i_0)$ are $\mathbb{R}$-valued random variables and $x^i_0$ is an $\mathbb{R}^d$-valued random variable. \\

\noindent
(2) We denote by $(\Omega^{0,i},\mathcal{F}^{0,i},\mathbb{P}^{0,i})$ ($i\in\mathbb{N}$) a complete probability space with $\Omega^{0,i} := \Omega^0 \times \Omega^i$ and with $(\mathcal{F}^{0,i},\mathbb{P}^{0,i})$, the completion of $(\mathcal{F}^0 \otimes \mathcal{F}^i,\mathbb{P}^{0}\otimes \mathbb{P}^{i})$. 
We denote by $\mathbb{F}^{0,i}:=(\mathcal{F}^{0,i}_t)_{t\in[0,T]}$ the complete and right-continuous augmentation of $(\mathcal{F}_t^0 \otimes \mathcal{F}_t^i)_{t\in[0,T]}$.\\

\noindent
(3) Let $(\Omega,\mathcal{F},\mathbb{P})$ be a complete probability space defined by $\Omega:=\prod_{i=0}^\infty\Omega^i$ and $(\mathcal{F},\mathbb{P})$, the completion of $\Bigl(\bigotimes_{i=0}^\infty\mathcal{F}^i,\bigotimes_{i=0}^\infty\mathbb{P}^i\Bigr)$. The filtration $\mathbb{F}:=(\mathcal{F}_t)_{t\in[0,T]}$ is the complete and right-continuous augmentation of $(\bigotimes_{i=0}^\infty\mathcal{F}^{i}_t)_{t\in[0,T]}$.\\

Let us first give a new assumption on the market as follows.
\begin{asm} 
    \label{asmEQG-market}~\\
    \textup{(i)} The risk-free interest rate is zero.\\
    \textup{(ii)} There are $n\in\mathbb{N}$ non-dividend paying risky stocks whose price dynamics, represented by an $n$-dimensional vector, are given by
    \begin{equation}
        \begin{split}
            \label{stock price-EQG}
            S_t&= S_0 + \int_0^t \mathrm{diag}(S_r)(\mu_rdr + \sigma_r dW^0_r),~~t\in[0,T],
        \end{split}
    \end{equation}
    for $S_0\in\mathbb{R}^n_{++}$, $\mu := (\mu_t)_{t\in[0,T]}\in\mathbb{H}^2(\mathbb{P}^{0},\mathbb{F}^0,\mathbb{R}^n)$ and $\sigma :=(\sigma_t)_{t\in[0,T]}\in\mathbb{L}^\infty(\mathbb{P}^{0},\mathbb{F}^0,\mathbb{R}^{n\times d_0})$. We also assume $n\leq d_0$.\\
    \textup{(iii)} The process $(\sigma_t)_{t\in[0,T]}$ is of the form $\sigma_t = (\hat{\sigma}_t, \check{\sigma}_t)$ for each $t\in[0,T]$, where $(\hat{\sigma}_t)_{t\in[0,T]}\in\mathbb{L}^\infty(\mathbb{P}^{0},\mathbb{F}^0,\mathbb{R}^{n\times n})$ is a process such that $\hat{\sigma}_t$ is invertible for all $t\in[0,T]$ and $(\check{\sigma}_t)_{t\in[0,T]}\in\mathbb{L}^\infty(\mathbb{P}^{0},\mathbb{F}^0,\mathbb{R}^{n\times (d_0-n)})$. Moreover, $(\sigma_t)_{t\in[0,T]}$ satisfies
    \[
        \underline{\lambda}I_n\leq (\sigma_t\sigma_t^\top)\leq\overline{\lambda}I_n,~~~~dt\otimes \mathbb{P}^0\text{-}\mathrm{a.e.}
    \]
    for some positive constants $0<\underline{\lambda}<\overline{\lambda}$ and $I_n$, an identity matrix of size $n$.\\
    \textup{(iv)} The risk premium process $\theta\in\mathbb{L}^0(\mathbb{F}^0,\mathbb{R}^{d_0})$, defined by $\theta_t = \sigma_t^\top(\sigma_t\sigma_t^\top)^{-1}\mu_t$ for $t\in[0,T]$, is a process such that the Dol\'{e}ans-Dade exponential $\displaystyle\Bigl\{\mathcal{E}\Bigl(-\int_0^\cdot \theta_s^\top dW^0_s\Bigr)_t; t\in[0,T]\Bigr\}$ is a martingale of class $\mathcal{D}$.
\end{asm}

\begin{rem}~\\
    \textup{(i)} Under Assumption \ref{asmEQG-market} (iii), the linear subspace $L_t$ defined in Definition \ref{subspace-L} are spanned by first $n$-standard bases of $\mathbb{R}^{1\times d_0}$ for all $t\in[0,T]$. We use the symbol $L$ instead of $L_t$ in this section. In addition, we denote by $\Pi$ the orthogonal projection of $\mathbb{R}^{1\times d_0}$ onto $L$.\\
    \textup{(ii)} Unlike Assumption \ref{asm1}, the process $\mu$ is no longer in $\mathbb{H}^2_{\mathrm{BMO}}$ and thus so is $\theta$. Despite this, the well-posedness of the stock price process $(S_t)_{t\in[0,T]}$ can still be shown by changing the original measure $\mathbb{P}^0$ to $\mathbb{Q}$, the risk neutral measure defined by \eqref{risk-neutral}, which is possible thanks to Assumption \ref{asmEQG-market} (iv).
\end{rem}

\begin{asm} ~\\
    \label{asm6}
    \textup{(i)} For each $i\in\mathbb{N}$, $\xi^i$ and $X^i_0$ are $\mathbb{R}$-valued, $\mathcal{F}^i_0$-measurable, and normally-distributed random variables representing agent-$i$'s initial wealth and initial consumption habit, respectively. $x^i_0$ is an $\mathbb{R}^d$-valued, $\mathcal{F}^i_0$-measurable, and normally-distributed random variable.
    
    \noindent
    \textup{(ii)} The random variables $\xi^i,X^i_0$ and $x_0^i$ are mutually independent for each $i\in\mathbb{N}$ and $(\xi^i,X^i_0,x_0^i)_{i\in\mathbb{N}}$ have the same distribution, i.e. they are independently and identically distributed on $(\Omega,\mathcal{F},\mathbb{P})$.
    
    \noindent
    \textup{(iii)} $(\gamma, \beta)\in\mathbb{R}_{++}\times\mathbb{R}_{++}$ are the coefficients of absolute risk aversion for agents' net wealth and consumption, respectively. In particular, they are common to all agents. 

    \noindent
    \textup{(iv)} The habit trend $\rho:[0,T]\to\mathbb{R}$ is a continuous function of time.

    \noindent
    \textup{(v)} For each $i\in\mathbb{N}$, the liability process $(F^i_t;t\in[0,T])_{i\in\mathbb{N}}$ is $\mathbb{R}$-valued and $\mathbb{F}^{0,i}$-progressively measurable, which is given by a quadratic form\footnote{The symbol $\langle \cdot,\cdot\rangle$ denotes the Euclidean inner product, i.e. $\langle x, y\rangle:=x^\top y$ for $x,y\in\mathbb{R}^n$.}
    \begin{equation}
        \begin{split}
        \label{EQG-F}
            F^i_t := \frac{1}{2} \langle A^{F}_{00}(t)x^0_t,x^0_t\rangle + \frac{1}{2} \langle A^{F}_{11}(t)x^i_t,x^i_t\rangle + \langle A^{F}_{10}(t)x^0_t,x^i_t\rangle + \langle B^{F}_0(t),x^0_t\rangle + \langle B_1^{F}(t),x^i_t\rangle + C^{F}(t),~~~t\in[0,T],
        \end{split}
    \end{equation}
    for $(A^{F}_{00}, A^{F}_{11}, A^{F}_{10},B^F_0,B^F_1,C^F)\in\mathcal{C}([0,T];\mathbb{M}_{d_0})\times\mathcal{C}([0,T];\mathbb{M}_{d})\times\mathcal{C}([0,T];\mathbb{R}^{d\times d_0})\times\mathcal{C}([0,T];\mathbb{R}^{d_0})\times\mathcal{C}([0,T];\mathbb{R}^{d})\times\mathcal{C}([0,T];\mathbb{R})$ and the Gaussian factor processes $(x^0,x^i)\in\mathbb{L}^0(\mathbb{F}^0,\mathbb{R}^{d_0})\times\mathbb{L}^0(\mathbb{F}^i,\mathbb{R}^{d})$ defined by
    \begin{equation*}
        \begin{split}
            x_t^0 = x^0_0 -\int_0^t K_0(x^0_s - m_0)ds + \Sigma_0 W_t^0,~~~x_t^i = x_0^i -\int_0^t K(x^i_s - m)ds + \Sigma W_t^i,~~~t\in[0,T]
        \end{split}
    \end{equation*}
    with\footnote{This method is still available with time-dependent deterministic and continuous coefficients $(m_0(t),m(t),K_0(t),K(t),\Sigma_0(t),\Sigma(t))$. For simplicity, however, we only consider the constant case in this paper.} $x^0_0\in\mathbb{R}^{d_0}$, $(K_0,K)\in\mathbb{R}_{++}\times \mathbb{R}_{++}$, $(m_0,m)\in\mathbb{R}^{d_0}\times \mathbb{R}^{d}$, and $(\Sigma_0,\Sigma)\in\mathbb{R}^{d_0\times d_0}\times \mathbb{R}^{d\times d}$.

    \noindent
    \textup{(vi)} Each agent is a price taker; agent-$i$ must accept the prevailing prices as he/she lacks the market share to impact the market price.
\end{asm}

\begin{rem}
    In this model, the agents are heterogeneous in the idiosyncratic noises $(W^i)_{i\in\mathbb{N}}$, initial wealths $(\xi^i)_{i\in\mathbb{N}}$, initial habits $(X^i_0)_{i\in\mathbb{N}}$, and initial conditions $(x^i_0)_{i\in\mathbb{N}}$ for the factor processes which affect the liabilities $(F^i)_{i\in\mathbb{N}}$.
\end{rem}

The agents' problems are modelled on the probability space $(\Omega,\mathcal{F},\mathbb{P},\mathbb{F})$. For each $i\in\mathbb{N}$, agent-$i$ solves the following utility maximization problem:
\begin{equation}
    \begin{split}
        \sup_{(\pi,c)\in\mathbb{A}^i_{\mathrm{EQG}}} {U}^i(\pi,c)
    \end{split}
\end{equation}
subject to
\[
    \mathcal{W}^{i,(\pi,c)}_t =\xi^i + \int_0^t (\pi_s^\top\sigma_s\theta_s - c_s)ds + \int_0^t \pi_s^\top\sigma_s dW_s^0,~~t\in[0,T],
\]
where $\mathbb{A}^i_{\mathrm{EQG}}$ is the admissible set for agent-$i$, whose definition is to be given. The utility function ${U}^i:\mathbb{A}^i_{\mathrm{EQG}}\to\mathbb{R}$ is defined by
\[
    {U}^i(\pi,c):=\mathbb{E}\Bigl[-\exp\Bigl(-\delta T-\gamma(\mathcal{W}^{i,(\pi,c)}_T-F^i_T)\Bigr) -a \int_0^T \exp\Bigl(-\delta t-\gamma(\mathcal{W}^{i,(\pi,c)}_t-F^i_t)-\beta(c_t-X_t^{i,c})\Bigr)dt\Bigr],
\]
with some common parameters $a,\delta>0$. The process $X^{i,c}$ represents the agent-$i$'s consumption habits and is defined by
\begin{equation}
    \begin{split} 
        \label{EQG-habit}
        X_t^{i,c} = X^i_0 + \int_0^t \{-\kappa(X^{i,c}_s-\rho_s) + b(c_s-\rho_s)\}ds, ~~t\in[0,T]
    \end{split}
\end{equation}
for some constants $\kappa,b>0$, which are also common to all agents. 
As usual, by setting $(p_t)_{t\in[0,T]}:=(\pi^\top_t\sigma_t)_{t\in[0,T]}$, the utility maximization problem can be equivalently written as
\begin{equation*}
    \begin{split}
        \sup_{(p,c)\in\mathcal{A}^i_{\mathrm{EQG}}} \widetilde{U}^i(p,c)
    \end{split}
\end{equation*}
subject to
\[
    \mathcal{W}^{i,(p,c)}_t =\xi^i + \int_0^t (p_s\theta_s - c_s)ds + \int_0^t p_s dW_s^0,~~t\in[0,T],
\]
where the set $\mathcal{A}^i_{\mathrm{EQG}}$ is defined by $\mathcal{A}_{\mathrm{EQG}}^i:=\{(p,c)=(\pi^\top\sigma,c);(\pi,c)\in\mathbb{A}^i_{\mathrm{EQG}}\}$ and the objective function $\widetilde{U}^i:\mathcal{A}^i_{\mathrm{EQG}}\to\mathbb{R}$ is defined by
\begin{equation}
    \label{tilde-U}
    \widetilde{U}^i(p,c):=\mathbb{E}\Bigl[-\exp\Bigl(-\delta T-\gamma(\mathcal{W}^{i,(p,c)}_T-F^i_T)\Bigr) -a \int_0^T \exp\Bigl(-\delta t-\gamma(\mathcal{W}^{i,(p,c)}_t-F^i_t)-\beta(c_t-X_t^{i,c})\Bigr)dt\Bigr].
\end{equation}

Under these assumptions, we define the process $R^{i,(p,c)}$ in analogy with Section 2.2 in the following way: for each $i\in\mathbb{N}$, we set
\begin{equation*}
    \begin{split}
        R^{i,(p,c)}_t := -\exp\Bigl(-\delta t-\gamma(\mathcal{W}^{i,(p,c)}_t-y^i_t-\zeta_tX_t^{i,c})\Bigr) -a \int_0^t \exp\Bigl(-\delta s-\gamma(\mathcal{W}^{i,(p,c)}_s-F^i_s)-\beta(c_s-X_s^{i,c})\Bigr)ds,~~t\in[0,T],
    \end{split}
\end{equation*}
where the process $(y^i_t)_{t\in[0,T]}$ is a solution to the BSDE \footnote{The solution is denoted by lower case letters in order to avoid confusion with the solution of the mean field BSDE \eqref{EQG-mfBSDE}, which is denoted by $(Y,Z^{i,0},Z^i)$.}:
\begin{equation}
    \begin{split}
        \label{std BSDE}
        y^i_t = F^i_T + \int_t^T \Bigl\{-z^{i,0\|}_s\theta_s - \frac{|\theta_s|^2}{2\gamma} + \frac{\gamma}{2}(|z^{i,0\perp}_s|^2 + |z^i_s|^2) -\frac{\gamma(1+b\zeta_s)}{\beta}y^i_s + g^i_s\Bigr\}ds - \int_t^T z^{i,0}_s dW^0_s - \int_t^T z^i_s dW^i_s
    \end{split}
  \end{equation}
with
\[
    g^i_s := - \frac{\delta}{\gamma} + (\kappa-b)\zeta_s\rho_s + \frac{1 + b\zeta_s}{\beta}\Bigl\{1 + \log\Bigl(\frac{a\beta}{\gamma(1 + b\zeta_s)}\Bigr)+\gamma F^i_s\Bigr\},
\]
and
\begin{equation*}
    \zeta_t := \frac{e^{(\delta^+-\delta^-)(T-t)}-1}{\delta^+-\delta^-e^{(\delta^+-\delta^-)(T-t)}},~~\delta^{\pm}:=-A\pm\sqrt{A^2+B},~~~A:=\frac{1}{2}\Bigl(\kappa-b+\frac{\gamma}{\beta}\Bigr),~~~B:=\frac{\gamma b}{\beta}
\end{equation*}
for $t\in[0,T]$. Moreover, we say that the process $R^{i,(p,c)}$ satisfies the condition-R if all conditions in Definition \ref{condition-R} with ``1'' replaced by ``$i$'' and $(\gamma^i,\beta^i)$ replaced by $(\gamma,\beta)$ hold. In order to work within this framework, we further need to modify the notion of admissibility.
\begin{dfn} (Admissible space for an EQG model)\\
    For each $i\in\mathbb{N}$, the admissible space $\mathbb{A}^i_{\mathrm{EQG}}$ is the set of $\mathbb{F}^{0,i}$-progressively measurable strategies $(\pi,c)\in\mathbb{H}^2(\mathbb{P}^{0,i},\mathbb{F}^{0,i},\mathbb{R}^{n})\times\mathbb{H}^2(\mathbb{P}^{0,i},\mathbb{F}^{0,i},\mathbb{R})$ which make the utility function finite (namely $U^i(\pi,c)>-\infty$) and the set $\{R^{i,(p,c)}_\tau;\tau\in\mathcal{T}^{0,i}\}$ uniformly integrable. 
\end{dfn}

We shall write the admissible space by $\mathcal{A}^i_{\mathrm{EQG}}(\theta)$ when we want to emphasize its dependence on the risk-premium process $\theta$. In a similar way as in Section 3, the market clearing condition motivates us to study the following mean field BSDE defined on the filtered probability space $(\Omega^{0,i},\mathcal{F}^{0,i},\mathbb{P}^{0,i},\mathbb{F}^{0,i})$ for each $i\in\mathbb{N}$:
\begin{equation}
    \begin{split}
        \label{EQG-mfBSDE}
            Y^i_t&= F^i_T + \int_t^T f^i(s,Y^i_s,Z^{i,0}_s,Z^i_s)ds - \int_t^T Z^{i,0}_s dW^0_s - \int_t^T Z^i_s dW^i_s,~~~t\in[0,T],
    \end{split}
  \end{equation}
where (note that we have $\hat\gamma = \gamma$ by Assumption \ref{asm6} (iii))
\begin{equation*}
    \begin{split}
        f^i(s,Y^i_s,Z^{i,0}_s,Z^i_s)
        &= 
        \gamma Z^{i,0\|}_s\mathbb{E}[Z^{i,0\|}_s|\mathcal{F}^0]^{\top} - \frac{\gamma}{2}|\mathbb{E}[Z^{i,0\|}_s|\mathcal{F}^0]|^2 + \frac{\gamma}{2}(|Z^{i,0\perp}_s|^2 + |Z^i_s|^2) -\frac{\gamma(1+b\zeta_s)}{\beta}Y^i_s + g^i_s.
    \end{split}
  \end{equation*}
By completing the square, the driver $f^i$ can be written as
\begin{equation*}
    \begin{split}
        f^i(s,Y^i_s,Z^{i,0}_s,Z^i_s)
        &=
        -\frac{\gamma}{2}\Bigl|\mathbb{E}[Z^{i,0\|}_s|\mathcal{F}^0] - Z^{i,0\|}_s\Bigr|^2 + \frac{\gamma}{2}(|Z^{i,0}_s|^2 + |Z^i_s|^2) -\frac{\gamma(1+b\zeta_s)}{\beta}Y^i_s + {g}^i_s.
    \end{split}
  \end{equation*}

\subsection{Mean field BSDE and the system of ODEs}
We now derive a system of ordinary differential equations (ODEs) which provides a solution of the mean field BSDE through the EQG modelling. 
Our approach basically follows Fujii \& Takahashi \cite{fujiiMakingMeanvarianceHedging2014} [Section 5], which proposes the method of associating the solution of the quadratic growth BSDE with the Riccati matrix equation. 
As a heuristic argument, if $Y^i$ is a quadratic form of $(x^0,x^i)$, its drift term is expected to be a quadratic form of $(x^0,x^i)$, and its diffusion terms are expected to be affine in $(x^0,x^i)$ by applying Ito formula. On the other hand, as the driver $f^i$ of the BSDE \eqref{EQG-mfBSDE} is quadratic in $(Z^{i,0},Z^i)$ and is linear in $Y^i$, it is anticipated that $f^i$ is a quadratic form of $(x^0,x^i)$ as well.
These observations imply that such an ansatz for $Y^i$ seems to be consistent and we thus search for a solution of the form:
\begin{equation}
    \label{Y-ansatz}
    Y^i_t = \frac{1}{2} \langle A^i_{00}(t)x^0_t,x^0_t\rangle + \frac{1}{2} \langle A^i_{11}(t)x^i_t,x^i_t\rangle + \langle A^i_{10}(t)x^0_t,x^i_t\rangle + \langle B^i_0(t),x^0_t\rangle + \langle B^i_1(t),x^i_t\rangle + C^i(t),~~t\in[0,T]
\end{equation}
for some processes $(A^i_{00},A^i_{11},A^i_{10},B^i_0,B^i_1,C^i):[0,T]\times\Omega\to\mathbb{M}_{d_0}\times\mathbb{M}_{d}\times\mathbb{R}^{d\times d_0}\times\mathbb{R}^{d_0}\times\mathbb{R}^{d}\times\mathbb{R}$, all of which are to be determined. At this moment, let us temporarily assume that $(A^i_{00},A^i_{11},A^i_{10},B^i_0,B^i_1,C^i)$ are once continuously time-differentiable and independent of $(\xi^i,X_0^i,x^i_0,W^0,W^i)$, i.e. they are deterministic functions of time common to all agents. After deriving the relevant ODEs, we shall verify this property. Since we search for functions common to all agents, we simply write $(A_{00},A_{11},A_{10},B_0,B_1,C)$ instead of $(A^i_{00},A^i_{11},A^i_{10},B^i_0,B^i_1,C^i)$ from now on. \par
As usual, we choose agent-1 as a representative agent and omit the superscript ``1" when there is no confusion. By applying Ito formula to \eqref{Y-ansatz}, we have
\begin{equation}
    \begin{split}
        \label{Y-ito}
        dY_t 
        &=
        \Bigl\{\Bigl\langle \Bigl(\frac{1}{2} \dot{A}_{00}(t) - K_0 A_{00}(t)\Bigr)x^0_t,x^0_t\Bigr\rangle + \Bigl\langle \Bigl(\frac{1}{2} \dot{A}_{11}(t) - K A_{11}(t)\Bigr)x^1_t,x^1_t\Bigr\rangle + \Bigl\langle \Bigl( \dot{A}_{10}(t) - (K_0+K)A_{10}(t)\Bigr)x^0_t,x^1_t\Bigr\rangle \Bigr.~~~~ \\
        &~~~+ \langle \dot{B}_0(t)-K_0B_0(t) + K_0A_{00}(t)m_0 + K A_{10}(t)^\top m,x^0_t\rangle + \langle \dot{B}_1(t)-K B_1(t) + K A_{11}(t)m + K_0A_{10}(t)m_0,x^1_t\rangle \\
        &~~~+ \Bigl.\dot{C}(t) + \langle K_0 B_0(t) ,m_0 \rangle + \langle K B_1(t) ,m \rangle +\frac{1}{2}\mathrm{tr}[A_{00}(t)\Sigma_0\Sigma_0^\top] + \frac{1}{2}\mathrm{tr}[A_{11}(t)\Sigma\Sigma^\top] \Bigr\}dt \\
        &~~~+   \langle\Sigma_0^\top (A_{00}(t)x^0_t + A_{10}(t)^\top x^1_t + B_0(t)) ,dW^0_t \rangle + \langle \Sigma^\top (A_{10}(t)x^0_t + A_{11}(t)x^1_t + B_1(t)) , dW^1_t \rangle.
    \end{split}
\end{equation}
In order for $Y$ given in \eqref{Y-ansatz} to be the solution to the mean field BSDE \eqref{EQG-mfBSDE}, we must have
\begin{equation*}
    \begin{split}
        Z_t^0 = \Bigl\{\Sigma_0^\top (A_{00}(t)x^0_t + A_{10}(t)^\top x^1_t + B_0(t))\Bigr\}^\top,~~~Z_t^1 = \Bigl\{\Sigma^\top (A_{10}(t)x^0_t + A_{11}(t)x^1_t + B_1(t))\Bigr\}^\top,~~~t\in[0,T].
    \end{split}
\end{equation*}
To deal with the process $Z^{0\|}$, let us write $\Sigma_0 = \hat{\Sigma}_0 + \check{\Sigma}_0$, where $\hat{\Sigma}_0, \check{\Sigma}_0 \in \mathbb{R}^{d_0\times d_0}$ are of the forms:
\[
    \hat{\Sigma}_0 = (\hat{\Sigma}^n_0~~ 0),~~\check{\Sigma}_0 = (0~~ \check{\Sigma}^{d_0-n}_0),
\]
for $\hat{\Sigma}^n_0\in\mathbb{R}^{d_0\times n}$ and $\check{\Sigma}^{d_0-n}_t\in\mathbb{R}^{d_0\times (d_0-n)}$, so that we have $\Pi(u^\top\Sigma_0)=u^\top \hat{\Sigma}_0$ for any $u\in\mathbb{R}^{d_0}$. In addition, it is easy to see
\begin{equation*}
    \begin{split}
        \mathbb{E}[x_t^0 |\mathcal{F}^0] = x_t^0,~~~\mu^1_t := \mathbb{E}[x_t^1 |\mathcal{F}^0] = \mathbb{E}[x_t^1] = \mathbb{E}[x_0^1]e^{-K t}+m(1-e^{-K t}),~~~t\in[0,T].
    \end{split}
\end{equation*}
Then we obtain:
\begin{equation*}
    \begin{split}
        Z_t^{0\|} = \Bigl\{\hat{\Sigma}_0^\top (A_{00}(t)x^0_t + A_{10}(t)^\top x^1_t + B_0(t))\Bigr\}^\top,~~~\mathbb{E}[Z_t^{0\|} |\mathcal{F}^0] = \Bigl\{\hat{\Sigma}_0^\top (A_{00}(t)x^0_t + A_{10}(t)^\top \mu^1_t + B_0(t)) \Bigr\}^\top,~~~t\in[0,T].
    \end{split}
\end{equation*}
Plugging these results into the driver $f$, we have: for $t\in[0,T]$,
\begin{equation}
    \begin{split}
        \label{f-EQG}
    &f(t,Y_t,Z^0_t,Z^1_t) \\
    &= 
    -\frac{\gamma}{2} \Bigl|\mathbb{E}[Z^{0\|}_t|\mathcal{F}^0] - Z^{0\|}_t\Bigr|^2 + \frac{\gamma}{2}(|Z^{0}_t|^2 + |Z^1_t|^2) -\frac{\gamma(1+b\zeta_t)}{\beta}(Y_t-F_t) + \widetilde{g}_t\\
    &=
    \Bigl\langle \Bigl\{\frac{\gamma}{2}\Bigl(A_{00}(t)\Sigma_0\Sigma_0^\top A_{00}(t) + A_{10}(t)^\top \Sigma\Sigma^\top A_{10}(t) \Bigr) -\frac{\gamma(1+b\zeta_t)}{2\beta}(A_{00}(t)-A^F_{00}(t))\Bigr\}x_t^0,x_t^0 \Bigr\rangle\\
    &~~~+\Bigl\langle \Bigl\{\frac{\gamma}{2}( A_{10}(t) \check{\Sigma}_0\check{\Sigma}_0^\top A_{10}(t)^\top + A_{11}(t) \Sigma\Sigma^\top A_{11}(t)) -\frac{\gamma(1+b\zeta_t)}{2\beta}(A_{11}(t)-A^F_{11}(t))\Bigr\} x_t^1 ,x_t^1 \Bigr\rangle \\
    &~~~+\Bigl\langle \Bigl\{\gamma( A_{10}(t)\Sigma_0\Sigma_0^\top A_{00}(t) + A_{11}(t) \Sigma\Sigma^\top A_{10}(t) )-\frac{\gamma(1+b\zeta_t)}{\beta}(A_{10}(t)-A^F_{10}(t)) \Bigr\}x_t^0,x_t^1 \Bigr\rangle \\
    &~~~+\Bigl\langle \gamma (A_{00}(t)\Sigma_0\Sigma_0^\top B_{0}(t) + A_{10}(t)^\top\Sigma\Sigma^\top B_{1}(t)) -\frac{\gamma(1+b\zeta_t)}{\beta}(B_{0}(t)-B^F_{0}(t)) ,x_t^0 \Bigr\rangle \\
    &~~~+\Bigl\langle \gamma (A_{10}(t)\hat{\Sigma}_0\hat{\Sigma}_0^\top A_{10}(t)^\top \mu_t^1 + A_{10}(t)\Sigma_0\Sigma_0^\top B_0(t) + A_{11}(t)\Sigma\Sigma^\top B_1(t)) - \frac{\gamma(1+b\zeta_t)}{\beta}(B_{1}(t)-B^F_{1}(t)),x_t^1 \Bigr\rangle\\
    &~~~- \frac{\gamma}{2}\langle A_{10}(t)\hat{\Sigma}_0\hat{\Sigma}_0^\top A_{10}(t)^\top \mu_t^1,\mu_t^1 \rangle + \frac{\gamma}{2}\langle \Sigma_0^\top B_0(t),\Sigma_0^\top B_0(t) \rangle + \frac{\gamma}{2}\langle \Sigma^\top B_1(t),\Sigma^\top B_1(t) \rangle-\frac{\gamma(1+b\zeta_t)}{\beta}(C(t)-C^F(t)) + \widetilde{g}_t,
    \end{split}
\end{equation}
where $\widetilde{g}$ is a deterministic and continuous function defined by:
\[
    \widetilde{g}_t := g_t - \frac{\gamma(1+b\zeta_t)}{\beta} F_t = - \frac{\delta}{\gamma} + (\kappa-b)\zeta_t\rho_t + \frac{1 + b\zeta_t}{\beta}\Bigl\{1 + \log\Bigl(\frac{a\beta}{\gamma(1 + b\zeta_t)}\Bigr)\Bigr\},~~~t\in[0,T].
\]

By matching \eqref{f-EQG} and the drift term of \eqref{Y-ito} with respect to the quadratic or linear coefficients of $(x^0,x^1)$ as well as the remaining constant terms, we obtain: for $t\in[0,T]$,
\begin{equation}
    \begin{split}
        \label{Riccati eqn}
        &\dot{A}_{00}(t) = -\gamma A_{00}(t)\Sigma_0\Sigma_0^\top A_{00}(t)  - \gamma A_{10}(t)^\top \Sigma\Sigma^\top A_{10}(t) + \Bigl(2K_0 + \frac{\gamma(1+b\zeta_t)}{\beta}\Bigr) A_{00}(t) - \frac{\gamma(1+b\zeta_t)}{\beta} A^F_{00}(t), \\
        &\dot{A}_{11}(t) = -\gamma A_{11}(t) \Sigma\Sigma^\top A_{11}(t)  - \gamma A_{10}(t) \check{\Sigma}_0\check{\Sigma}_0^\top A_{10}(t)^\top + \Bigl(2K + \frac{\gamma(1+b\zeta_t)}{\beta}\Bigr) A_{11}(t) - \frac{\gamma(1+b\zeta_t)}{\beta} A^F_{11}(t), \\
        &\dot{A}_{10}(t)  = -\gamma A_{10}(t)\Sigma_0\Sigma_0^\top A_{00}(t) - \gamma A_{11}(t) \Sigma\Sigma^\top A_{10}(t) + \Bigl((K_0+K)+\frac{\gamma(1+b\zeta_t)}{\beta}\Bigr)A_{10}(t) - \frac{\gamma(1+b\zeta_t)}{\beta} A^F_{10}(t),\\
        &\dot{B}_0(t)=\Bigl(- \gamma A_{00}(t)\Sigma_0\Sigma_0^\top  + \frac{\gamma(1+b\zeta_t)}{\beta} + K_0\Bigr)B_{0}(t) - \gamma A_{10}(t)^\top\Sigma\Sigma^\top B_{1}(t) - \frac{\gamma(1+b\zeta_t)}{\beta}B^F_{0}(t) - K_0A_{00}(t)m_0 - KA_{10}(t)^\top m,\\
        &\dot{B}_1(t)= \Bigl(-\gamma A_{11}(t)\Sigma\Sigma^\top + \frac{\gamma(1+b\zeta_t)}{\beta} + K\Bigr)B_1(t) - \gamma \Bigl(A_{10}(t)\hat{\Sigma}_0\hat{\Sigma}_0^\top A_{10}(t)^\top \mu_t^1 + A_{10}(t)\Sigma_0\Sigma_0^\top B_0(t)\Bigr) \\
        &~~~~~~~~~~~~~- \frac{\gamma(1+b\zeta_t)}{\beta}B^F_{1}(t) - KA_{11}(t)m - K_0A_{10}(t)m_0, \\
        &\dot{C}(t)= \frac{\gamma(1+b\zeta_t)}{\beta}C(t) - \frac{\gamma(1+b\zeta_t)}{\beta}C^F(t) - \frac{\gamma}{2}\langle \Sigma_0^\top B_0(t),\Sigma_0^\top B_0(t) \rangle - \frac{\gamma}{2} \langle \Sigma^\top B_1(t),\Sigma^\top B_1(t) \rangle - \langle K_0 B_0(t) ,m_0 \rangle - \langle K B_1(t) ,m \rangle \\
        &~~~~~~~~~~~~~+ \frac{\gamma}{2}\langle A_{10}(t)\hat{\Sigma}_0\hat{\Sigma}_0^\top A_{10}(t)^\top \mu_t^1,\mu_t^1 \rangle - \frac{1}{2}\mathrm{tr}[A_{00}(t)\Sigma_0\Sigma_0^\top] - \frac{1}{2}\mathrm{tr}[A_{11}(t)\Sigma\Sigma^\top] - \widetilde{g}_t,\\
        &A_{00}(T)=A_{00}^F(T),~~ A_{11}(T)=A_{11}^F(T), ~~A_{10}(T)=A_{10}^F(T), ~~B_0(T)=B_0^F(T), ~~B_1(T)=B_1^F(T), ~~C(T)=C^F(T).
    \end{split}
\end{equation}
Here, the terminal conditions for $(A_{00},A_{11},A_{10},B_0,B_1,C)$ are set to satisfy $Y_T=F_T$.

\begin{rem}~\\
    \textup{(i)}  The equations for $(A_{00},A_{11},A_{10})$ are of Riccati type. In this paper, however, we do not delve into the general well-posedness result due to its complexity.\\
    \textup{(ii)} Since the coefficients appeared in \eqref{Riccati eqn} are all deterministic and in particular, independent of $(\xi^i,X_0^i,x^i_0,W^0,W^i)$, we deduce that $(A_{00},A_{11},A_{10},B_0,B_1,C)$ are deterministic function of time and common to all agents if they exists.\\
    \textup{(iii)} By the local Lipschitz condition, the equation \eqref{Riccati eqn} has a locally unique solution. Furthermore, by making $|\Sigma_0|$ and $|\Sigma|$ sufficiently small, we expect to have also a global solution since the Riccati equation for $(A_{00},A_{11},A_{10})$ becomes approximately linear.\\
    \textup{(iv)} We may possibly allow heterogeneity among the coefficients of risk aversion $(\gamma^i,\beta^i)_{i\in\mathbb{N}}$ as in Section 3. However, in this case, the system of equations \eqref{Riccati eqn} becomes mean field type and checking its well-posedness would be much harder.
\end{rem}

These observations result in the following theorem. 
\begin{thm}
    \label{EQG solution}
    Let Assumption \ref{asmEQG-market} and \ref{asm6} be in force. In addition, assume that the equation \eqref{Riccati eqn} has a global solution $(A_{00},A_{11},A_{10},B_0,B_1,C)\in\mathcal{C}^1([0,T];\mathbb{M}_{d_0})\times\mathcal{C}^1([0,T];\mathbb{M}_{d})\times\mathcal{C}^1([0,T];\mathbb{R}^{d\times d_0})\times\mathcal{C}^1([0,T];\mathbb{R}^{d_0})\times\mathcal{C}^1([0,T];\mathbb{R}^{d})\times\mathcal{C}^1([0,T];\mathbb{R})$. 
    Then, for each $i\in\mathbb{N}$, the process $(Y^i,Z^{i,0},Z^i)\in\mathbb{S}^2(\mathbb{P}^{0,i},\mathbb{F}^{0,i},\mathbb{R}) \times \mathbb{S}^2(\mathbb{P}^{0,i},\mathbb{F}^{0,i},\mathbb{R}^{1\times d_0}) \times \mathbb{S}^2(\mathbb{P}^{0,i},\mathbb{F}^{0,i},\mathbb{R}^{1\times d})$, defined by
    \begin{equation}
        \begin{split}
            \label{EQG solution YZ}
            &Y^i_t  = \frac{1}{2} \langle A_{00}(t)x^0_t,x^0_t\rangle + \frac{1}{2} \langle A_{11}(t)x^i_t,x^i_t\rangle + \langle A_{10}(t)x^0_t,x^i_t\rangle + \langle B_0(t),x^0_t\rangle + \langle B_1(t),x^i_t\rangle + C(t), \\
            &Z_t^{i,0} = \Bigl\{\Sigma_0^\top (A_{00}(t)x^0_t + A_{10}(t)^\top x^i_t + B_0(t))\Bigr\}^\top,~~~Z_t^i = \Bigl\{\Sigma^\top (A_{10}(t)x^0_t + A_{11}(t)x^i_t + B_1(t))\Bigr\}^\top,
        \end{split}
    \end{equation}
    for $t\in[0,T]$ solves the mean field BSDE \eqref{EQG-mfBSDE}. The solution is unique among those with the quadratic Gaussian form.
\end{thm}

\subsection{Optimality, verification and asymptotic equilibrium}
Let $(Y_t^i,Z_t^{i,0},Z_t^i;t\in[0,T])_{i\in\mathbb{N}}$ be processes defined by \eqref{EQG solution YZ} and suppose that they are well-defined. Then the process $\vartheta$, defined by\footnote{Since $(Z^{i,0}_t;t\in[0,T])_{i\in\mathbb{N}}$ have the same distribution, we can, without loss of generality, choose $Z^{1,0}$ to define $\vartheta$.}
\begin{equation}
    \begin{split}
    \label{EQG risk premium}
    \vartheta_t 
    &:=
    -\gamma \mathbb{E}[Z_t^{1,0\|} |\mathcal{F}^0]^\top,~~~t\in[0,T],
    \end{split}
\end{equation}
is expected to be the market-clearing risk premium process in the large population limit in analogy with Section 3. However, we have
\begin{equation*}
    \begin{split}
    \vartheta_t 
    =
    -\gamma \hat{\Sigma}_0^\top \Bigl(A_{00}(t)x^0_t + A_{10}(t)^\top \mu^1_t + B_0(t) \Bigr)
    =
    -\gamma \hat{\Sigma}_0^\top \Bigl(A_{00}(t)\mathbb{E}[x^0_t] + A_{10}(t)^\top\mu^1_t + B_0(t)\Bigr)-\gamma \hat{\Sigma}_0^\top A_{00}(t)\Sigma_0\int_0^t e^{-K_0(t-s)}dW^0_s
    \end{split}
\end{equation*}
for $t\in[0,T]$, which implies that $\vartheta$ is a Gaussian process and thus $\vartheta\notin\mathbb{H}^2_{\mathrm{BMO}}$. Furthermore, since $Y^i$ and $F^i$ are given by quadratic forms of $x^0$ and $x^i$, they are unbounded processes. 
Therefore, this EQG model does not fulfil the assumptions of Section 3. Despite this, if $|\mathrm{Var}(x^i_0)|$, $|\Sigma_0|$ and $|\Sigma|$ are small enough, we shall see that we can still obtain the well-posedness. Here, $\mathrm{Var}(x^i_0)$ is a covariance matrix of $x^i_0$, defined by $\mathrm{Var}(x^i_0):=\mathbb{E}[(x^i_0-\mathbb{E}[x^i_0])(x^i_0-\mathbb{E}[x^i_0])^\top]$.
The following result is well known.
\begin{lem}
    \label{crisan}
    Let $(\Omega,\mathcal{F},\mathbb{P},\mathbb{F}~(:=(\mathcal{F}_t)_{t\in[0,T]}))$ be a filtered probability space with usual conditions and $W:=(W_t)_{t\in[0,T]}$ be a standard $k$-dimensional $(\mathbb{F},\mathbb{P})$-Brownian motion. Also, let $\mathscr{X}$ be an $m$-dimensional $\mathbb{F}$-adapted process defined by
    \[
        \mathscr{X}_t = \mathscr{X}_0 + \int_0^t B(\mathscr{X}_s)ds + \int_0^t \Xi (\mathscr{X}_s)dW_s,~~t\in[0,T],
    \]
    where $B:\mathbb{R}^{m}\to\mathbb{R}^{m}$ and $\Xi :\mathbb{R}^{m}\to\mathbb{R}^{m\times k}$ are Lipschitz continuous functions and $\mathscr{X}_0\in\mathbb{L}^2(\mathbb{P},\mathcal{F}_0,\mathbb{R}^m)$. Moreover, let $h:\mathbb{R}^{m}\to\mathbb{R}^{k}$ be a Borel-measurable function satisfying $|h(x)|^2 \leq C(1+|x|^2)$ for all $x\in\mathbb{R}^{m}$ and some constant $C>0$. Then, the Dol\'{e}ans-Dade exponential $\displaystyle\Bigl\{\mathcal{E}\Bigl(\int_0^\cdot h(\mathscr{X}_s)^\top dW_s\Bigr)_t;t\in[0,T]\Bigr\}$ is a martingale of class $\mathcal{D}$.
\end{lem}
\noindent
\textbf{\textit{proof}}\\
See Bain \& Crisan \cite{bain_fundamentals_2009} [Exercise 3.11]. $\square$

\begin{prop}
    Let Assumption \ref{asmEQG-market} and \ref{asm6} be in force. In addition, assume that the equation \eqref{Riccati eqn} has a global solution $(A_{00},A_{11},A_{10},B_0,B_1,C)\in\mathcal{C}^1([0,T];\mathbb{M}_{d_0})\times\mathcal{C}^1([0,T];\mathbb{M}_{d})\times\mathcal{C}^1([0,T];\mathbb{R}^{d\times d_0})\times\mathcal{C}^1([0,T];\mathbb{R}^{d_0})\times\mathcal{C}^1([0,T];\mathbb{R}^{d})\times\mathcal{C}^1([0,T];\mathbb{R})$. 
    Then, the Dol\'{e}ans-Dade exponential $\displaystyle\Bigl\{\mathcal{E}\Bigl(-\int_0^\cdot \vartheta_s^\top dW_s^0\Bigr)_t;t\in[0,T]\Bigr\}$ is a martingale of class $\mathcal{D}$, where the process $\vartheta\in\mathbb{L}^0(\mathbb{F}^0,\mathbb{R}^{d_0})$ is defined by \eqref{EQG risk premium}.
\end{prop}
\noindent
\textbf{\textit{proof}}\\
This is a direct result of Lemma \ref{crisan}. $\square$

This proposition particularly shows that the process $\vartheta$ is consistent with Assumption \ref{asmEQG-market} as a risk premium process. With these preparations, we can recover the corresponding result of Section 2.
\begin{thm} (Optimality and verification)\\
    \label{EQG-verification}
    Let Assumption \ref{asmEQG-market} and \ref{asm6} be in force. Assume further that the equation \eqref{Riccati eqn} has a global solution $(A_{00},A_{11},A_{10},B_0,B_1,C)\in\mathcal{C}^1([0,T];\mathbb{M}_{d_0})\times\mathcal{C}^1([0,T];\mathbb{M}_{d})\times\mathcal{C}^1([0,T];\mathbb{R}^{d\times d_0})\times\mathcal{C}^1([0,T];\mathbb{R}^{d_0})\times\mathcal{C}^1([0,T];\mathbb{R}^{d})\times\mathcal{C}^1([0,T];\mathbb{R})$. 
    Then, there exists a constant $\varsigma > 0$ such that, as long as $|\Sigma_0|^2\lor |\Sigma|^2 \lor |\mathrm{Var}(x^1_0)| < \varsigma$, the process $(p^{i,*},c^{i,*})$, defined by
    \begin{equation}
        \begin{split}
            \label{EQG-optimal}
            p^{i,*}_t &:= (\pi^{i,*}_t)^\top\sigma_t := Z^{i,0\|}_t + \frac{\vartheta_t^\top}{\gamma},~~~t\in[0,T], \\
            c^{i,*}_t &= X^{i,c^*}_t + \frac{1}{\beta}\Bigl\{\log\Bigl(\frac{a\beta}{\gamma(1+b\zeta_t)}\Bigr)-\gamma(Y^i_t-F^i_t+\zeta_tX^{i,c^*}_t)\Bigr\},~~~t\in[0,T],
        \end{split}
    \end{equation}
    belongs to $\mathcal{A}^i_{\mathrm{EQG}}(\vartheta)$ and is an optimal strategy for agent-$i$ for each $i\in\mathbb{N}$. 
    Here, the process $X^{i,c}$ represents the agent's consumption habit \eqref{EQG-habit}, the process $(Y^i,Z^{i,0},Z^i)\in\mathbb{S}^2(\mathbb{P}^{0,i},\mathbb{F}^{0,i},\mathbb{R}) \times \mathbb{S}^2(\mathbb{P}^{0,i},\mathbb{F}^{0,i},\mathbb{R}^{1\times d_0}) \times \mathbb{S}^2(\mathbb{P}^{0,i},\mathbb{F}^{0,i},\mathbb{R}^{1\times d})$ is given by \eqref{EQG solution YZ} and the market risk premium process $\vartheta$ is defined by \eqref{EQG risk premium}.
\end{thm}
\begin{rem}
    Note that the strategy $(p^{i,*},c^{i,*})$ given above may not be the unique optimal strategy for agent-$i$ under the risk premium process $\vartheta$. This is because the BSDE \eqref{std BSDE} may have a solution outside of the quadratic Gaussian form.
\end{rem}
\noindent
\textbf{\textit{proof}}\\
In this proof, we denote the general nonnegative constant by $\widetilde{C}$ to avoid confusion with the function $C$, which is a part of the solution to the ODE \eqref{Riccati eqn}. By the definition of the process $R^{i,(p,c)}$ and the argument in the proof of Theorem \ref{verification}, the process $R^{i,(p^*,c^*)}$ is a local martingale and thus the optimality follows once $(p^{i,*},c^{i,*})\in\mathcal{A}^i_{\mathrm{EQG}}(\vartheta)$ is achieved. 
It then suffices to show that the function $\widetilde{U}^i(p^{i,*},c^{i,*})$ defined in \eqref{tilde-U} has a finite value and that the process $R^{i,(p^*,c^*)}$ is of class $\mathcal{D}$.

Let us write
\begin{equation*}
    \begin{split}
        \phi^0_t := \int_0^t e^{-K_0(t-s)}dW^0_s,~~~\phi^i_t := \int_0^t e^{-K(t-s)}dW^i_s,~~t\in[0,T].
    \end{split}
\end{equation*}
Then we have
\begin{equation*}
    \begin{split}
    x^0_t =  x_0^0e^{-K_0 t}+m_0(1-e^{-K_0 t}) + \Sigma_0 \phi^0_t,~~~x^i_t =  x^i_0e^{-Kt} + m(1-e^{-Kt}) + \Sigma \phi^i_t,~~t\in[0,T],
    \end{split}
\end{equation*}
and in particular, $|x^0_t|^2 + |x^i_t|^2 \leq \widetilde{C}(1+|x^i_0|^2+|\Sigma_0|^2|\phi^0_t|^2+|\Sigma|^2|\phi^i_t|^2)$ for all $t\in[0,T]$. We shall show that there exists a constant $\eta>0$ such that
\begin{equation}
    \label{u.i.-EQG}
    \sup_{t\in[0,T]}\mathbb{E}\Bigl[\exp\Bigl(-(1+\eta)\gamma \mathcal{W}^{i,(p^*,c^*)}_t + (1+\eta)M \Bigl\{|F^i_t| + |Y^i_t| + |c^{i,*}_t| + |X^{i,c^*}_t|\Bigr\}\Bigr)\Bigr] <\infty,
\end{equation}
where $M$ is a constant satisfying $M\geq \max\{\gamma, \beta, \sup_{t\in[0,T]}|\gamma\zeta_t|\}$. If this is the case, we clearly have $\widetilde{U}^i(p^{i,*},c^{i,*})>-\infty$. 
Moreover, Jensen's inequality and Doob's submartingale inequality yields
\[
    \mathbb{E}\Bigl[\sup_{t\in[0,T]}|R^{i,(p^*,c^*)}_t|\Bigr]^{1+\eta}\leq \mathbb{E}\Bigl[\sup_{t\in[0,T]}|R^{i,(p^*,c^*)}_t|^{1+\eta}\Bigr]\leq \tilde{C}\mathbb{E}\Bigl[|R^{i,(p^*,c^*)}_T|^{1+\eta}\Bigr]= \tilde{C}\sup_{t\in[0,T]}\mathbb{E}\Bigl[|R^{i,(p^*,c^*)}_t|^{1+\eta}\Bigr]<\infty.
\]
This implies that the process $R^{i,(p^*,c^*)}$ is dominated by an integrable random variable $\sup_{t\in[0,T]}|R^{i,(p^*,c^*)}_t|$. Using Medvegyev \cite{Medvegyev} [Corollary 1.145], we deduce that $R^{i,(p^*,c^*)}$ is a martingale of class $\mathcal{D}$.

Without loss of generality, we set $i=1$ and omit the superscript ``1'' when obvious. As $(A_{00},A_{11},A_{10},B_0,B_1,C)$ is a global solution and hence is bounded, we have, from \eqref{EQG-F} and \eqref{EQG solution YZ},
\begin{equation}
    \begin{split}
    \label{estimate for p}
    |\vartheta_t|\leq \widetilde{C}(1+|x^0_t|),~~~ |p^{*}_t| \leq \widetilde{C}(1+|x^0_t| + |x^1_t|),~~~ |Y_t| \leq \widetilde{C}(1 + |x^0_t|^2 + |x^1_t|^2),~~~ |F_t| \leq \widetilde{C}(1 + |x^0_t|^2 + |x^1_t|^2)
    \end{split}
\end{equation}
for all $t\in[0,T]$. Moreover, by Gronwall's inequality, we have $\displaystyle |X^{c^*}_t| \leq  |X_0| + \widetilde{C} + \widetilde{C}\int_0^t |c^{*}_s| ds$. Using this, we get
\begin{equation*}
    \begin{split}
    |c^{*}_t|\leq \widetilde{C}(1 + |X^{c^*}_t| + |Y_t| + |F_t|) \leq  \widetilde{C}(1 + |X_0| + |Y_t| + |F_t|) + \widetilde{C}\int_0^t |c^{*}_s| ds.
\end{split}
\end{equation*}
and then $|c^{*}_t| \leq  \widetilde{C}(1 + |X_0| + |Y_t| + |F_t|)$, again, by Gronwall's inequality. Together with \eqref{estimate for p}, we obtain
\begin{equation*}
    \begin{split}
    &|c^{*}_t| \leq  \widetilde{C}(1 + |X_0| + |x^0_t|^2 + |x^1_t|^2), \\
    &|X^{c^*}_t| \leq  |X_0| + \widetilde{C} + \widetilde{C}\int_0^t |c^{*}_s| ds  \leq \widetilde{C}\Bigl(1+ |X_0| + \int_0^t (|x^0_s|^2 + |x^1_s|^2) ds\Bigr),
\end{split}
\end{equation*}
for each $t\in[0,T]$.
Using these estimates, we have
\begin{equation*}
    \begin{split}
        \mathcal{W}^{(p^*,c^*)}_t
        &=
        \xi + \int_0^t (p^{*}_s\vartheta_s - c^{*}_s) ds  + \int_0^t p^{*}_s dW^0_s\\
        &\geq
        - |\xi| - \int_0^t (|p^{*}_s||\vartheta_s| + |c^{*}_s|) ds - \gamma(1+\eta)\int_0^t |p^{*}_s|^2 ds + \gamma(1+\eta)\int_0^t |p^{*}_s|^2 ds + \int_0^t p^{*}_s dW^0_s \\
        &\geq
        - \widetilde{C} (1+|\xi|+|X_0|) - \widetilde{C}\int_0^t (|x^0_s|^2 + |x^1_s|^2) ds + \gamma(1+\eta)\int_0^t |p^{*}_s|^2 ds +  \int_0^t p^{*}_s dW^0_s,~~t\in[0,T].
    \end{split}
\end{equation*}
Putting these together, it follows that, for all $t\in[0,T]$,
\begin{equation*}
    \begin{split}
        &-\gamma \mathcal{W}^{(p^*,c^*)}_t + M (|F_t| + |Y_t| + |c^{*}_t| + |X^{c^*}_t|)\\
        &\leq
        \widetilde{C} (1+|\xi|+ |X_0|) + \widetilde{C}(|x^0_t|^2 + |x^1_t|^2) + \widetilde{C}\int_0^t (|x^0_t|^2 + |x^1_t|^2) ds - \gamma^2(1+\eta)\int_0^t |p^{*}_s|^2 ds -\gamma  \int_0^t p^{*}_s dW^0_s \\
        &\leq
        \widetilde{C} (1+|\xi|+ |X_0|) + \widetilde{C}(|\Sigma_0|^2|\phi^0_t|^2 + |\Sigma|^2|\phi^1_t|^2 + |x^1_0|^2) + \widetilde{C}\int_0^t (|\Sigma_0|^2|\phi^0_s|^2 + |\Sigma|^2|\phi^1_s|^2 + |x^1_0|^2) ds - \gamma^2(1+\eta)\int_0^t |p^{*}_s|^2 ds -\gamma  \int_0^t p^{*}_s dW^0_s.
    \end{split}
\end{equation*}
Then, for all $t\in[0,T]$,
\begin{equation*}
    \begin{split}
        &\mathbb{E}\Bigl[\exp\Bigl(-(1+\eta)\gamma \mathcal{W}^{(p^*,c^*)}_t + (1+\eta)M (|F_t| + |Y_t| + |c^{*}_t| + |X^{c^*}_t|)\Bigr)\Bigr]\\
        &\leq
        \widetilde{C}\mathbb{E}\Bigl[\exp\Bigl(\widetilde{C}(|\xi| + |X_0|) + \widetilde{C}\Bigl((|\Sigma_0|^2|\phi^0_t|^2 + |\Sigma|^2|\phi^1_t|^2 + |x^1_0|^2) + \int_0^T (|\Sigma_0|^2|\phi^0_s|^2 + |\Sigma|^2|\phi^1_s|^2 + |x^1_0|^2) ds\Bigr)\Bigr.\Bigr.\\
        &~~~~~~~~~~~~~~~~~~~~~~\Bigl.\Bigl. - \gamma^2(1+\eta)^2\int_0^t |p^{*}_s|^2 ds -\gamma (1+\eta) \int_0^t p^{*}_s dW^0_s \Bigr)\Bigr]\\
        &\leq
        \widetilde{C}\mathbb{E}\Bigl[\exp\Bigl(\widetilde{C}(|\xi| + |X_0|) + \widetilde{C}(|\Sigma_0|^2|\phi^0_t|^2 + |\Sigma|^2|\phi^1_t|^2 + |x^1_0|^2)\Bigr)\Bigr]^{\frac{1}{4}}\\
        &~~~~~~~~~~~~~~~~~~\times \mathbb{E}\Bigl[\exp\Bigl(\widetilde{C}\int_0^T (|\Sigma_0|^2|\phi^0_s|^2 + |\Sigma|^2|\phi^1_s|^2 + |x^1_0|^2)ds \Bigr)\Bigr]^{\frac{1}{4}} \mathbb{E}\Bigl[\mathcal{E}\Bigl(-2\gamma(1+\eta)\int_0^\cdot p^{*}_s dW^0_s\Bigr)_t\Bigr]^{\frac{1}{2}}\\
        &=
        \widetilde{C}\mathbb{E}\Bigl[\exp\Bigl(\widetilde{C}(|\xi| + |X_0|)\Bigr)\Bigr]^{\frac{1}{4}}\mathbb{E}\Bigl[\exp\Bigl(\widetilde{C}(|\Sigma_0|^2|\phi^0_t|^2 + |\Sigma|^2|\phi^1_t|^2 + |x^1_0|^2)\Bigr)\Bigr]^{\frac{1}{4}}\\
        &~~~~~~~~~~~~~~~~~~\times \mathbb{E}\Bigl[\exp\Bigl(\widetilde{C}\int_0^T (|\Sigma_0|^2|\phi^0_s|^2 + |\Sigma|^2|\phi^1_s|^2 + |x^1_0|^2)ds \Bigr)\Bigr]^{\frac{1}{4}} \mathbb{E}\Bigl[\mathcal{E}\Bigl(-2\gamma(1+\eta)\int_0^\cdot p^{*}_s dW^0_s\Bigr)_t\Bigr]^{\frac{1}{2}}
    \end{split}
\end{equation*}
by using Holder's inequality.\par 
As $\xi$ and $X_0$ are independent and normally distributed, we have $\mathbb{E}\Bigl[\exp\Bigl(\widetilde{C}(|\xi| + |X_0|)\Bigr)\Bigr] = \mathbb{E}\Bigl[e^{\widetilde{C}|\xi|}\Bigr]\mathbb{E}\Bigl[e^{\widetilde{C}|X_0|}\Bigr] <\infty$.
By Lemma \ref{crisan} and \eqref{estimate for p}, we deduce
\[
    \sup_{t\in[0,T]}\mathbb{E}\Bigl[\mathcal{E}\Bigl(-2\gamma(1+\eta)\int_0^\cdot p^{*}_s dW^0_s\Bigr)_t\Bigr] < \infty.
\]
Furthermore, since the random variables $\phi^0_t$, $\phi^1_t$ and $x^1_0$ are mutually independent, we have
\begin{equation*}
    \begin{split}
        \mathbb{E}\Bigl[\exp\Bigl(\widetilde{C}\Bigl\{(|\Sigma_0|^2|\phi^0_t|^2 + |\Sigma|^2|\phi^1_t|^2 + |x^1_0|^2)\Bigr\}\Bigr)\Bigr]
        &=
        \mathbb{E}\Bigl[\exp\Bigl(\widetilde{C}|\Sigma_0|^2|\phi^0_t|^2\Bigr)\Bigr]\mathbb{E}\Bigl[\exp\Bigl(\widetilde{C}|\Sigma|^2|\phi^1_t|^2\Bigr)\Bigr]\mathbb{E}\Bigl[\exp\Bigl(\widetilde{C}|x^1_0|^2\Bigr)\Bigr]\\
        &\leq
        \widetilde{C}\mathbb{E}\Bigl[\exp\Bigl(\widetilde{C}|\Sigma_0|^2v^0_t Z^2\Bigr)\Bigr]\mathbb{E}\Bigl[\exp\Bigl(\widetilde{C}|\Sigma|^2v^1_t Z^2\Bigr)\Bigr]\mathbb{E}\Bigl[\exp\Bigl(\widetilde{C}|\mathrm{Var}(x^1_0)|Z^2\Bigr)\Bigr],
    \end{split}
\end{equation*}
where $Z\sim N(0,1)$ and 
\begin{equation*}
    \begin{split}
        v^0_t := \int_0^t e^{-2K_0(t-s)} ds = \frac{1}{2K_0}(1-e^{-2K_0 t}) < \frac{1}{2K_0},~~~v^1_t := \int_0^t e^{-2K(t-s)} ds = \frac{1}{2K}(1-e^{-2K t}) < \frac{1}{2K}
    \end{split}
\end{equation*}
for $t\in[0,T]$. Therefore, we have
\[
    \mathbb{E}\Bigl[\exp\Bigl(\widetilde{C}\Bigl\{(|\Sigma_0|^2|\phi^0_t|^2 + |\Sigma|^2|\phi^1_t|^2 + |x^1_0|^2)\Bigr\}\Bigr)\Bigr]<\infty
\]
if and only if 
\[
    \widetilde{C}(|\Sigma_0|^2v^0_t\lor |\Sigma|^2v^1_t \lor |\mathrm{Var}(x^1_0)|)<\frac{1}{2}.
\]
Similarly, we have
\[
    \mathbb{E}\Bigl[\exp\Bigl(\widetilde{C}\int_0^T (|\Sigma_0|^2|\phi^0_s|^2 + |\Sigma|^2|\phi^1_s|^2 + |x^1_0|^2)ds \Bigr)\Bigr] < \infty
\]
if and only if
\[
    \widetilde{C}\Bigl(|\Sigma_0|^2\int_0^T v^0_tdt \lor |\Sigma|^2\int_0^T v^1_tdt \lor |\mathrm{Var}(x^1_0)|T\Bigr)<\frac{1}{2}.
\]

Above all, if
\begin{equation}
    \label{var-sigma}
    |\Sigma_0|^2\lor |\Sigma|^2 \lor |\mathrm{Var}(x^1_0)| < \widetilde{C}^{-1}(1\land T^{-1})(K_0\land K \land 2^{-1}) =: \varsigma
\end{equation}
holds, we get \eqref{u.i.-EQG}, which implies $(p^{*},c^{*})\in\mathcal{A}^1_{\mathrm{EQG}}(\vartheta)$. $\square$

\begin{rem}
    If the quadratic form of $(\phi^0,\phi^i)$ in the exponential function of $R^{i,(p^*,c^*)}$ happens to be negative semidefinite, we need no constraints on the diffusion coefficients.
\end{rem}

This result also recovers the corresponding asymptotic properties of Theorem \ref{Asymp}. To be specific, the process $\vartheta$ satisfies the market clearing condition in the large population limit.

\begin{thm} (Asymptotic equilibrium in the EQG model)\\
    Let Assumption \ref{asmEQG-market} and \ref{asm6} be in force. Assume further that the equation \eqref{Riccati eqn} has a global solution $(A_{00},A_{11},A_{10},B_0,B_1,C)\in\mathcal{C}^1([0,T];\mathbb{M}_{d_0})\times\mathcal{C}^1([0,T];\mathbb{M}_{d})\times\mathcal{C}^1([0,T];\mathbb{R}^{d\times d_0})\times\mathcal{C}^1([0,T];\mathbb{R}^{d_0})\times\mathcal{C}^1([0,T];\mathbb{R}^{d})\times\mathcal{C}^1([0,T];\mathbb{R})$ and that 
    \[
        |\Sigma_0|^2\lor |\Sigma|^2 \lor |\mathrm{Var}(x^1_0)| < \varsigma,
    \]
    where $\varsigma>0$ is a positive constant specified in \eqref{var-sigma}.
    Then, as long as each agent adopts \eqref{EQG-optimal} as his/her optimal strategy, the process $\vartheta$ defined by \eqref{EQG risk premium} clears the financial market in large population limit, i.e. the agents' optimal trading strategies $(\pi^{i,*}_t;t\in[0,T])_{i\in\mathbb{N}}$ given by \eqref{EQG-optimal} satisfy
    \begin{equation}
        \lim_{N\to\infty}\mathbb{E}\int_0^T \Bigl|\frac{1}{N}\sum_{i=1}^N \pi^{i,*}_t\Bigr|^2dt = 0.
    \end{equation}
\end{thm}
\begin{rem}
    Since the optimal strategy (under market risk premium process $\vartheta$) is not shown to be unique, this statement is only valid for the specific choice of optimal strategies \eqref{EQG-optimal}.
\end{rem}
\noindent
\textbf{\textit{proof}}\\
Again, we denote the general constant by $\widetilde{C}$ to avoid misinterpretation with the function $C$. Since $\pi^{i,*}_t=(\sigma_t\sigma_t^\top)^{-1}\sigma_t (p^{i,*}_t)^\top$ for $t\in[0,T]$ and $|(\sigma_t\sigma_t^\top)^{-1}\sigma_t|\leq \widetilde{C}$ uniformly in $t$ by Assumption \ref{asmEQG-market}, we have
\begin{equation*}
    \begin{split}
        \mathbb{E}\int_0^T \Bigl|\frac{1}{N}\sum_{i=1}^N \pi^{i,*}_t\Bigr|^2dt
        &\leq
        \widetilde{C}\mathbb{E}\int_0^T \Bigl|\frac{1}{N}\sum_{i=1}^N p^{i,*}_t\Bigr|^2dt.
    \end{split}
\end{equation*}
By \eqref{EQG solution YZ}, the process $p^{i,*}$ can be written as
\begin{equation*}
    \begin{split}
        p^{i,*}_t
        :=
        Z^{i,0\|}_t + \frac{\vartheta_t^\top}{\gamma}
        =
        Z^{i,0\|}_t - \mathbb{E}[Z_t^{1,0\|} |\mathcal{F}^0]
        =
        (x^i_t-\mu^1_t)^\top A_{10}(t) \hat\Sigma_0
    \end{split}
\end{equation*}
for every $i\in\mathbb{N}$ and $t\in[0,T]$. It is then easy to see
\begin{equation*}
    \begin{split}
       \mathbb{E}\int_0^T \Bigl|\frac{1}{N}\sum_{i=1}^N \pi^{i,*}_t\Bigr|^2dt
       \leq
       \widetilde{C}\mathbb{E}\int_0^T \Bigl|\frac{1}{N}\sum_{i=1}^N \Bigl(x^i_t-\mu^1_t\Bigr)\Bigr|^2 dt 
       \leq
       \frac{\widetilde{C}}{N^2}\sum_{i=1}^N\mathbb{E}\int_0^T \Bigl|x^i_t-\mu^1_t\Bigr|^2 dt 
       \leq
       \frac{\widetilde{C}}{N}
       \to
       0~~~~(N\to\infty),
    \end{split}
\end{equation*}
where we have used the fact that $(x^i_t;t\in[0,T])_{i\in\mathbb{N}}$ are mutually independent and that $\mathbb{E}[x^i_t]=\mu^1_t$ for every $i\in\mathbb{N}$ and $t\in[0,T]$. $\square$

\section{Conclusion and discussions} \label{Section 5}
In this paper, we have studied a theoretical model of asset pricing among heterogeneous agents with habit formation in consumption preferences using the mean field game theory. In Section 3, the market clearing condition has motivated us to study the mean field BSDE \eqref{MF-BSDE1}, which 
was shown to have a bounded solution under additional assumptions on the size of the parameters. Furthermore, we have proved that the solution of this equation does indeed characterize the market equilibrium in the large population limit.
In addition, Section 4 introduced an exponential Gaussian model, in which an unbounded solution of the mean field BSDE can be obtained in a semi-analytic form, characterized by a system of ODEs, under appropriate assumptions. Subsequently, we have verified an optimal strategy and the asymptotic market equilibrium in the large population limit within the EQG framework. 

    The result of Section 4 helps us to conduct a numerical analysis in future studies. The solutions of \eqref{Riccati eqn} can be calculated by the standard Euler method. A solution to the mean field BSDE can be then obtained by \eqref{EQG solution YZ} with pathwise simulations. 
The numerical analysis can provide a visualization of our equilibrium model, allowing us to investigate the distribution of wealth and the effect of the habit formation. We can also expect an application to the empirical study using the market data.
See also Carmona \& Delarue \cite{carmonaProbabilisticTheoryMean2018} [Section 3.5] for linear quadratic mean field games and [Section 3.6] for numerical results.

    There still remain some possible extensions. We may possibly impose some additional conditions on the habit trend $\rho$, such as $\rho_t=\mathbb{E}[c^{1,*}_t|\mathcal{F}^0]$ for $t\in[0,T]$, so that the model allows us to investigate the consumption behavior under the relative performance criteria. We may also generalize the information structure or add a jump process into the security price process for describing the possibility of default, for example.
Furthermore, as noted in the previous work Fujii \& Sekine \cite{fujiiMeanFieldEquilibriumPrice2023a}, the general solvability of the mean field BSDE \eqref{MF-BSDE1} still remains open. As long as working within our framework, such an equation is likely to appear in possibly more generalized forms.

\section*{Acknowledgement}
This research is supported by Grant-in-Aid for JSPS Research Fellows, Grant Number JP23KJ0648.

\appendix
\section*{Appendix}
\section{Proof of Theorem \ref{sec2-well-posed} (Continued)}

We introduce a smooth convex function $\phi:\mathbb{R}\to\mathbb{R}_+$ satisfying $\phi(0)=\phi'(0)=0$, whose concrete form is to be determined. Let us denote $\Delta Y^{n,m}:=Y^n-Y^m$ and $\Delta Z^{n,m,i}:=Z^{n,i}-Z^{m,i}$ for $i=0,1$. Notice that $\Delta Y^{n,m}\geq 0 $ when $m\geq n$. 

Using Ito formula, we have
\begin{equation*}
    \begin{split}
        &\mathbb{E}^{\mathbb{Q}}[\phi(\Delta Y^{n,m}_0)] + \frac{1}{2}\mathbb{E}^{\mathbb{Q}}\Bigl[\int_0^T \phi''(\Delta Y^{n,m}_s)(|\Delta Z^{n,m,0}_s|^2 + |\Delta Z^{n,m,1}_s|^2) ds  \Bigr] \\
        &=
        \mathbb{E}^{\mathbb{Q}}\Bigl[\int_0^T \phi'(\Delta Y^{n,m}_s)\Bigl\{- \frac{|\theta_s|^2\land n}{2\gamma} + \frac{|\theta_s|^2\land m}{2\gamma} + \frac{\gamma}{2}(|Z^{n,0\perp}_s|^2 + |Z^{n,1}_s|^2) - \frac{\gamma}{2}(|Z^{m,0\perp}_s|^2 + |Z^{m,1}_s|^2) -\frac{\gamma(1+b\zeta_s)}{\beta}\Delta Y^{n,m}_s \Bigr\} ds  \Bigr]\\
        &\leq
        \mathbb{E}^{\mathbb{Q}}\Bigl[\int_0^T \phi'(\Delta Y^{n,m}_s)\Bigl\{\frac{|\theta_s|^2}{2\gamma} + \frac{\gamma}{2}(|Z^{n,0}_s|^2 + |Z^{n,1}_s|^2)\Bigr\} ds  \Bigr]\\
        &\leq
        \mathbb{E}^{\mathbb{Q}}\Bigl[\int_0^T C_0\phi'(\Delta Y^{n,m}_s)\Bigl\{|\theta_s|^2 + | Z^{n,0}_s - Z^{0}_s|^2 + |Z^{n,1}_s-Z^{1}_s|^2 + |Z^{0}_s|^2 + |Z^{1}_s|^2\Bigr\} ds  \Bigr],
    \end{split}
  \end{equation*}
where $C_0$ is a positive constant satisfying $C_0\geq \dfrac{1}{2}(\underline{\gamma}^{-1}+2\overline{\gamma})$. Then,
  \begin{equation*}
    \begin{split}
        \frac{1}{2}\mathbb{E}^{\mathbb{Q}}\Bigl[\int_0^T \phi''(\Delta Y^{n,m}_s)(|\Delta Z^{n,m,0}_s|^2 + |\Delta Z^{n,m,1}_s|^2) ds  \Bigr]
        \leq
        \mathbb{E}^{\mathbb{Q}}\Bigl[\int_0^T C_0\phi'(\Delta Y^{n,m}_s)\Bigl\{|\theta_s|^2 + | Z^{n,0}_s - Z^{0}_s|^2 + |Z^{n,1}_s-Z^{1}_s|^2 + |Z^{0}_s|^2 + |Z^{1}_s|^2\Bigr\} ds  \Bigr].
    \end{split}
  \end{equation*}
Now set $\phi$ as
  \begin{equation}
    \begin{split}
      \label{phi}
      \phi(y):=\frac{1}{2C_0^2}(e^{2C_0 y}-2C_0 y-1),
        \end{split}
  \end{equation}
  then $\phi(0)=\phi'(0)=0$ and
  \begin{equation*}
    \begin{split}
      \phi'(y)=\frac{1}{C_0}(e^{2C_0 y}-1),~~~\phi''(y)=2e^{2C_0 y}.
      \end{split}
  \end{equation*}
In particular, $\phi''(y)=2C_0\phi'(y) + 2$. With these relations, we get
  \begin{equation*}
    \begin{split}
        &\mathbb{E}^{\mathbb{Q}}\Bigl[\int_0^T (C_0\phi'(\Delta Y^{n,m}_s) + 1)(|\Delta Z^{n,m,0}_s|^2+|\Delta Z^{n,m,1}_s|^2) ds  \Bigr]\\
        &\leq
        \mathbb{E}^{\mathbb{Q}}\Bigl[\int_0^T C_0\phi'(\Delta Y^{n,m}_s)\Bigl\{|\theta_s|^2 + | Z^{n,0}_s - Z^{0}_s|^2 + |Z^{n,1}_s-Z^{1}_s|^2 + |Z^{0}_s|^2 + |Z^{1}_s|^2\Bigr\}  ds  \Bigr].
    \end{split}
\end{equation*}
Since $\sqrt{\phi'(\Delta Y^{n,m})+1}\Delta Z^{n,m,i}$ ($i=0,1$) is weakly convergent to $\sqrt{\phi'(Y^{n}-Y)+1}\Delta Z^{n,i}_s$ in $\mathbb{H}^2$ as $m\to\infty$, we obtain
  \begin{equation*}
    \begin{split}
        &\mathbb{E}^{\mathbb{Q}}\Bigl[\int_0^T (C_0\phi'( Y^{n}_s - Y_s) + 1)(|Z^{n,0}_s-Z^0_s|^2 + |Z^{n,1}_s-Z^1_s|^2) ds  \Bigr] \\
        &\leq
        \liminf_{m\to\infty}\mathbb{E}^{\mathbb{Q}}\Bigl[\int_0^T (C_0\phi'(\Delta Y^{n,m}_s) + 1) (|\Delta Z^{n,m,0}_s|^2 +|\Delta Z^{n,m,1}_s|^2) ds  \Bigr]\\
        &\leq
        \mathbb{E}^{\mathbb{Q}}\Bigl[\int_0^T C_0\phi'(Y^{n}_s - Y_s)\Bigl\{|\theta_s|^2 + | Z^{n,0}_s - Z^{0}_s|^2 + |Z^{n,1}_s-Z^{1}_s|^2 + |Z^{0}_s|^2 + |Z^{1}_s|^2\Bigr\}  ds  \Bigr].
    \end{split}
  \end{equation*}
This implies
\begin{equation*}
    \begin{split}
        \mathbb{E}^{\mathbb{Q}}\Bigl[\int_0^T | Z^{n,0}_s-Z^0_s|^2 + | Z^{n,1}_s-Z^1_s|^2 ds  \Bigr]
        \leq
        \mathbb{E}^{\mathbb{Q}}\Bigl[\int_0^T C_0\phi'(Y^{n}_s - Y_s)\Bigl\{|\theta_s|^2 + |Z^{0}_s|^2 + |Z^{1}_s|^2\Bigr\}  ds  \Bigr]
        \to
        0~~~(n\to\infty)
    \end{split}
\end{equation*}
by the dominated convergence theorem. Thus,
\[
    Z^{n,0}\to Z^0,~~~~~Z^{n,1}\to Z^1,~~~~(n\to\infty)
\]
strongly in $\mathbb{H}^2$. Taking a subsequence if necessary, it is now straightforward to see, for $\mathbb{Q}$-a.s.,
\begin{equation*}
    \begin{split}
        \sup_{t\in[0,T]}|Y_t-Y^n_t| \to 0,~~~\sup_{t\in[0,T]}\Bigl|\int_0^t (Z^0_s-Z^{n,0}_s)dW^{0,\mathbb{Q}}_s\Bigr| + \sup_{t\in[0,T]}\Bigl|\int_0^t (Z^1_s-Z^{n,1}_s)dW^{1,\mathbb{Q}}_s\Bigr| \to 0
    \end{split}
\end{equation*}
as $n\to\infty$. Hence $(Y,Z^0,Z^1)\in\mathbb{S}^\infty\times\mathbb{H}^2_{\mathrm{BMO}}\times\mathbb{H}^2_{\mathrm{BMO}}$ is a bounded solution to the BSDE \eqref{qgBSDE-Q}. $\square$

% \bibliographystyle{abbrv}
% \bibliography{Habit_formation.bib}

\end{document}